\documentclass{JFM-FLM_Au}
\usepackage{placeins}

\lefttitle{M. Kim}
\righttitle{Journal of Fluid Mechanics}

\title{Linear stability of electroconvection in a viscoelastic electrolyte: polymeric-work pathways and eigenmode selection}

\author{Minyoung Kim}

\affiliation{Department of Radiology, Northwestern University, Chicago IL 60611, USA}

\corresau{Minyoung Kim, \email{minyoung@northwestern.edu}}
\begin{document}
\maketitle

\begin{abstract}
Electroconvective (EC) instability near ion-selective surfaces plays a key role in electrochemical transport beyond the diffusion-limited regime. While EC instability in Newtonian electrolytes has been extensively investigated, the influence of viscoelasticity on the instability mechanism remains less understood. In the present study, we investigate the viscoelastic modulation of EC instability using linear stability, eigenspectrum, and mechanical-work analyses, supported by asymptotic pathway ordering. The analysis reveals a non-monotonic dependence on the Weissenberg number \(Wi\), with the maximum growth rate and minimum critical voltage occurring near \(Wi\sim O(1)\). The mechanical-work balance separates electric, solvent-viscous, and polymeric contributions, and further decomposes the polymeric work into normal-stress-difference and shear-stress pathways according to their constitutive origins. For \(Wi\ll1\), the polymer stress recovers Newtonian viscous dissipation at leading order. As \(Wi\) approaches unity, the phase efficiencies of the Newtonian-like polymeric pathways decrease substantially, causing the negative polymeric work to collapse and the instability to intensify. At intermediate \(Wi\), an additional minimum in the critical voltage appears at low wavenumbers. Eigenspectrum analysis attributes this feature to the selection of a slowly propagating mode that retains substantial negative polymeric work. At large \(Wi\), this mode no longer determines the instability threshold, and the low-wavenumber minimum disappears. The normal-stress-difference and shear-stress work then nearly cancel, accompanied by stronger ESC-localized shear deformation and predominantly solvent-viscous dissipation. Thus, the non-monotonic response is governed by stress–deformation phase relations, solvent dissipation, and eigenmode selection rather than stress magnitude or a single polymeric contribution.
\end{abstract}


\section{Introduction}
\label{sec:headings}
When an electric potential is applied across an electrolyte adjacent to an ion-selective surface, ion transport is initially sustained by one-dimensional diffusion and electromigration. As the applied potential increases, ion depletion develops near the ion-selective surface and the current approaches the limiting-current regime. Under sufficiently strong electric fields, the space-charge region extends beyond the equilibrium electric double layer, leading to the formation of a nonequilibrium extended space-charge (ESC) layer where strong electric fields are localized~\citep{Rubinstein2010a,Rubinstein2010b,Mani2020}. Small perturbations of ion concentration and electric potential within the ESC can then trigger hydrodynamic instability, leading to electroconvection and enhanced ion transport.

Electroconvective (EC) instability near ion-selective surfaces has important consequences in electrodialysis~\citep{Strathmann2010,Kwak2013}, desalination~\citep{Kim2010,Kim2016}, electrochemical separations and concentrations~\citep{Kwak2011,Kwak2016,Han2016}, and energy-storage devices~\citep{Fleury1993,Huth1995}. The resulting enhancement of ion transport beyond the diffusion-limited regime produces the so-called overlimiting current~\citep{Nikonenko2010,Dydek2011}.   In membrane-based desalination systems, electroconvection can enhance ion transport and improve current efficiency~\citep{Kwak2013}, whereas in electrochemical energy-storage systems, uncontrolled EC may generate localized current hotspots that promote dendritic growth and electrochemical degradation~\citep{Huth1995,Zheng2021}. With the increasing global demand for clean water and renewable energy technologies, electrochemical systems for water treatment and energy storage have attracted growing attention as promising alternatives to conventional approaches~\citep{Alkhadra2022,Sim2024,Mani2020}. Understanding the onset and modulation of EC instability is therefore important for enhancing electrohydrodynamic transport and suppressing undesirable electrochemical instabilities according to the application.

Fundamental studies by~\citet{Rubinstein2000,Rubinstein2001} established the theoretical foundation of electroconvective instability driven by the nonequilibrium ESC layer near ion-selective surfaces. In these studies, the instability problem was formulated by considering perturbations only within the electroneutral bulk region, while the effect of the strong voltage drop across the ESC layer was incorporated indirectly through an electroosmotic slip condition imposed at the edge of the space charge layer. Consequently, the linear stability analysis was performed only in the electroneutral bulk using momentum and ion-transport equations that did not explicitly include charge-density and electric-field effects.

This framework, often referred to as the classical bulk analysis, successfully predicted the existence of a critical condition for instability onset and clarified the destabilizing role of electroosmotic slip in driving electroconvection. However, because the ESC and electric double layer (EDL) structures were not explicitly resolved, the analysis remained only qualitatively accurate. In particular, it underestimated the critical voltage for instability onset and exhibited the so-called short-wave catastrophe, in which the instability growth rate diverges at large wavenumbers~\citep{Li2021}. To overcome these limitations, subsequent studies performed full linear stability analyses where the entire physical domain, including the electroneutral bulk region, ESC layer, and EDL, was explicitly resolved~\citep{Li2021,Li2022,Zaltzman2007}. These studies demonstrated that high-wavenumber perturbations are dissipated within the ESC before they can propagate into the bulk region, which removes the short-wave catastrophe predicted by the bulk analysis. Moreover, the predicted critical conditions from the full analysis showed quantitative agreement with direct numerical simulations~\citep{Demekhin2013,Urtenov2013, Nikonenko2016}.

More recently,~\citet{Kim2026} further clarified the physical origin of EC instability through detailed analysis of perturbation eigenfunctions within the ESC layer. Unlike previous studies that primarily focused on perturbations in the electroneutral bulk region, their analysis demonstrated that the instability originates directly within the ESC layer itself. Localized perturbations inside the ESC reduce the local ion concentration, resulting in a stronger local electric field and a bulk-directed perturbation flow through the electric body force.
This ESC-localized electrohydrodynamic feedback mechanism was shown to govern instability onset across different ion valences, diffusivities, electrolyte concentrations, viscosities, and membrane types. 

Despite substantial progress in Newtonian electrolytes, the influence of viscoelasticity on electroconvective instability remains less understood. Previous studies have shown that polymer additives can suppress electroconvection through enhanced viscous and elastic stresses~\citep{Tikekar2018,Li2019,Sharma2023}. Direct numerical simulations have further demonstrated that viscoelastic stresses can significantly alter electrohydrodynamic flow structures, nonlinear transitions, and ion transport~\citep{Li2019}. 

Similar viscoelastic effects have been extensively studied in other hydrodynamic instabilities, including Rayleigh--B\'enard convection~\citep{Alves2016,Martinez-Mardones2000,Yao2025} and inertial and inertialess Poiseuille-flow instabilities \citep{Garg2018,Khalid2021a,Khalid2021b,Khalid2025}. In Rayleigh--B\'enard convection, in particular, polymer additives have been shown to either enhance or reduce heat transport and to substantially modify the underlying flow structures and stability characteristics, depending on the flow regime and polymer properties~\citep{Benzi2010,Benzi2012,Benzi2016,Dubief2020,Wei2012,Xie2015,Xu2025a,Xu2025b}. However, compared with these classical instability problems, viscoelastic electroconvection has received relatively limited attention because of its coupled multiphysics and multiscale nature involving ion transport, electric fields, polymeric stress, ESC dynamics, and bulk hydrodynamics. A key unresolved question is how viscoelasticity modifies electroconvective instability across different Weissenberg-number regimes, especially whether viscoelasticity acts solely as a stabilizing mechanism or can also enhance instability at finite $Wi$.

In this study, we investigate how viscoelasticity modifies the onset of
electroconvective instability across different Weissenberg-number
regimes. A full-domain linear stability analysis is performed for an
Oldroyd-B electrolyte, explicitly resolving the EDL, ESC, and
electroneutral bulk regions. The analysis examines the dependence of the
growth rate, critical voltage, selected wavenumber, perturbation
structure, and eigenspectrum on $Wi$. To quantify the mechanical role of
the polymer stress, we formulate a mechanical-work-rate balance that
separates the electric, solvent-viscous, and polymeric contributions. The
polymeric work is then decomposed into normal-stress-difference and
shear-stress contributions and their constitutive pathways. Asymptotic
ordering, spatial work distributions, regional integrals, and phase
relations are used to interpret these pathways. 

The paper is organized as follows. Section 2 presents the governing
equations, problem formulation, and linear stability framework. Section 3
examines the instability characteristics and underlying physical
mechanisms, including the ESC-localized instability origin,
mechanical-work-rate balance, constitutive pathway decomposition,
asymptotic interpretation, eigenspectrum structure, and robustness across
parameter ranges. Concluding remarks are given in \S 4.

\begin{figure}
\centering
  \centerline{\includegraphics[scale=0.8]{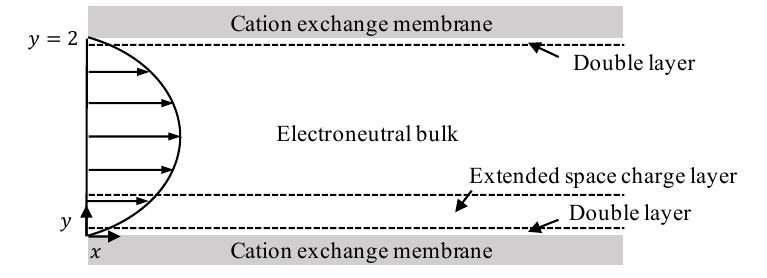}} 
  \caption{Schematic of the computational domain for electroconvective instability between two cation-exchange membranes under an applied electric field. A pressure-driven flow is imposed in the streamwise ($x$) direction. The thicknesses of the electric double layer and ESC layer are enlarged for clarity and are not drawn to scale.
}
\label{fig:fig1}
\end{figure}

\section{Methods}

\subsection{Problem formulation}

We consider a binary electrolyte confined between two parallel cation-selective surfaces located at $y=0$ and $y=2$, as illustrated in figure~\ref{fig:fig1}. A pressure-driven flow is imposed in the streamwise ($x$) direction, while an electric potential difference is applied across the channel in the wall-normal ($y$) direction.

Near the ion-selective surface at $y=0$, ion depletion occurs. Under a sufficiently strong applied potential, the space charge layer near the ion-selective surface extends into the bulk, forming an ESC layer in which a strong electric field is established. Within this region, the combination of low ion concentration and strong electric field generates a localized electric body force, which drives perturbation velocities directed away from the wall. As established in previous studies~\citep{Li2021,Kim2026}, the instability originates within the ESC and subsequently propagates into the bulk region. In the absence of an externally imposed magnetic field, the self-induced magnetic field and associated Lorentz force are neglected under the low-magnetic-Reynolds-number approximation, appropriate to electrolyte flows in microfluidic channels~\citep{Cai2023,Qin2011}.

The governing equations consist of a coupled system describing incompressible fluid flow, ionic transport, electric potential, and viscoelastic stress. The total stress includes both solvent and polymeric contributions, and the polymeric stress is modeled using the Oldroyd-B constitutive equation. The dimensionless governing equations are given by

\begin{equation}
0 = -\bnabla p
+ \beta \nabla^2 \boldsymbol{u}
+ \frac{1-\beta}{Wi} (\bnabla \bcdot \mathsfbi{A})
+ \frac{Ra_E}{Pe} \nabla^2 \phi \bnabla \phi,
\label{eq:fullmomentum}
\end{equation}

\begin{equation}
\frac{\partial c_{+}}{\partial t}
+ \boldsymbol{u} \bcdot \bnabla c_{+}
=
\frac{1}{Pe}
\left[
V \left(
\bnabla c_{+} \bcdot \bnabla \phi
+
c_{+} \nabla^2 \phi
\right)
+
\nabla^2 c_{+}
\right],
\label{eq:fullc+}
\end{equation}

\begin{equation}
\frac{\partial c_{-}}{\partial t}
+ \boldsymbol{u} \bcdot \bnabla c_{-}
=
-\frac{1}{Pe}
\left[
V \left(
\bnabla c_{-} \bcdot \bnabla \phi
+
c_{-} \nabla^2 \phi
\right)
-
\nabla^2 c_{-}
\right],
\label{eq:fullc-}
\end{equation}

\begin{equation}
-2\varepsilon^2 V \nabla^2 \phi
=
c_{+} - c_{-},
\label{eq:fullphi}
\end{equation}

\begin{equation}
\frac{\partial \mathsfbi{A}}{\partial t}
+
\boldsymbol{u} \bcdot \bnabla \mathsfbi{A}
-
(\bnabla \boldsymbol{u})^{T} \bcdot \mathsfbi{A}
-
\mathsfbi{A} \bcdot \bnabla \boldsymbol{u}
=
-\frac{1}{Wi}
(\mathsfbi{A}-\mathsfbi{I}).
\label{eq:fullA}
\end{equation}

Here, $\boldsymbol{u}$, $p$, $\phi$, and $\mathsfbi{A}$ denote the dimensionless velocity, pressure, electric potential, and conformation tensor, respectively. $c_{+}$ and $c_{-}$ represent the dimensionless cation and anion concentrations. The viscosity ratio is defined as $\beta=\mu_s/(\mu_s+\mu_p)$, where $\mu_s$ and $\mu_p$ are the solvent and polymer viscosities. The parameter $V=V_A/V_T$ denotes the dimensionless applied potential, defined as the ratio of the applied voltage $V_A$ to the thermal voltage $V_T=RT/F$. The dimensionless Debye length is defined as $\varepsilon = \sqrt{\delta RT/(2 F^2 C_0)}/H$, where $C_0$ is the bulk ion concentration, $\delta$ is the permittivity, $R$ is the gas constant, $T$ is the temperature, $F$ is the Faraday constant and $H$ is the channel half-width. The electrolyte is assumed to be symmetric and univalent, and the diffusivities of the cation and anion are taken to be equal, $D_{+}=D_{-}=D$. The electric Rayleigh number is given by $Ra_E=\delta V_A^2/\mu D$, which characterizes the strength of the electric forcing relative to viscous and diffusive effects. The Peclet number and Weissenberg number are defined as $Pe=U_m H/D, Wi=\lambda U_m/H$, where $D$ is the ionic diffusivity and $\lambda$ is the polymer relaxation time. All variables are nondimensionalized using the channel half-width $H$ as the characteristic length scale, the maximum velocity of the base Poiseuille flow $U_m$ as the velocity scale, the applied potential $V_A$ as the electric potential scale, and the bulk ion concentration $C_0$ as the concentration scale. Time is scaled by $H/U_m$.

\subsection{Linear stability analysis}

To investigate the onset of instability, we perform a linear stability analysis of a steady base state. The base state corresponds to a one-dimensional solution in the wall-normal direction, incorporating the ESC structure and the associated electric field.

All variables are decomposed into a steady base component and a small perturbation. Specifically, any field variable $q$ (representing $\boldsymbol{u}$, $c^{\pm}$, $\phi$, and $\mathsfbi{A}$) is expressed as $q(x,y,t)=\bar{q}(y)+q'(x,y,t)$, where $\bar{q}$ denotes the base-state solution and $q'$ represents the perturbation field. The perturbation is assumed to take the form $q'(x,y,t)=\hat{q}(y)e^{ikx+\sigma t}$, where $\hat{q}$ is the perturbation eigenfunction, $k$ is the real streamwise wavenumber, and $\sigma=\sigma_r+i\sigma_i$ is the complex eigenvalue. The real part $\sigma_r$ corresponds to the temporal growth rate, and the imaginary part $\sigma_i$ determines the wave propagation speed $-\sigma_i/k$. The flow is linearly unstable when $\sigma_r>0$.

The base state is governed by the one-dimensional Nernst--Planck and Poisson equations,

\begin{equation}
\frac{\partial^2 \bar{c}_{+}}{\partial y^2}
+
V
\left(
\frac{\partial \bar{c}_{+}}{\partial y}
\frac{\partial \bar{\phi}}{\partial y}
+
\bar{c}_{+}
\frac{\partial^2 \bar{\phi}}{\partial y^2}
\right)
=0,
\label{eq:basec+}
\end{equation}

\begin{equation}
\frac{\partial^2 \bar{c}_{-}}{\partial y^2}
-
V
\left(
\frac{\partial \bar{c}_{-}}{\partial y}
\frac{\partial \bar{\phi}}{\partial y}
+
\bar{c}_{-}
\frac{\partial^2 \bar{\phi}}{\partial y^2}
\right)
=0,
\label{eq:basec-}
\end{equation}

\begin{equation}
-2\varepsilon^2 V
\frac{\partial^2 \bar{\phi}}{\partial y^2}
=
\bar{c}_{+}
-
\bar{c}_{-}.
\label{eq:basephi}
\end{equation}

These equations are to be solved subject to 
 $\bar{c}_{+}|_{y=0,2}=2.5$, $\int_{0}^{2} \bar{c}_{-}\,dy = 2$,
$\bar{\phi}|_{y=0}=0$ and $\bar{\phi}|_{y=2}=1$.

Based on the Poiseuille base flow,
$\bar{U}(y)=y(2-y),$
the conformation tensor components can be obtained analytically as

\begin{equation}
\bar{A}_{xx}
=
2[(2-2y)Wi]^2 +1,
\label{eq:baseAxx}
\end{equation}

\begin{equation}
\bar{A}_{xy}
=
(2-2y)Wi,
\label{eq:baseAxy}
\end{equation}

\begin{equation}
\bar{A}_{yy}
=
1.
\label{eq:baseAyy}
\end{equation}

Substituting the normal-mode expansions into the governing equations and retaining only linear perturbation terms yield the perturbation equations. The perturbation equations for the linear stability analysis are written as

\begin{equation}
\sigma \hat{A}_{xx}
=
-
\left(
ik\bar{U}+\frac{1}{Wi}
\right)
\hat{A}_{xx}
+
2\frac{\partial \bar{U}}{\partial y}\hat{A}_{xy}
+
\left[
2\bar{A}_{xy}\frac{i}{k}\frac{\partial^2}{\partial y^2}
-
2\bar{A}_{xx}\frac{\partial}{\partial y}
-
\frac{\partial \bar{A}_{xx}}{\partial y}
\right]
\hat{v},
\label{eq:perturbAxx}
\end{equation}

\begin{equation}
\sigma \hat{A}_{xy}
=
-
\left(
ik\bar{U}+\frac{1}{Wi}
\right)
\hat{A}_{xy}
+
\frac{\partial \bar{U}}{\partial y}\hat{A}_{yy}
+
\left[
\bar{A}_{yy}\frac{i}{k}\frac{\partial^2}{\partial y^2}
-
\frac{\partial \bar{A}_{xy}}{\partial y}
+
ik\bar{A}_{xx}
\right]
\hat{v},
\label{eq:perturbAxy}
\end{equation}

\begin{equation}
\sigma \hat{A}_{yy}
=
-
\left(
ik\bar{U}+\frac{1}{Wi}
\right)
\hat{A}_{yy}
+
2
\left[
ik\bar{A}_{xy}
+
\bar{A}_{yy}\frac{\partial}{\partial y}
\right]
\hat{v}.
\label{eq:perturbAyy}
\end{equation}

\begin{equation}
-2\varepsilon^2V
\left(
\frac{\partial^2 \hat{\phi}}{\partial y^2}
-
k^2\hat{\phi}
\right)
=
\hat{c}_{+}
-
\hat{c}_{-}.
\label{eq:perturbphi}
\end{equation}

\begin{equation}
\begin{aligned}
&
\beta
\left(
\frac{\partial^4 \hat{v}}{\partial y^4}
-
2k^2\frac{\partial^2 \hat{v}}{\partial y^2}
+
k^4\hat{v}
\right)
+
\frac{1-\beta}{Wi}
\left[
k^2\frac{\partial \hat{A}_{xx}}{\partial y}
-
ik
\left(
\frac{\partial^2}{\partial y^2}
+
k^2
\right)
\hat{A}_{xy}
-
k^2\frac{\partial \hat{A}_{yy}}{\partial y}
\right]
\\
&\hspace{4.8cm}=
k^2\frac{Ra_E}{Pe}
\left(
\frac{\partial^2 \hat{\phi}}{\partial y^2}
-
k^2\hat{\phi}
\right)
\frac{\partial \bar{\phi}}{\partial y}
-
k^2\frac{Ra_E}{Pe}
\hat{\phi}
\frac{\partial^3 \bar{\phi}}{\partial y^3}.
\end{aligned}
\label{eq:perturbmomentum}
\end{equation}

\begin{equation}
\begin{aligned}
\sigma \hat{c}_{+}
+
ik\bar{U}\hat{c}_{+}
+
\hat{v}
\frac{\partial \bar{c}_{+}}{\partial y}
=
\frac{1}{Pe}
\Bigg[
\frac{\partial^2 \hat{c}_{+}}{\partial y^2}
-
k^2\hat{c}_{+}
-
k^2V\bar{c}_{+}\hat{\phi}
+
V
\frac{\partial}{\partial y}
\left(
\bar{c}_{+}
\frac{\partial \hat{\phi}}{\partial y}
+
\hat{c}_{+}
\frac{\partial \bar{\phi}}{\partial y}
\right)
\Bigg],
\end{aligned}
\label{eq:perturbc+}
\end{equation}

\begin{equation}
\begin{aligned}
\sigma \hat{c}_{-}
+
ik\bar{U}\hat{c}_{-}
+
\hat{v}
\frac{\partial \bar{c}_{-}}{\partial y}
=
\frac{1}{Pe}
\Bigg[
\frac{\partial^2 \hat{c}_{-}}{\partial y^2}
-
k^2\hat{c}_{-}
+
k^2V\bar{c}_{-}\hat{\phi}
-
V
\frac{\partial}{\partial y}
\left(
\bar{c}_{-}
\frac{\partial \hat{\phi}}{\partial y}
+
\hat{c}_{-}
\frac{\partial \bar{\phi}}{\partial y}
\right)
\Bigg].
\end{aligned}
\label{eq:perturbc-}
\end{equation}
with boundary conditions
$
\hat{c}_{+}
=
\hat{J}_{-}
=
\hat{v}
=
\hat{v}/\partial y
=
\hat{\phi}
=
0$
at
$y=0,2,
$
where
$
\hat{J}_{-}=
\partial \hat{c}_{-}/\partial y
-
\left(
\bar{c}_{-}
\partial \hat{\phi}/\partial y
+
\hat{c}_{-}
\partial \bar{\phi}/\partial y
\right)
$
is the dimensionless anion flux associated with the perturbation state.

The resulting generalized eigenvalue problem is solved numerically to determine the complex growth rate $\sigma$ as a function of the streamwise wavenumber $k$ and the governing dimensionless parameters. The accuracy of the linear stability analysis is verified through comparisons with previous studies and grid-refinement tests; details are provided in Appendix~A.

The parameter ranges considered in this study are chosen to be consistent with experimentally relevant microfluidic electrokinetic systems. Typical systems involve channel half-widths
$H \sim 10^{-2}$--$1\,\mathrm{mm}$,
characteristic velocities
$U \sim 10^{-2}$--$1\,\mathrm{mm/s}$,
dynamic viscosities
$\mu \sim 1$--$10\,\mathrm{mPa\,s}$,
ionic diffusivities
$D \sim 10^{-9}\,\mathrm{m}^2/\mathrm{s}$,
bulk ion concentrations
$C_0 \sim 1$--$100\,\mathrm{mM}$,
applied voltages
$V_A \sim 0.5$--$5\,\mathrm{V}$,
and polymer relaxation times
$\lambda \sim 10^{-2}$--$10\,\mathrm{s}$.
These values correspond approximately to the dimensionless parameter ranges
$Pe \sim O(1$--$100)$,
$Wi \sim O(10^{-2}$--$10^{2})$,
$\beta \sim 0.3$--$0.8$,
$\varepsilon \sim 10^{-5}$--$10^{-3}$,
$V \sim O(10$--$100)$,
and
$Ra_E \sim O(10^2$--$10^3)$
considered in the present study.

\begin{figure}
\centerline{\includegraphics[scale=0.45]{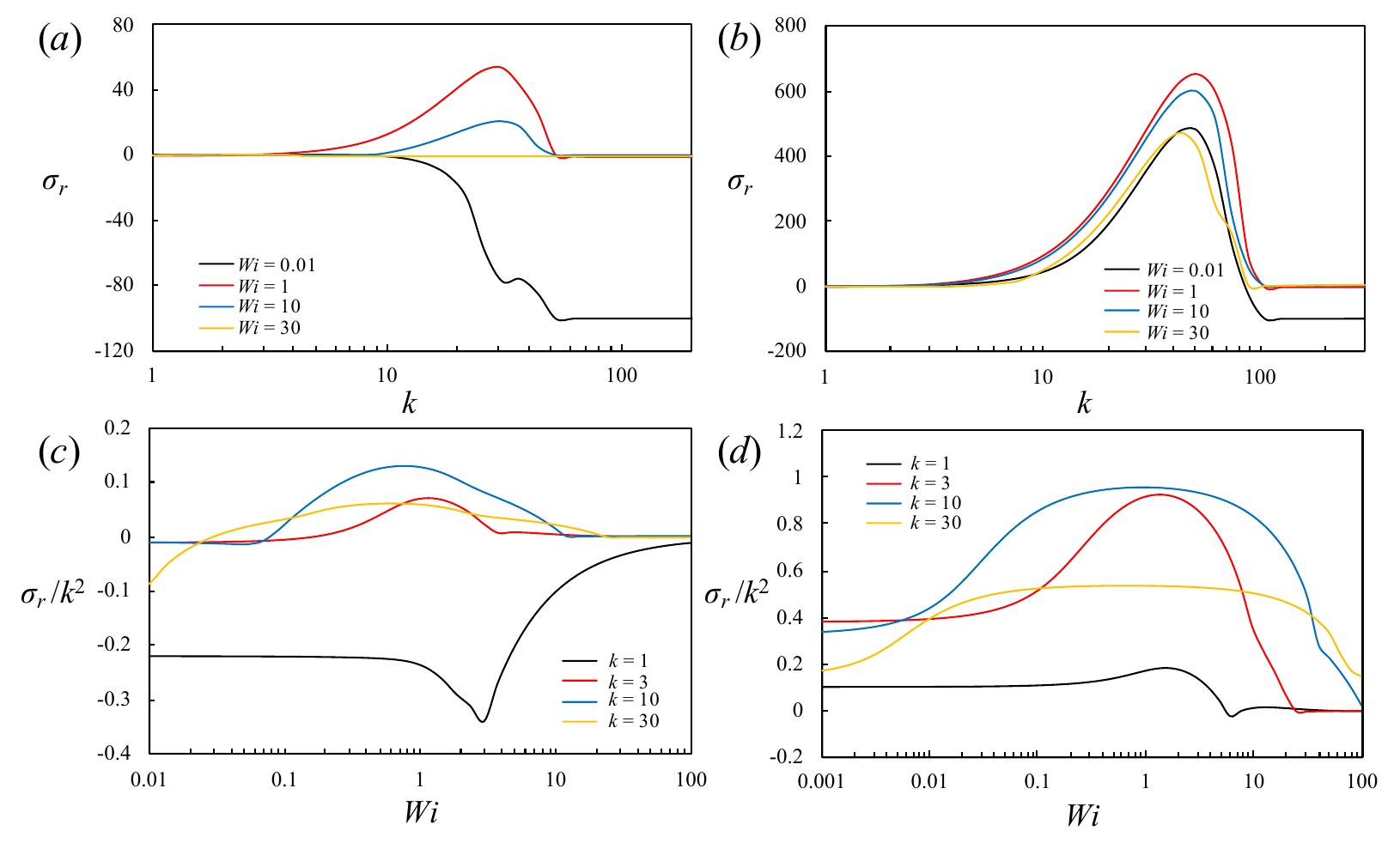}}
\caption{Growth rate of the electroconvective instability for different Weissenberg numbers and wavenumbers at $Pe=10$, $\beta=0.5$, and $\varepsilon=10^{-4}$. Panels ($a$) and ($c$) show results at $V=30$ ($Ra_E=105$), and panels ($b$) and ($d$) show results at $V=40$ ($Ra_E=187$). ($a,b$) Growth rate $\sigma_r$ as a function of $k$ for representative values of $Wi$. ($c,d$) Normalized growth rate $\sigma_r/k^2$ as a function of $Wi$ for representative wavenumbers. At both applied voltages, the instability is maximized at $Wi\sim O(1)$, demonstrating the non-monotonic effect of viscoelasticity on electroconvective instability.}
\label{fig:fig2}
\end{figure}

\section{Results and discussion}

\subsection{Non-monotonic effect of $Wi$ on instability}

We first examine the effect of viscoelasticity on the growth rate of electroconvective instability. Figure~\ref{fig:fig2} compares the growth rates obtained at $V=30$ and $40$ for the reference parameter set $Pe=10$, $\beta=0.5$, and $\varepsilon=10^{-4}$. The results at $V=30$ provide a comparison closer to the instability threshold, and those at $V=40$ show the corresponding response under strongly supercritical conditions.

Figures~\ref{fig:fig2}($a,b$) show the growth rate as a function of wavenumber for representative values of $Wi$. At small wavenumbers, corresponding to long-wave disturbances, the instability is suppressed because of confinement effects imposed by the channel geometry, consistent with previous studies \citep{Li2021,Kim2026}. In contrast, at large wavenumbers, corresponding to short-wave disturbances, the instability is attenuated because perturbations are dissipated within the ESC layer and therefore cannot effectively propagate into the bulk flow \citep{Li2021,Kim2026}. As a result, the instability reaches a maximum growth rate at an intermediate wavelength associated with the ESC length scale. Importantly, at sufficiently large wavenumbers, the growth rate for $Wi=0.01$ decreases to approximately $\sigma_r \approx -100$, whereas the other cases remain close to zero. This behaviour corresponds to the continuous-spectrum eigenvalues~\citep{Khalid2021a} and will be discussed further in \S 3.5.4 and Appendix~A.

Figures~\ref{fig:fig2}($c,d$) show $\sigma_r/k^2$ as a function of $Wi$. At both $V=30$ and $40$, the curves for $k=3$, $10$, and $30$ reach their maximum values near $Wi\sim O(1)$. The growth rates are lower at both smaller and larger values of $Wi$, demonstrating the non-monotonic effect of viscoelasticity.  The trend suggests that viscoelasticity plays a dual role: it enhances instability at intermediate $Wi$ but suppresses it when elastic effects become dominant.

To determine whether this fixed-voltage trend is also reflected in the onset condition, we next calculate the neutral stability curves defined by $\sigma_r(k,Wi,V_{\mathrm{cr}})=0$.

\begin{figure}
\centerline{\includegraphics[scale=0.5]{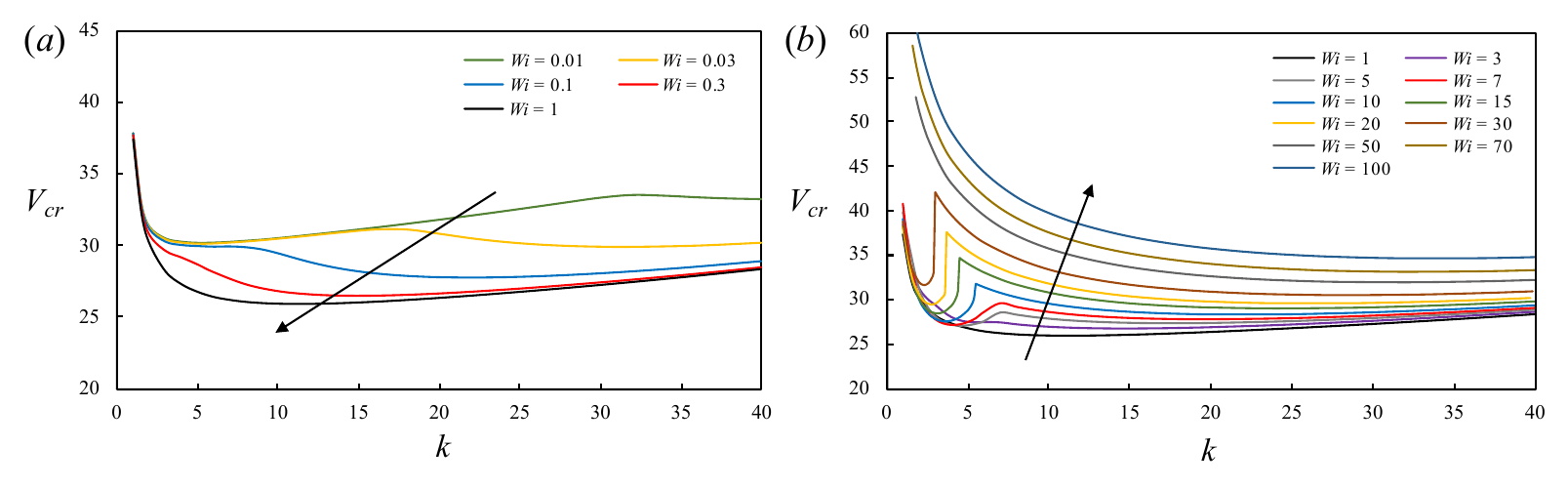}}
\caption{Neutral stability curves of the electroconvective instability
at $Pe=10$, $\beta=0.5$, and $\varepsilon=10^{-4}$.
($a$) Critical voltage $V_{\mathrm{cr}}$ as a function of $k$ for
$Wi=0.01$, $0.03$, $0.1$, $0.3$, and $1$.
($b$) Corresponding neutral curves for
$Wi=1$, $3$, $5$, $7$, $10$, $15$, $20$, $30$, $50$, $70$, and $100$.
The arrows indicate increasing $Wi$.
The neutral boundary shifts downward as $Wi$ approaches $O(1)$ and
rises again at larger $Wi$.
A distinct low-wavenumber minimum develops over the intermediate range
$3\leq Wi\leq30$.}
\label{fig:fig3}
\end{figure}

Figure~\ref{fig:fig3} shows the critical voltage $V_{\mathrm{cr}}$ as a function of wavenumber. For $Wi\leq1$ (figure~\ref{fig:fig3}$a$), increasing $Wi$ progressively lowers the neutral boundary. The smallest critical voltages occur near $Wi \sim O(1)$. For $Wi\geq1$ (figure~\ref{fig:fig3}$b$), the critical voltage generally increases with $Wi$. The neutral curves therefore confirm that the non-monotonic growth-rate response observed at fixed voltage corresponds to a minimum in the instability threshold near $Wi \sim O(1)$.

An additional feature is a distinct low-wavenumber minimum in the neutral boundary over $3\leq Wi\leq30$. The location of this local minimum shifts toward smaller $k$ as $Wi$ increases. The low-wavenumber minimum weakens and ultimately disappears at still larger $Wi$, identifying it as an intermediate-$Wi$ feature.

To investigate the origin of these features, we first examine the spatial organization of the electroconvective perturbations in §§3.2 and 3.3. Section~3.4 then introduces the polymeric-work formulation used to interpret the results. This framework is applied to the fixed-wavenumber mode at $k=10$ in §3.5 and to the modes associated with the intermediate-$Wi$ low-wavenumber minimum in §3.6.

\begin{figure}
  \centerline{\includegraphics[scale=0.3]{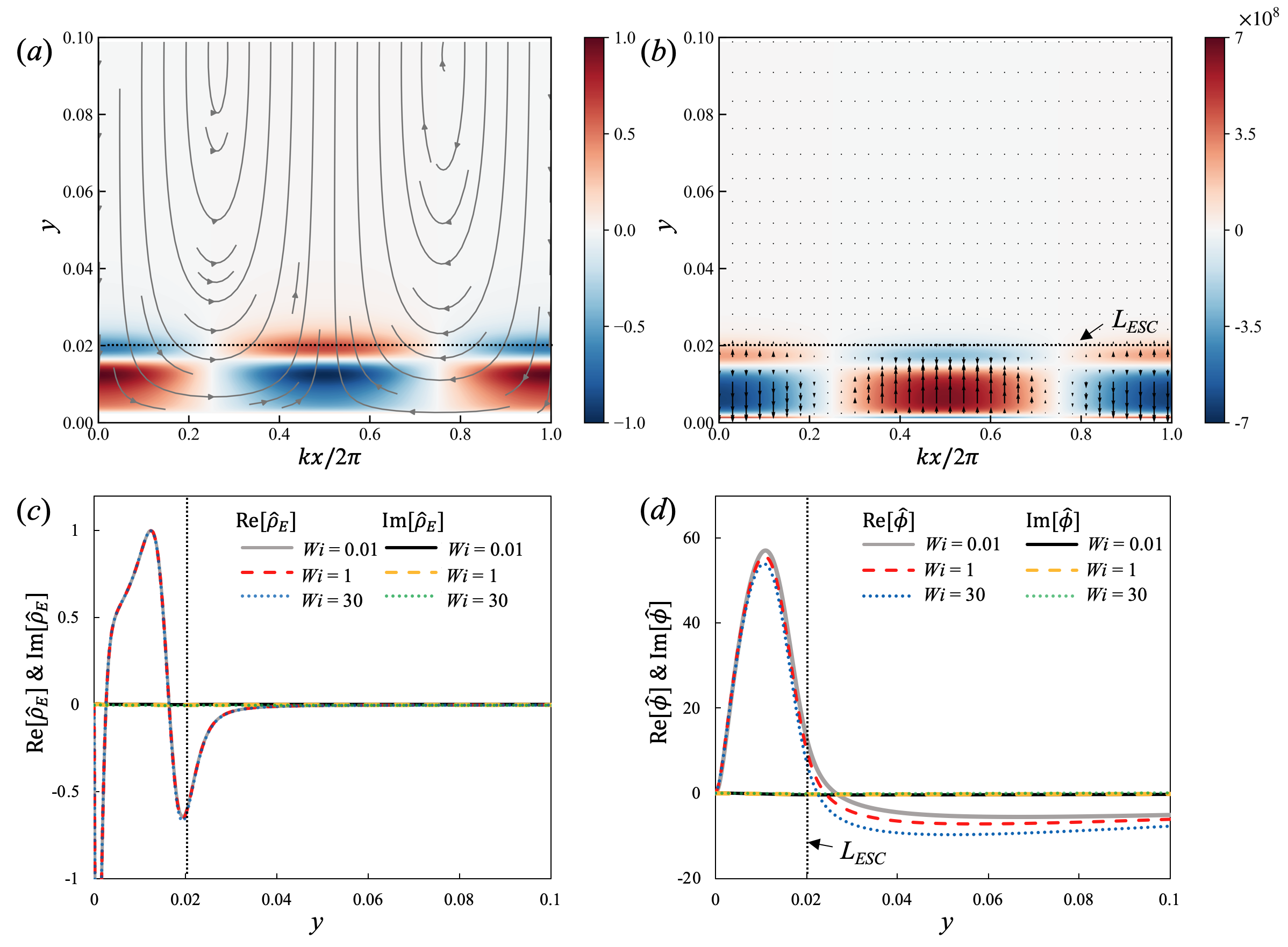}} 
  \caption{
  Structure of the perturbation charge-density and electric-force eigenfunctions near the ESC region. Dotted lines represent the ESC boundary. ($a,b$) Results at $Pe=10$, $Wi=1$, $\beta=0.5$, $\varepsilon =10^{-4}$, $Ra_E=105$ ($V=30$),  and $k=10$. ($a$) Two-dimensional distribution of the perturbation charge density, $\Real[\hat{\rho}_E(y)e^{ikx}]$, together with perturbation velocity streamlines. ($b$) Vertical component of the perturbation electric body force, $\Real[\hat{f}(y)e^{ikx}]$, and corresponding electric-force vectors. The electric force is predominantly oriented in the wall-normal direction within the ESC layer. ($c,d$) Profiles of the perturbation charge density and electric potential for representative values of $Wi=0.01$, $1$, and $30$ at $k=10$. The nearly identical profiles indicate that the electric forcing structure remains largely independent of viscoelasticity.
}
\label{fig:fig4_}
\end{figure}

\subsection{ESC localization of the instability mechanism}

The spatial origin of the electroconvective driving is examined using reconstructed charge, electric-potential, force, and velocity perturbations. The results show that the destabilizing electric forcing is strongly concentrated within the ESC, consistent with the mechanism previously reported by \citet{Kim2026}.

We reconstruct the two-dimensional perturbation fields from the normal-mode eigenfunctions according to $q'(x,y,t)=\hat{q}(y)e^{ikx+\sigma t}$ and visualize the perturbation structure at $t=0$ (i.e., $\Real[q'(x,y,0)]=\Real[\hat{q}(y)e^{ikx}]$). The perturbation charge density is defined as $\hat{\rho}_E=\hat{c}_{+}-\hat{c}_{-}$. The eigenfunctions are normalized such that the maximum magnitude of $\hat{\rho}_E$ within the ESC region is unity, and the global phase is chosen such that  $\hat{\rho}_E$ at this location has zero phase angle. The ESC boundary is denoted by the dotted lines in figure~\ref{fig:fig4_} and corresponds to the ESC thickness $L_{ESC}=1.1\varepsilon^{2/3}(V+2/3\ln\varepsilon)^{2/3}$ \citep{Kim2026}. Figure~\ref{fig:fig4_}($a$) presents the reconstructed perturbation charge density, $\Real[\hat{\rho}_E(y)e^{ikx}]$, together with streamlines of the perturbation velocity field. Within the ESC layer, the perturbation charge density is dominated by the cation concentration because co-ions are nearly depleted, leading to $\hat{\rho}_E \approx \hat{c}_{+}$ and $\hat{c}_{-}\approx0$. Regions of reduced cation concentration correspond to a local increase in electrical resistance, which intensifies the electric field and produces a localized electric body force that drives upward flow. This coupling between charge depletion, enhanced electric field, and induced flow forms the core electroconvective feedback mechanism~\citep{Kim2026}.

Figure~\ref{fig:fig4_}($b$) shows the vertical component of the electric force together with the corresponding two-dimensional force vectors. The direction of the arrows clearly indicates that, although both streamwise and wall-normal force components are present, the force is predominantly oriented in the wall-normal direction. This demonstrates that the instability is primarily driven by vertical electric forcing localized within the ESC.

Figures~\ref{fig:fig4_}($c,d$) compare the perturbation charge density and electric potential for $Wi=0.01$, $1$, and $30$. The spatial distributions remain nearly identical across these cases, suggesting that the electric forcing structure is essentially independent of viscoelasticity. In addition, both the perturbation charge density and electric potential remain nearly in phase across all values of $Wi$, with no appreciable phase lag observed in the ESC region. This further confirms that the electric forcing mechanism is largely unaffected by viscoelasticity.
This observation is consistent with the perturbation equation~(\ref{eq:perturbmomentum}), where the electric body force depends only on the base and perturbed charge density and electric potential fields and does not directly involve $Wi$. Therefore, the non-monotonic dependence on $Wi$ does not originate from modifications to the electric driving mechanism itself.

\begin{figure}
\centerline{\includegraphics[scale=0.23]{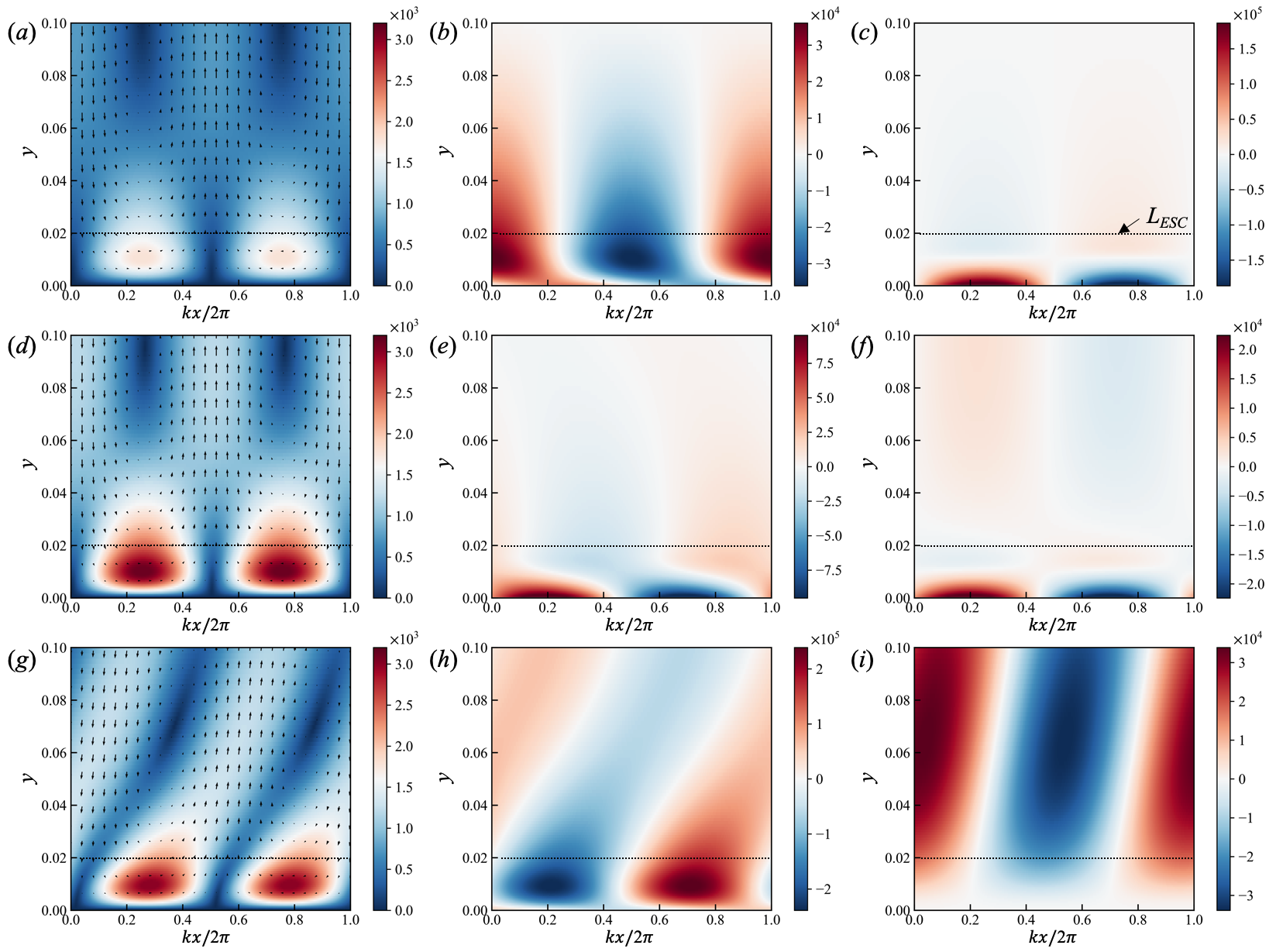}}
\caption{Perturbation velocity and polymer-stress fields for the eigenmode at $k=10$, $Pe=10$, $\beta=0.5$, $\varepsilon=10^{-4}$, and  $Ra_E=105$ ($V=30$). ($a$--$c$) $Wi=0.01$, ($d$--$f$) $Wi=1$, and ($g$--$i$) $Wi=30$. Panels ($a,d,g$) show the perturbation velocity magnitude as contours and the velocity field as vectors. Panels ($b,e,h$) show the polymeric normal stress difference, and panels ($c,f,i$) show the polymeric shear stress. The dotted line indicates the outer edge of the ESC. For each eigenmode, the maximum magnitude of the perturbation charge density within the ESC is normalized to unity.}
\label{fig:fig5}
\end{figure}

\subsection{Viscoelastic modification of the velocity and polymer-stress fields}
\label{sec:stress_fields}

We next examine how viscoelasticity modifies the coupled perturbation velocity and polymer-stress fields. Figure~\ref{fig:fig5} compares the eigenmodes at $k=10$ for $Wi=0.01$, $1$, and $30$. The eigenfunctions are normalized by setting the maximum magnitude of the perturbation charge density in the ESC to unity. The first column shows the perturbation velocity magnitude and velocity vectors. The second and third columns show the polymeric normal stress difference and shear stress, respectively.

From $Wi=0.01$ to $Wi=1$, the magnitude of the perturbation velocity increases substantially, but its overall spatial organization remains similar (figures~\ref{fig:fig5}$a,d$). In both cases, the strongest velocity perturbations occur near the ESC, and the associated flow extends into the bulk. At $Wi=0.01$, the normal stress difference is distributed primarily within the ESC (figure~\ref{fig:fig5}$b$). At $Wi=1$, its peak magnitude increases and the stress becomes more strongly localized near the membrane surface (figure~\ref{fig:fig5}$e$). Outside the ESC, its magnitude remains comparable to or smaller than that at $Wi=0.01$. The shear stress retains a similar alternating spatial pattern from $Wi=0.01$ to $Wi=1$, with its strongest variations concentrated near the surface (figures~\ref{fig:fig5}$c,f$).

A substantial structural change occurs between $Wi=1$ and $Wi=30$. At $Wi=30$, the perturbation velocity develops an inclined spatial organization that differs from the pattern observed at the two smaller values of $Wi$ (figure~\ref{fig:fig5}$g$). The normal stress difference becomes much larger and is concentrated primarily within the ESC (figure~\ref{fig:fig5}$h$). The shear stress undergoes a different spatial redistribution. Its alternating positive and negative regions extend well beyond the ESC and occupy a substantial portion of the bulk (figure~\ref{fig:fig5}$i$). The large-$Wi$ eigenmode is therefore characterized by strong normal-stress variations near the surface and an extended shear-stress response in the bulk.

These results show that increasing $Wi$ modifies both the polymer-stress distribution and the velocity eigenfunction of the fully coupled mode. The stress contours alone, however, do not determine whether a given stress component supplies mechanical power to the perturbation or extracts it. That mechanical action depends on the local stress--deformation product and their relative phase. Section~\ref{sec:mechanical_work} therefore formulates the mechanical-work balance, separates the normal-stress-difference and shear-stress contributions, and decomposes them into their constitutive pathways. The resulting framework is applied to the $k=10$ and low-wavenumber eigenmodes in \S\S~\ref{sec:k10_pathways} and \ref{sec:low_k}, respectively.

\subsection{Mechanical-work formulation and asymptotic pathway ordering}
\label{sec:mechanical_work}

To determine how the polymer stress modifies the electroconvective instability, we formulate a mechanical-work-rate balance for each linear eigenmode and decompose the polymeric contribution into individual constitutive pathways. The pathway ordering is then examined in the limits $Wi\ll1$ and $Wi\gg1$. Sections~3.5 and 3.6 use this framework to interpret the pathway-resolved work of the $k=10$ and low-wavenumber modes, respectively.

\subsubsection{Mechanical-work-rate balance}
\label{sec:work_balance}

The Fourier-transformed linearized momentum equation can be written in vector form as
\begin{equation}
0
=
-\bnabla_k \hat{p}
+\bnabla_k\bcdot\widehat{\boldsymbol{\tau}}_{\mathrm{sol}}
+\bnabla_k\bcdot\widehat{\boldsymbol{\tau}}_p
+\widehat{\boldsymbol{f}}_e,
\label{eq:linear_momentum_work}
\end{equation}
where $\widehat{\boldsymbol{u}}=(\hat{u},\hat{v})$, $\bnabla_k=(ik,D)$, and $D\equiv{\mathrm{d}}/{\mathrm{d}y}$. The perturbation solvent stress, polymer stress, and rate-of-strain tensor are defined by
\begin{equation}
\widehat{\boldsymbol{\tau}}_{\mathrm{sol}}
=2\beta\widehat{\mathsfbi{E}},
\qquad
\widehat{\boldsymbol{\tau}}_p
=\frac{1-\beta}{Wi}\widehat{\mathsfbi{A}},
\qquad
\widehat{\mathsfbi{E}}
=\frac{1}{2}
\left[
\bnabla_k\widehat{\boldsymbol{u}}
+(\bnabla_k\widehat{\boldsymbol{u}})^T
\right].
\label{eq:stress_perturbations_work}
\end{equation}

The linearized electric body force is
\begin{equation}
\widehat{\boldsymbol{f}}_e
=
\frac{Ra_E}{Pe}
\left[
(D^2\bar{\phi})\bnabla_k\hat{\phi}
+(D^2-k^2)\hat{\phi}\,D\bar{\phi}\,\boldsymbol{e}_y
\right].
\label{eq:electric_force_work}
\end{equation}

Taking the inner product of equation~\eqref{eq:linear_momentum_work} with $\widehat{\boldsymbol{u}}^{*}=(\hat{u}^{*},\hat{v}^{*})$, integrating over an arbitrary interval $y_1\leq y\leq y_2$, and retaining the real part gives
\begin{equation}
0=W_{\mathrm{pr}}+W_{\mathrm{sol}}+W_p+W_e.
\label{eq:mechanical_work_balance}
\end{equation}

Here and below, $W$ denotes a mechanical-work rate. Because the present flow is inertialess, equation~\eqref{eq:mechanical_work_balance} contains no perturbation kinetic-energy storage term. It is therefore the mechanical-power balance satisfied by an eigenmode and does not constitute a separate evolution equation for its growth rate.

Incompressibility reduces the pressure contribution to boundary work,
\begin{equation}
W_{\mathrm{pr}}
=
-\Real\left[\hat{v}^{*}\hat{p}\right]_{y_1}^{y_2}.
\label{eq:pressure_work}
\end{equation}
Integration by parts similarly separates the solvent and polymer contributions into surface and internal parts:
\begin{subequations}
\begin{align}
W_{\mathrm{sol}}
&=
\Real\left[
\widehat{\boldsymbol{u}}^{*}\bcdot
\widehat{\boldsymbol{\tau}}_s\bcdot\boldsymbol{e}_y
\right]_{y_1}^{y_2}
-2\beta\int_{y_1}^{y_2}
\widehat{\mathsfbi{E}}^{*}:\widehat{\mathsfbi{E}}\,\mathrm{d}y
\equiv W_{\mathrm{sol}}^{\mathrm{surf}}+W_{\mathrm{sol}}^{\mathrm{int}},
\label{eq:solvent_work}\\
W_p
&=
\Real\left[
\widehat{\boldsymbol{u}}^{*}\bcdot
\widehat{\boldsymbol{\tau}}_p\bcdot\boldsymbol{e}_y
\right]_{y_1}^{y_2}
-\Real\int_{y_1}^{y_2}
(\bnabla_k\widehat{\boldsymbol{u}})^{*}:
\widehat{\boldsymbol{\tau}}_p\,\mathrm{d}y
\equiv W_p^{\mathrm{surf}}+W_p^{\mathrm{int}}.
\label{eq:polymer_work}
\end{align}
\end{subequations}
Thus, $W_{\mathrm{sol}}^{\mathrm{int}}\leq0$ represents solvent-viscous dissipation, but $W_p^{\mathrm{int}}$ is not sign definite. The electric contribution is retained as body-force work,
\begin{equation}
W_e
=
\Real\int_{y_1}^{y_2}
\widehat{\boldsymbol{u}}^{*}\bcdot
\widehat{\boldsymbol{f}}_e\,\mathrm{d}y.
\label{eq:electric_work}
\end{equation}
Detailed componentwise derivations of equations~\eqref{eq:pressure_work}--\eqref{eq:electric_work} are provided in Appendix~B.

Defining the total traction-mediated surface work as
$W_{\mathrm{tot}}^{\mathrm{surf}}
=W_{\mathrm{pr}}+W_s^{\mathrm{surf}}+W_p^{\mathrm{surf}}$,
the balance becomes
\begin{equation}
0
=
W_{\mathrm{tot}}^{\mathrm{surf}}
+W_{\mathrm{sol}}^{\mathrm{int}}
+W_p^{\mathrm{int}}
+W_e.
\label{eq:regional_work_balance}
\end{equation}
For the full domain, $0\leq y\leq2$, the no-slip boundary conditions make all surface contributions vanish. When the domain is partitioned into the ESC and bulk regions, however, the traction work at their common interface is generally non-zero and satisfies
\begin{equation}
W_{\mathrm{tot}}^{\mathrm{surf,ESC}}
=
-W_{\mathrm{tot}}^{\mathrm{surf,bulk}}.
\label{eq:interface_work_cancellation}
\end{equation}
It therefore represents mechanical-power transfer between the two regions, not an additional source or sink in the full-domain balance.

To separate the internal polymeric work into normal-stress-difference and shear-stress contributions, we define
\begin{equation}
\hat{\gamma}=D\hat{u}+ik\hat{v}=2\widehat{E}_{xy}.
\label{eq:shear_rate_work}
\end{equation}
Using conjugate incompressibility, $D\hat{v}^{*}=ik\hat{u}^{*}$, equation~\eqref{eq:polymer_work} gives
\begin{equation}
W_p^{\mathrm{int}}
=
\Real\int_{y_1}^{y_2}
\left[
ik\hat{u}^{*}
(\hat{\tau}_{p,xx}-\hat{\tau}_{p,yy})
-\hat{\gamma}^{*}\hat{\tau}_{p,xy}
\right]\mathrm{d}y
\equiv W_N+W_S,
\label{eq:polymer_normal_shear_work}
\end{equation}
where
\begin{subequations}
\begin{align}
W_N
&=
\Real\int_{y_1}^{y_2}
ik\hat{u}^{*}
(\hat{\tau}_{p,xx}-\hat{\tau}_{p,yy})\,\mathrm{d}y,
\label{eq:normal_work}\\
W_S
&=
-\Real\int_{y_1}^{y_2}
\hat{\gamma}^{*}\hat{\tau}_{p,xy}\,\mathrm{d}y.
\label{eq:shear_work}
\end{align}
\end{subequations}
Positive values of $W_N$ or $W_S$ indicate mechanical-power input to the perturbation velocity field, and negative values indicate extraction. Unless otherwise stated, all local work densities, regional work rates, and magnitude capacities reported below are normalized by
\begin{equation}
\mathcal{W}_e
=
\left|W_e^{\mathrm{total}}\right|,
\label{eq:work_normalization}
\end{equation}
evaluated for the corresponding eigenmode. The same full-domain normalization is used for the ESC, bulk, and total contributions, thereby removing the arbitrary eigenfunction amplitude while preserving work signs and additive regional balances. The following subsection decomposes $W_N$ and $W_S$ into their constitutive pathways.

\subsubsection{Constitutive pathways of polymer work}
\label{sec:constitutive_pathways}

To identify the constitutive origin of the normal-stress-difference and shear-stress work defined in \S~\ref{sec:work_balance}, we decompose the linearized conformation equations into individual pathways. We introduce
\begin{subequations}
\begin{align}
\Sigma(y)&=\sigma+ik\bar{U}(y),\\
\Lambda(y)&=\Sigma(y)+\frac{1}{Wi},
\end{align}
\label{eq:sigma_lambda}
\end{subequations}
and denote the base shear rate by
\begin{equation}
g(y)=D\bar{U}=2-2y.
\label{eq:base_shear_rate}
\end{equation}
Here, $\Sigma$ combines the temporal response of the mode with advection by the base flow, whereas $Wi^{-1}$ represents polymer relaxation.

Subtracting equation~\eqref{eq:perturbAyy} from equation~\eqref{eq:perturbAxx} and using incompressibility gives
\begin{equation}
\Lambda(\hat A_{xx}-\hat A_{yy})
=N_1+N_2+N_3+N_4,
\label{eq:normal_pathway_balance}
\end{equation}
where
\begin{subequations}
\begin{align}
N_1&=2g\hat A_{xy},
\label{eq:N1}\\
N_2&=2\bar A_{xy}(D\hat u-ik\hat v),
\label{eq:N2}\\
N_3&=-2(\bar A_{xx}+\bar A_{yy})D\hat v,
\label{eq:N3}\\
N_4&=-\hat v D\bar A_{xx}.
\label{eq:N4}
\end{align}
\label{eq:normal_pathways}
\end{subequations}
The $N_1$ pathway represents the action of the base shear on the perturbed shear conformation. The $N_2$ pathway couples the base shear conformation to the perturbation velocity gradients. The $N_3$ pathway represents deformation of the base normal conformation by the perturbation normal strains, and $N_4$ represents wall-normal advection of the base streamwise-conformation gradient.

The shear-conformation equation can similarly be written as
\begin{equation}
\Lambda\hat A_{xy}
=S_1+S_2+S_3+S_4,
\label{eq:shear_pathway_balance}
\end{equation}
where
\begin{subequations}
\begin{align}
S_1&=g\hat A_{yy},
\label{eq:S1}\\
S_2&=\bar A_{yy}D\hat u,
\label{eq:S2}\\
S_3&=-\hat v D\bar A_{xy},
\label{eq:S3}\\
S_4&=\bar A_{xx}ik\hat v.
\label{eq:S4}
\end{align}
\label{eq:shear_pathways}
\end{subequations}
The $S_1$ pathway is generated by the action of the base shear on the perturbed wall-normal conformation. The $S_2$ pathway couples the base wall-normal conformation to $D\hat u$, while $S_3$ represents wall-normal advection of the base shear-conformation gradient. Finally, $S_4$ couples the stretched streamwise base conformation to $ik\hat v$.

The stress contributions generated by these sources are defined as
\begin{subequations}
\begin{align}
\hat\tau_N^{(j)}
&=\frac{1-\beta}{Wi}\frac{N_j}{\Lambda},
\\
\hat\tau_S^{(j)}
&=\frac{1-\beta}{Wi}\frac{S_j}{\Lambda},
\qquad j=1,\ldots,4.
\end{align}
\label{eq:pathway_stresses}
\end{subequations}
They satisfy the exact additive relations
\begin{subequations}
\begin{align}
\hat\tau_{p,xx}-\hat\tau_{p,yy}
&=\sum_{j=1}^{4}\hat\tau_N^{(j)},\\
\hat\tau_{p,xy}
&=\sum_{j=1}^{4}\hat\tau_S^{(j)}.
\end{align}
\label{eq:pathway_stress_sums}
\end{subequations}
The factor $\Lambda^{-1}$ converts each constitutive source into a conformation response and changes both its magnitude and phase. The additional factor $(1-\beta)/Wi$ converts that response into polymer stress.

Substitution into equations~\eqref{eq:normal_work} and \eqref{eq:shear_work} gives
\begin{subequations}
\begin{align}
W_{Nj}
&=\Real\int_{y_1}^{y_2}
ik\hat u^*\hat\tau_N^{(j)}\,\mathrm{d}y,
\\
W_{Sj}
&=-\Real\int_{y_1}^{y_2}
\hat\gamma^*\hat\tau_S^{(j)}\,\mathrm{d}y,
\end{align}
\label{eq:pathway_work_rates}
\end{subequations}
and hence
\begin{subequations}
\begin{align}
W_N&=\sum_{j=1}^{4}W_{Nj},\\
W_S&=\sum_{j=1}^{4}W_{Sj}.
\end{align}
\label{eq:pathway_work_sums}
\end{subequations}

Each pathway identifies the constitutive term that generates a particular contribution to the polymer stress. The mechanical work of that stress contribution is determined by its interaction with the corresponding deformation field. For example, $S_4$ is generated by the constitutive source $\bar A_{xx}ik\hat v$, but the resulting shear stress performs work against the complete shear deformation $\hat\gamma=D\hat u+ik\hat v$. All pathway contributions are computed from the same fully coupled eigenmode and do not represent independent body forces or stress fields obtained from separate stability problems. Their sum therefore recovers the internal polymeric work exactly.

The relative importance of the pathways changes with $Wi$ through both the base conformation and the response factor $\Lambda^{-1}$. The following subsection identifies the leading and subleading stress contributions in the limits $Wi\ll1$ and $Wi\gg1$. 

\subsubsection{Asymptotic interpretation}
\label{sec:asymptotic_pathways}

The non-monotonic dependence of the instability on the Weissenberg number can be interpreted by examining how the polymeric-stress pathways change between the small- and large-$Wi$ limits. For fixed $k$ and $\Sigma=O(1)$, the orderings below describe their explicit constitutive dependence on $Wi$. For the representative $k=10$ and low-wavenumber eigenmodes analysed in \S\S~3.5 and 3.6, $|\Sigma|$ is typically $O(0.1-10)$ away from convective matching positions. Accordingly, $Wi=0.01-0.03$ approximately corresponds to $Wi|\Sigma|\ll1$, $Wi=0.1-3$ spans the crossover regime, and $Wi\geq10$ generally approaches $Wi|\Sigma|\gg1$. These classifications are not spatially uniform because $\Sigma$ depends on $y$.

When $Wi|\Sigma|\ll1$,
\begin{equation}
\frac{1}{\Lambda}
=Wi\left[1-Wi\Sigma+O(Wi^2)\right].
\label{eq:small_Wi_response}
\end{equation}
Using $\bar A_{xx}=1+2Wi^2g^2$, $\bar A_{xy}=Wig$, and $\bar A_{yy}=1$ in equations~\eqref{eq:normal_pathways}--\eqref{eq:pathway_stresses} gives
\begin{equation}
\hat\tau_N^{(3)}=O(1),\qquad
\hat\tau_N^{(1)},\hat\tau_N^{(2)}=O(Wi),\qquad
\hat\tau_N^{(4)}=O(Wi^2).
\label{eq:small_Wi_normal_ordering}
\end{equation}
Thus, $N_3$ generates the leading normal-stress difference. The first correction contains the $O(Wi)$ response-factor correction to $N_3$ together with the leading $N_1$ and $N_2$ contributions; $N_4$ enters one order later.

For the shear stress,
\begin{equation}
\hat\tau_S^{(2)},\hat\tau_S^{(4)}=O(1),
\qquad
\hat\tau_S^{(1)},\hat\tau_S^{(3)}=O(Wi).
\label{eq:small_Wi_shear_ordering}
\end{equation}
The leading $S_2$ and $S_4$ stresses combine to give
\begin{equation}
\hat\tau_{p,xy}
=(1-\beta)(D\hat u+ik\hat v)+O(Wi)
=(1-\beta)\hat\gamma+O(Wi).
\label{eq:small_Wi_shear_stress}
\end{equation}
Although $S_2$ and $S_4$ have the same formal order in $Wi$, $S_2$ provides most of the computed shear work at small $Wi$ because the $k=10$ and low-$k$ eigenfunctions are dominated by $D\hat u$ in the mechanically active ESC. This observed work dominance is therefore an eigenfunction property, not an additional asymptotic power of $Wi$.

Combining the leading normal and shear stresses gives
\begin{subequations}
\begin{align}
w_N
&=-4(1-\beta)k^2|\hat u|^2+O(Wi),
\\
w_S
&=-(1-\beta)|\hat\gamma|^2+O(Wi).
\end{align}
\label{eq:small_Wi_work}
\end{subequations}
The polymer contribution consequently recovers Newtonian viscous dissipation at leading order. The $O(Wi)$ corrections are not sign definite because they depend on the complex response factor and on the phase relations within the eigenmode.

In the opposite limit, $Wi|\Sigma|\gg1$,
\begin{equation}
\frac{1}{\Lambda}
=
\frac{1}{\Sigma}
\left[
\frac{1}{1+(Wi\Sigma)^{-1}}
\right]
=
\frac{1}{\Sigma}
\left[
1-\frac{1}{Wi\Sigma}
+\frac{1}{Wi^2\Sigma^2}
+O(Wi^{-3})
\right].
\label{eq:large_Wi_response}
\end{equation}

The normal pathways then satisfy
\begin{equation}
\hat\tau_N^{(1)},\hat\tau_N^{(3)},\hat\tau_N^{(4)}
=O(Wi),
\qquad
\hat\tau_N^{(2)}=O(1).
\label{eq:large_Wi_normal_ordering}
\end{equation}
Hence, $N_1$, $N_3$, and $N_4$ form the leading normal-stress response. The leading $N_2$ term and the response-factor corrections to the other three pathways constitute the first subleading, $O(1)$, response. The presence of $N_3$ at leading order in stress magnitude does not guarantee leading-order work: its mechanical contribution can become small when its phase efficiency decreases.

For the shear pathways,
\begin{equation}
\hat\tau_S^{(4)}=O(Wi),\qquad
\hat\tau_S^{(1)},\hat\tau_S^{(3)}=O(1),\qquad
\hat\tau_S^{(2)}=O(Wi^{-1}).
\label{eq:large_Wi_shear_ordering}
\end{equation}
The $S_4$ pathway is therefore the leading shear-stress response, $S_1$ and $S_3$ provide the first subleading contributions, and the Newtonian-like $S_2$ pathway becomes asymptotically weak.

These limits identify a transition from a Newtonian-like polymer response at small $Wi$ to a response governed by the highly stretched streamwise base conformation at large $Wi$. They do not, however, determine the signed work at $Wi=O(1)$, where neither expansion is uniformly valid. Moreover, the large-$Wi$ ordering requires $Wi|\Sigma|\gg1$ throughout the mechanically active region. For a near-neutral travelling mode,
\begin{equation}
\Sigma(y)
=\sigma_r+ik[\bar{U}(y)-c],
\qquad
c=-\frac{\sigma_i}{k}.
\label{eq:convective_response}
\end{equation}
The expansion can therefore become non-uniform near a convective matching position $\bar{U}(y_c)=c$. This limitation is important for the slowly propagating low-wavenumber modes examined in \S~\ref{sec:low_k}.

Finally, equations~\eqref{eq:small_Wi_normal_ordering}--\eqref{eq:large_Wi_shear_ordering} rank stress amplitudes rather than signed work. The relative phase between each pathway stress and its associated deformation determines whether that pathway supplies or extracts mechanical power.

\subsubsection{Phase dependence of mechanical work}
\label{sec:phase_control}

For complex amplitudes $a$ and $b$ with phase angles
$\theta_a$ and $\theta_b$,
\begin{equation}
\Real(a^*b)
=
|a||b|\cos(\theta_b-\theta_a).
\label{eq:complex_phase_identity}
\end{equation}
It is therefore convenient to introduce the effective deformation amplitudes
\begin{equation}
\hat d_N=-ik\hat u,
\qquad
\hat d_S=-\hat\gamma.
\label{eq:effective_deformations}
\end{equation}
The local pathway work densities can then be written without an additional sign convention:
\begin{subequations}
\begin{align}
w_{Nj}
&=
\Real\left[(\hat d_N)^*\hat\tau_N^{(j)}\right]
=
|\hat d_N||\hat\tau_N^{(j)}|
\cos\left(\theta_{\hat \tau,Nj}-\theta_{\hat d,N}\right),
\\
w_{Sj}
&=
\Real\left[(\hat d_S)^*\hat\tau_S^{(j)}\right]
=
|\hat d_S||\hat\tau_S^{(j)}|
\cos\left(\theta_{\hat\tau,Sj}-\theta_{\hat d,S}\right).
\end{align}
\label{eq:local_phase_work}
\end{subequations}
A pathway performs little signed work when the stress and effective
deformation approach quadrature, even if the stress magnitude is large.

To separate the available stress--deformation amplitude from its phase
utilization, we define the magnitude capacities
\begin{subequations}
\begin{align}
\mathcal M_{Nj}
&=
\int_{y_1}^{y_2}
|\hat d_N||\hat\tau_N^{(j)}|\,\mathrm{d}y,
\\
\mathcal M_{Sj}
&=
\int_{y_1}^{y_2}
|\hat d_S||\hat\tau_S^{(j)}|\,\mathrm{d}y,
\end{align}
\label{eq:pathway_magnitude_capacity}
\end{subequations}
and the integrated phase efficiencies
\begin{equation}
C_{\alpha j}
=
\frac{W_{\alpha j}}{\mathcal M_{\alpha j}},
\qquad
-1\leq C_{\alpha j}\leq1,
\qquad
\alpha=N,S.
\label{eq:pathway_phase_efficiency}
\end{equation}

Here, $C_{\alpha j}>0$ denotes net mechanical-power input, $C_{\alpha j}<0$ denotes net extraction, and $C_{\alpha j}\simeq0$ denotes weak coherent work owing to phase cancellation. These quantities diagnose the mechanical action of a pathway within a selected coupled eigenmode; the growth or decay of that mode is still determined by the eigenvalue of the full electrohydrodynamic--viscoelastic system.

\subsection{Polymeric-work pathways and eigenmode evolution at $k=10$}
\label{sec:k10_pathways}
For the reference parameter set $Pe=10$, $\beta=0.5$, and $\varepsilon=10^{-4}$, the critical voltage at $k=10$ decreases from $V_{\mathrm{cr}}=30.50$ at $Wi=0.01$ to $V_{\mathrm{cr}}=25.98$ at $Wi=1$ and subsequently increases to $V_{\mathrm{cr}}=33.40$ at $Wi=30$. To identify the mechanism underlying this non-monotonic variation, we evaluate the polymeric-work pathways using the neutral eigenmode at each $Wi$.

The following analysis applies the framework developed in \S\S~3.4.2--3.4.4. The normal-stress-difference pathways $N_1$--$N_4$ and the shear-stress pathways $S_1$--$S_4$ are assessed using their asymptotic orderings, spatial work densities, regional integrals, and phase efficiencies. These quantities are considered together because the signed mechanical work depends on both the stress magnitude and its phase relation with the corresponding perturbation deformation. We first examine the normal-stress-difference pathways, followed by the shear-stress pathways and their combined contribution to the polymeric work. The final subsection examines the evolution of the $k=10$ eigenspectrum at the fixed voltage $V=30$, matching the fixed-voltage condition used in \S\S~3.1--3.3 and allowing the spectral changes with $Wi$ to be compared under the same operating condition.

\subsubsection{Normal-stress-difference pathways}
\label{sec:k10_normal}

\begin{figure}
\centerline{\includegraphics[scale=0.45]{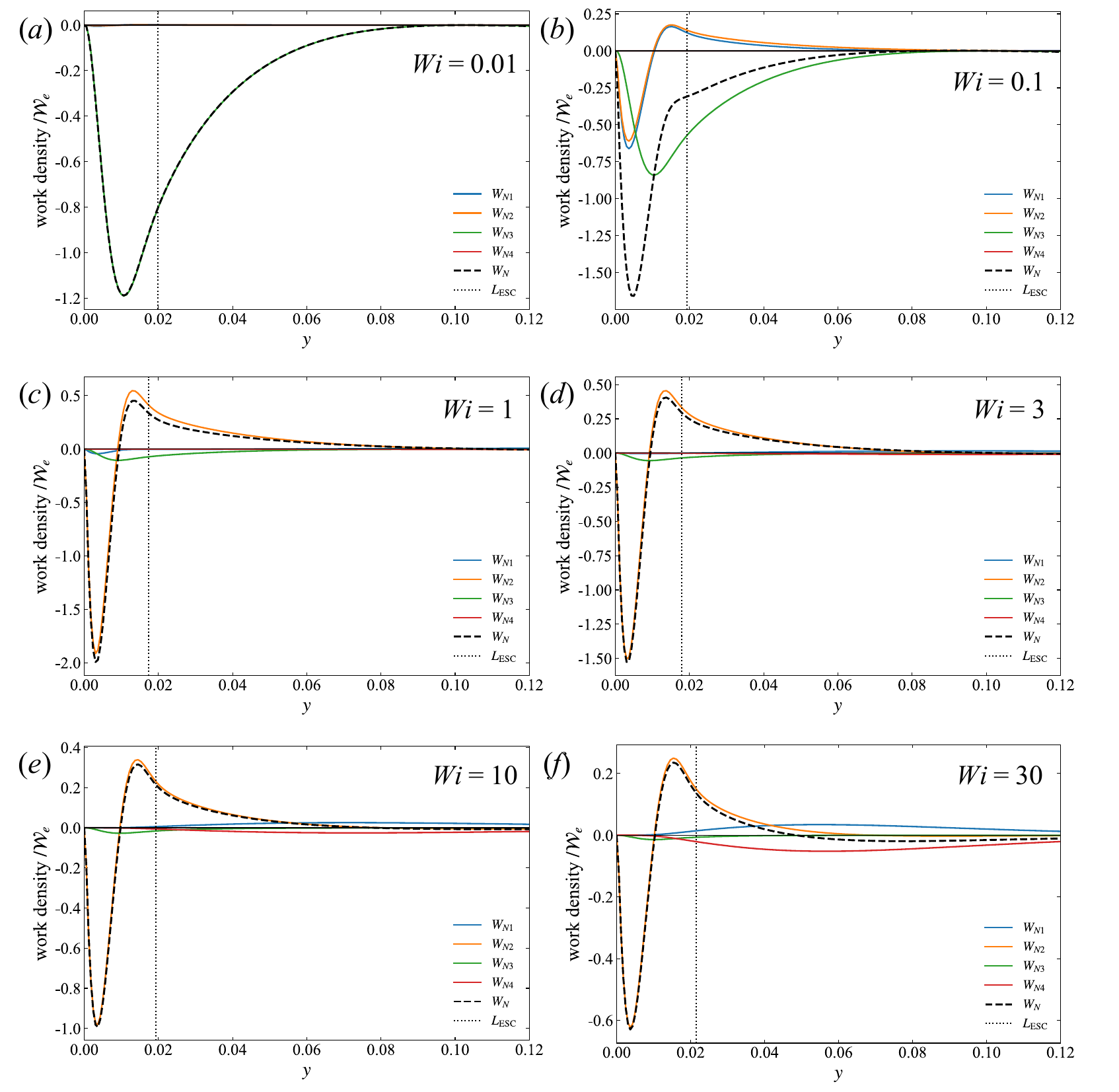}}
\caption{Wall-normal profiles of the normalized normal-stress pathway work densities, $W_{N_1}$, $W_{N_2}$, $W_{N_3}$, and $W_{N_4}$, for the $k=10$ mode at ($a$) $Wi=0.01$, ($b$) $Wi=0.1$, ($c$) $Wi=1$, ($d$) $Wi=3$, ($e$) $Wi=10$, and ($f$) $Wi=30$. All work densities are normalized by the magnitude of the full-domain electric work, $\mathcal{W}_e$, for the corresponding eigenmode.}
\label{fig:k10_normal_work}
\end{figure}

The eigenfunctions obtained from the linear stability analysis are rescaled according to (\ref{eq:work_normalization}), such that the magnitude of the total electric work is unity. Figure~\ref{fig:k10_normal_work} shows the resulting wall-normal profiles of the normalized normal-stress pathway work densities. Their integrals over the ESC, bulk, and full domain are reported in table~\ref{tab:k10_normal_work}. The corresponding phase efficiencies are reported in table~\ref{tab:k10_phase_efficiency} in Appendix~C.

Figure~\ref{fig:k10_normal_work} and table~\ref{tab:k10_normal_work} show that the full-domain normal-stress-difference work remains negative over the investigated range, although the pathway balance changes substantially with $Wi$. At $Wi=0.01$, $N_3$ accounts for nearly all of $W_N$ in the ESC, bulk, and full domain (figure~\ref{fig:k10_normal_work}$a$). At $Wi=0.1$, $N_3$ remains dominant, and the local amplitudes of the $N_1$ and $N_2$ contributions become appreciable (figure~\ref{fig:k10_normal_work}$b$). This behaviour agrees with equation~\eqref{eq:small_Wi_normal_ordering}: $N_3$ provides the leading Newtonian-like normal response, and $N_1$ and $N_2$ enter at $O(Wi)$.

Near $Wi=O(1)$, the individual normal pathways become appreciable but cancel strongly across both pathways and spatial regions. In particular, table~\ref{tab:k10_normal_work} shows that $N_2$ produces large contributions of opposite sign in the ESC and bulk at $Wi=1$ and $3$. The small full-domain value of $W_N/\mathcal{W}_e$ therefore results from cancellation and does not indicate uniformly weak local stress--deformation coupling. The phase efficiency of the small-$Wi$ $N_3$ pathway also decreases sharply. Its full-domain value changes from $C_{N3}=-1.000$ at $Wi=0.01$ to $-0.770$ at $Wi=0.1$, $-0.123$ at $Wi=1$, and $-0.03$ at $Wi=3$ (table~\ref{tab:k10_phase_efficiency}). The $N_3$ stress consequently performs progressively less negative work as its phase relation with the normal deformation becomes less effective.

At $Wi=10$ and $30$, the full-domain balance is controlled primarily by positive $N_1$ work and negative $N_4$ work. Their work densities become more prominent than that of $N_3$ in figures~\ref{fig:k10_normal_work}($e,f$). The magnitude of the negative $N_4$ contribution exceeds that of the positive $N_1$ contribution, and the combined balance maintains $W_N/\mathcal{W}_e<0$. The $N_2$ contribution remains negative within the ESC and positive in the bulk. These regional contributions almost cancel in the full-domain integral, as shown in table~\ref{tab:k10_normal_work}. The pathway governing the local ESC work is therefore not necessarily the pathway governing the full-domain balance.

The normal-stress results demonstrate the distinction between stress-amplitude and mechanical-work dominance. The large-$Wi$ stress expansion contains a leading $N_3$ contribution, but its phase efficiency approaches zero as $Wi$ increases. The signed normal work consequently shifts toward the $N_1$ and $N_4$.

\begin{table}
  \begin{center}
  \begin{tabular}{ccrrrrr}
    $Wi$ & Region & $W_N/\mathcal{W}_e$ & $W_{N_1}/\mathcal{W}_e$ & $W_{N_2}/\mathcal{W}_e$ & $W_{N_3}/\mathcal{W}_e$ & $W_{N_4}/\mathcal{W}_e$ \\[3pt]

    0.01 & ESC   & $-1.657\times10^{-2}$ & $-1.112\times10^{-5}$ & $-4.386\times10^{-6}$ & $-1.656\times10^{-2}$ & $ 5.877\times10^{-8}$ \\
         & Bulk  & $-1.992\times10^{-2}$ & $ 1.958\times10^{-5}$ & $ 1.273\times10^{-5}$ & $-1.995\times10^{-2}$ & $ 1.667\times10^{-8}$ \\
         & Total & $-3.649\times10^{-2}$ & $ 8.458\times10^{-6}$ & $ 8.345\times10^{-6}$ & $-3.651\times10^{-2}$ & $ 7.544\times10^{-8}$ \\[3pt]

    0.1  & ESC   & $-1.673\times10^{-2}$ & $-2.943\times10^{-3}$ & $-2.428\times10^{-3}$ & $-1.136\times10^{-2}$ & $ 3.798\times10^{-6}$ \\
         & Bulk  & $-1.047\times10^{-2}$ & $ 2.555\times10^{-3}$ & $ 2.588\times10^{-3}$ & $-1.561\times10^{-2}$ & $-6.916\times10^{-6}$ \\
         & Total & $-2.720\times10^{-2}$ & $-3.883\times10^{-4}$ & $ 1.595\times10^{-4}$ & $-2.697\times10^{-2}$ & $-3.118\times10^{-6}$ \\[3pt]

    1    & ESC   & $-8.068\times10^{-3}$ & $-3.088\times10^{-4}$ & $-6.476\times10^{-3}$ & $-1.289\times10^{-3}$ & $ 6.424\times10^{-6}$ \\
         & Bulk  & $ 5.474\times10^{-3}$ & $ 1.622\times10^{-3}$ & $ 7.148\times10^{-3}$ & $-2.581\times10^{-3}$ & $-7.150\times10^{-4}$ \\
         & Total & $-2.594\times10^{-3}$ & $ 1.313\times10^{-3}$ & $ 6.719\times10^{-4}$ & $-3.870\times10^{-3}$ & $-7.085\times10^{-4}$ \\[3pt]

    3    & ESC   & $-5.730\times10^{-3}$ & $-5.224\times10^{-5}$ & $-4.984\times10^{-3}$ & $-6.907\times10^{-4}$ & $-2.761\times10^{-6}$ \\
         & Bulk  & $ 4.934\times10^{-3}$ & $ 2.519\times10^{-3}$ & $ 5.219\times10^{-3}$ & $-1.070\times10^{-3}$ & $-1.734\times10^{-3}$ \\
         & Total & $-7.958\times10^{-4}$ & $ 2.466\times10^{-3}$ & $ 2.349\times10^{-4}$ & $-1.760\times10^{-3}$ & $-1.737\times10^{-3}$ \\[3pt]

    10   & ESC   & $-3.403\times10^{-3}$ & $ 2.225\times10^{-5}$ & $-3.029\times10^{-3}$ & $-3.639\times10^{-4}$ & $-3.253\times10^{-5}$ \\
         & Bulk  & $ 2.514\times10^{-3}$ & $ 3.006\times10^{-3}$ & $ 3.067\times10^{-3}$ & $-4.052\times10^{-4}$ & $-3.155\times10^{-3}$ \\
         & Total & $-8.897\times10^{-4}$ & $ 3.029\times10^{-3}$ & $ 3.832\times10^{-5}$ & $-7.691\times10^{-4}$ & $-3.188\times10^{-3}$ \\[3pt]

    30   & ESC   & $-1.917\times10^{-3}$ & $ 8.615\times10^{-5}$ & $-1.673\times10^{-3}$ & $-2.016\times10^{-4}$ & $-1.288\times10^{-4}$ \\
         & Bulk  & $-1.629\times10^{-4}$ & $ 2.914\times10^{-3}$ & $ 1.657\times10^{-3}$ & $-1.528\times10^{-4}$ & $-4.581\times10^{-3}$ \\
         & Total & $-2.080\times10^{-3}$ & $ 3.000\times10^{-3}$ & $-1.592\times10^{-5}$ & $-3.544\times10^{-4}$ & $-4.710\times10^{-3}$
  \end{tabular}
  \caption{Normalized normal-stress pathway work integrated over the ESC, bulk, and full domain for the $k=10$ mode. The eigenfunctions are normalized by the
  magnitude of the full-domain electric work $\mathcal{W}_e$, and $W_N=\sum_{j=1}^{4}W_{N_j}$.}
  \label{tab:k10_normal_work}
  \end{center}
\end{table}

\subsubsection{Shear-stress pathways}
\label{sec:k10_shear}

\begin{figure}
\centerline{\includegraphics[scale=0.45]{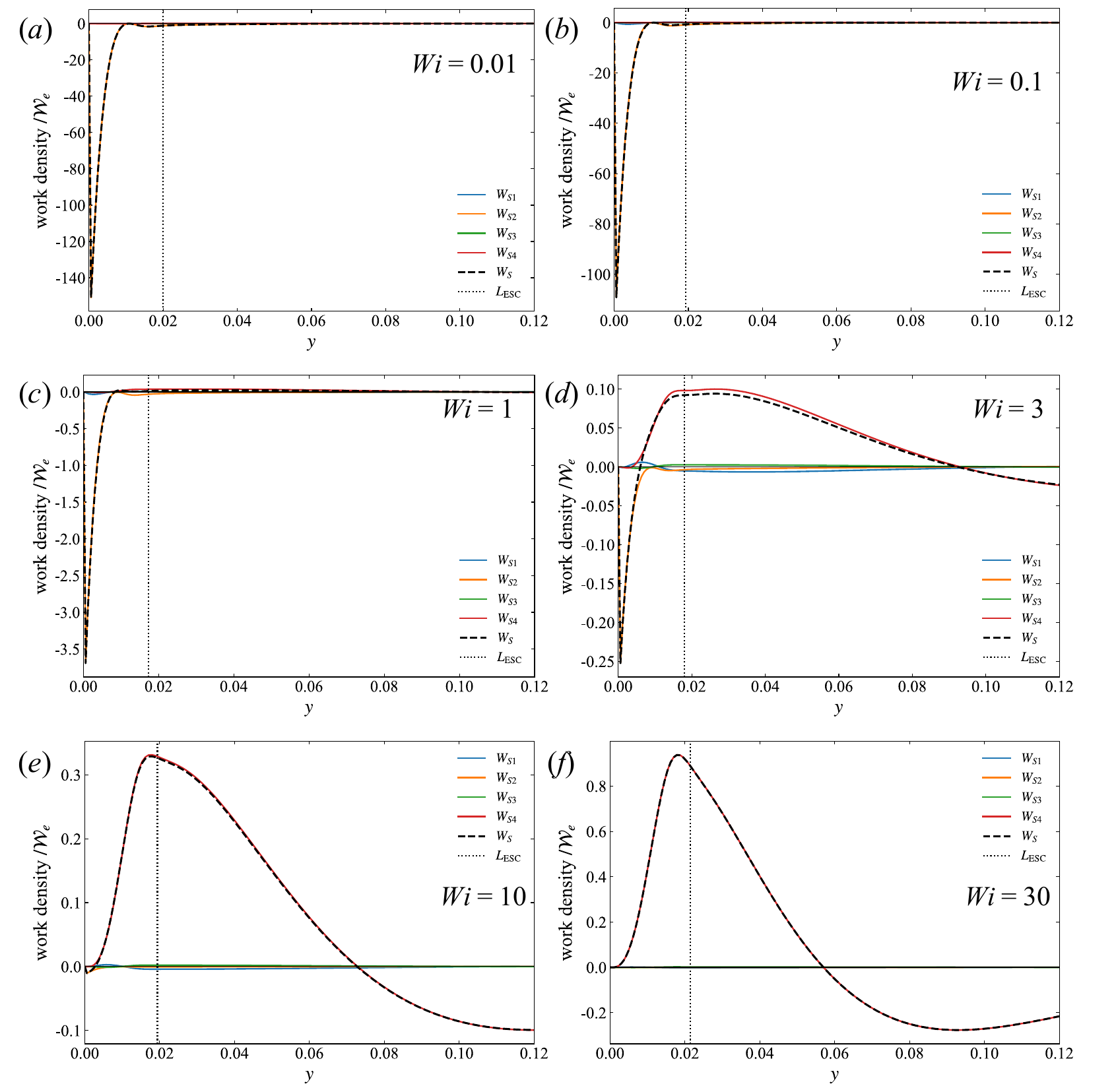}}
\caption{Wall-normal profiles of the normalized shear-stress pathway work densities, $w_{S1}$, $w_{S2}$, $w_{S3}$, and $w_{S4}$, for the $k=10$ mode at ($a$) $Wi=0.01$, ($b$) $Wi=0.1$, ($c$) $Wi=1$, ($d$) $Wi=3$, ($e$) $Wi=10$, and ($f$) $Wi=30$. All work densities are normalized by the magnitude of the full-domain electric work, $\mathcal{W}_e$, for the corresponding eigenmode.}
\label{fig:k10_shear_work}
\end{figure}

Figure~\ref{fig:k10_shear_work} shows the resulting wall-normal profiles of the normalized shear-stress pathway work densities, and their integrals over the ESC, bulk, and full domain are reported in table~\ref{tab:k10_shear_work}. The corresponding phase efficiencies are reported in table~\ref{tab:k10_phase_efficiency} in Appendix~C. These results show that the full-domain shear work changes from negative at small $Wi$ to positive at larger $Wi$.

The small-$Wi$ expansion gives $\hat{\tau}_S^{(2)},\hat{\tau}_S^{(4)}=O(1)$, as shown in equation~\eqref{eq:small_Wi_shear_ordering}. This formal ordering does not determine their relative contributions to the signed work. The full-domain phase efficiencies of $S_2$ and $S_4$ are $-0.997$ and $-0.029$, respectively, at $Wi=0.01$, and $-0.731$ and $-0.089$ at $Wi=0.1$ (table~\ref{tab:k10_phase_efficiency}). Thus, despite having the same formal $O(1)$ ordering, $S_2$ couples much more effectively to the deformation field and provides the dominant negative shear work, producing the Newtonian-like shear response.

This suppressing action weakens rapidly as $Wi$ approaches unity. The full-domain phase efficiency of $S_2$ decreases to $-0.101$ at $Wi=1$ and $-0.026$ at $Wi=3$, accompanied by a substantial reduction in its negative work. At $Wi=3$, $S_4$ performs positive work in the ESC and negative work in the bulk (figure~\ref{fig:k10_shear_work}$d$). The positive ESC contribution is larger, giving positive full-domain $S_4$ work. This contribution nearly balances the remaining negative $S_2$ work, resulting in a small negative value of $W_S/\mathcal{W}_e$. The small full-domain shear work at $Wi=3$ therefore arises from cancellation between pathways and spatial regions.

At $Wi=10$ and $30$, $S_4$ dominates the shear-work density and the regional integrals (figures~\ref{fig:k10_shear_work}$e,f$). Its work is positive in the ESC and negative in the bulk, with the ESC contribution providing the larger magnitude. The resulting full-domain shear work is positive. The increasing prominence of $S_4$ agrees with the large-$Wi$ scaling $\hat{\tau}_S^{(4)}=O(Wi)$ in equation~\eqref{eq:large_Wi_shear_ordering}. Its full-domain phase efficiency is positive, with values of $0.029$ at $Wi=10$ and $0.026$ at $Wi=30$ (table~\ref{tab:k10_phase_efficiency}). The large magnitude of the $S_4$ contribution and its positive phase efficiency therefore produce positive full-domain shear work.

The shear-work transition is thus characterized by the loss of the mechanically effective $S_2$ response and the emergence of a large $S_4$ contribution with positive full-domain work.

\begin{table}
  \begin{center}
  \begin{tabular}{ccrrrrr}
    $Wi$ & Region & $W_S/\mathcal{W}_e$ & $W_{S_1}/\mathcal{W}_e$ & $W_{S_2}/\mathcal{W}_e$ & $W_{S_3}/\mathcal{W}_e$ & $W_{S_4}/\mathcal{W}_e$ \\[3pt]

    0.01 & ESC   & $-4.415\times10^{-1}$ & $-3.160\times10^{-5}$ & $-4.417\times10^{-1}$ & $-1.795\times10^{-8}$ & $ 2.312\times10^{-4}$ \\
         & Bulk  & $-2.181\times10^{-2}$ & $ 1.158\times10^{-5}$ & $-2.128\times10^{-2}$ & $ 3.166\times10^{-7}$ & $-5.457\times10^{-4}$ \\
         & Total & $-4.633\times10^{-1}$ & $-2.002\times10^{-5}$ & $-4.630\times10^{-1}$ & $ 2.987\times10^{-7}$ & $-3.144\times10^{-4}$ \\[3pt]

    0.1  & ESC   & $-3.110\times10^{-1}$ & $-2.734\times10^{-3}$ & $-3.085\times10^{-1}$ & $ 3.093\times10^{-6}$ & $ 1.813\times10^{-4}$ \\
         & Bulk  & $-1.439\times10^{-2}$ & $ 2.630\times10^{-3}$ & $-1.587\times10^{-2}$ & $ 8.121\times10^{-6}$ & $-1.159\times10^{-3}$ \\
         & Total & $-3.254\times10^{-1}$ & $-1.041\times10^{-4}$ & $-3.244\times10^{-1}$ & $ 1.121\times10^{-5}$ & $-9.775\times10^{-4}$ \\[3pt]

    1    & ESC   & $-9.088\times10^{-3}$ & $-5.708\times10^{-5}$ & $-9.276\times10^{-3}$ & $ 7.715\times10^{-6}$ & $ 2.373\times10^{-4}$ \\
         & Bulk  & $-5.887\times10^{-4}$ & $ 6.610\times10^{-4}$ & $-3.644\times10^{-4}$ & $-7.592\times10^{-5}$ & $-8.093\times10^{-4}$ \\
         & Total & $-9.676\times10^{-3}$ & $ 6.039\times10^{-4}$ & $-9.640\times10^{-3}$ & $-6.820\times10^{-5}$ & $-5.720\times10^{-4}$ \\[3pt]

    3    & ESC   & $ 1.980\times10^{-4}$ & $ 1.038\times10^{-5}$ & $-6.637\times10^{-4}$ & $ 8.574\times10^{-6}$ & $ 8.427\times10^{-4}$ \\
         & Bulk  & $-2.969\times10^{-4}$ & $-5.384\times10^{-5}$ & $-4.224\times10^{-6}$ & $ 2.699\times10^{-5}$ & $-2.658\times10^{-4}$ \\
         & Total & $-9.897\times10^{-5}$ & $-4.346\times10^{-5}$ & $-6.679\times10^{-4}$ & $ 3.557\times10^{-5}$ & $ 5.768\times10^{-4}$ \\[3pt]

    10   & ESC   & $ 3.202\times10^{-3}$ & $-1.092\times10^{-5}$ & $-3.329\times10^{-5}$ & $ 9.904\times10^{-6}$ & $ 3.237\times10^{-3}$ \\
         & Bulk  & $-1.449\times10^{-3}$ & $-1.359\times10^{-4}$ & $ 4.220\times10^{-6}$ & $ 6.680\times10^{-5}$ & $-1.384\times10^{-3}$ \\
         & Total & $ 1.753\times10^{-3}$ & $-1.468\times10^{-4}$ & $-2.907\times10^{-5}$ & $ 7.671\times10^{-5}$ & $ 1.852\times10^{-3}$ \\[3pt]

    30   & ESC   & $ 1.043\times10^{-2}$ & $-1.343\times10^{-5}$ & $-3.130\times10^{-6}$ & $ 1.132\times10^{-5}$ & $ 1.044\times10^{-2}$ \\
         & Bulk  & $-7.126\times10^{-3}$ & $-7.212\times10^{-5}$ & $ 1.581\times10^{-6}$ & $ 5.271\times10^{-5}$ & $-7.109\times10^{-3}$ \\
         & Total & $ 3.304\times10^{-3}$ & $-8.555\times10^{-5}$ & $-1.549\times10^{-6}$ & $ 6.404\times10^{-5}$ & $ 3.327\times10^{-3}$
  \end{tabular}
  \caption{Normalized shear-stress pathway work integrated over the ESC, bulk, and full domain for the $k=10$ mode. The eigenfunctions are normalized by the magnitude of the full-domain electric work $\mathcal{W}_e$, and $W_S=\sum_{j=1}^{4}W_{Sj}$.}
  \label{tab:k10_shear_work}
  \end{center}
\end{table}

\subsubsection{Overall mechanical-work transition}
\label{sec:k10_overall}
The minimum in $V_{\mathrm{cr}}$ near $Wi\sim O(1)$ is associated with a crossover between the small- and large-$Wi$ polymeric responses. In the $Wi\ll1$ limit, the negative polymeric work is dominated by the mechanically effective Newtonian-like $N_3$ and $S_2$ pathways. As $Wi$ approaches unity, their phase efficiencies decrease substantially. Moreover, the $N_1$, $N_4$, and $S_4$ pathways characteristic of the large-$Wi$ balance remain weak over this range. Consequently, the magnitude of the negative polymeric work decreases sharply as $Wi$ approaches unity. This reduction coincides with the decrease in $V_{\mathrm{cr}}$ to its minimum at $Wi=1$. At $Wi=3$, both $W_N$ and $W_S$ remain negative but small because of cancellation among the individual pathways and between the ESC and bulk contributions, marking the crossover to the large-$Wi$ pathway structure.

At $Wi=10$ and $30$, the normal- and shear-stress contributions perform opposing mechanical roles. The $N_1$--$N_4$ balance maintains $W_N<0$, and the $S_4$ pathway produces $W_S>0$. The positive shear contribution exceeds the negative normal contribution, resulting in $W_p^{\mathrm{int}}/\mathcal W_e=8.6\times10^{-4}$ at $Wi=10$ and $1.22\times10^{-3}$ at $Wi=30$. The net polymeric work is therefore weakly positive, so the increase in $V_{\mathrm{cr}}$ cannot be attributed to direct mechanical-power extraction by the polymer stress.

Accordingly, a complete assessment of the mechanical effect of viscoelasticity requires consideration of the fully coupled neutral eigenmode. Each value of $Wi$ corresponds to a different neutral voltage, base state, and fully coupled eigenmode. Changes in the polymeric response can therefore be accompanied by changes in the velocity-deformation field and the associated solvent-viscous dissipation. We next quantify these changes.

Using incompressibility, the solvent deformation measure can be separated into normal and shear contributions:
\begin{subequations}
\label{eq:k10_solvent_deformation}
\begin{align}
\mathcal D
&=\mathcal D_N+\mathcal D_S, \\
\mathcal D_N
&=\int_0^2 2|ik\hat u|^2\,\mathrm{d}y, \\
\mathcal D_S
&=\int_0^2 \frac{1}{2}|D\hat u+ik\hat v|^2\,\mathrm{d}y.
\end{align}
\end{subequations}
The corresponding solvent-work contributions are
\begin{equation}
W_{\mathrm{sol},N}^{\mathrm{int}}
=-2\beta\mathcal D_N,
\qquad
W_{\mathrm{sol},S}^{\mathrm{int}}
=-2\beta\mathcal D_S,
\qquad
W_{\mathrm{sol}}^{\mathrm{int}}
=W_{\mathrm{sol},N}^{\mathrm{int}}
+W_{\mathrm{sol},S}^{\mathrm{int}}.
\label{eq:k10_solvent_work_decomposition}
\end{equation}
To distinguish changes in the velocity-field structure from changes in its overall amplitude, we introduce
\begin{equation}
K_u
=
\int_0^2
\left(
|\hat u|^2+|\hat v|^2
\right)\mathrm{d}y.
\label{eq:velocity_amplitude_measure}
\end{equation}
The quantity $K_u$ provides a scalar measure of the overall amplitude of the perturbation velocity eigenfunction. Normalization by $\mathcal W_e$ measures the contribution of each term to the mechanical-work balance. Normalization by $K_u$ measures the deformation relative to the squared $L^2$-norm of the perturbation velocity. The latter normalization therefore distinguishes changes in the spatial deformation structure from changes caused solely by the overall eigenfunction amplitude.

Table~\ref{tab:k10_solvent_work} shows that shear deformation accounts for $92.7$--$94.6\%$ of the full-domain solvent dissipation over $0.01\leq Wi\leq30$. Moreover, $92.5$--$95.9\%$ of this shear dissipation occurs within the ESC layer. Thus, the solvent dissipation is dominated by shear deformation localized within the ESC region. 

 The full-domain value of $-W_{\mathrm{sol}}^{\mathrm{int}}/K_u$ increases progressively from $1346$ at $Wi=0.01$ to $2734$ at $Wi=30$, corresponding to a factor of $2.03$. Over the same range, $K_u$ and the total solvent deformation $\mathcal D$ increase by factors of $2.26$ and $4.59$, respectively. The increase in $\mathcal D/K_u$ demonstrates that the increase in solvent dissipation cannot be attributed solely to amplification of the perturbation velocity under the charge-density normalization. The coupled eigenmode develops a more strongly shear-deforming velocity structure relative to the overall velocity magnitude.

Consistent with this structural change, $W_{\mathrm{sol}}^{\mathrm{int}}/\mathcal W_e$ decreases from $-0.50$ at $Wi=0.01$ to approximately $-1.00$ at $Wi=10$ and $30$. The net polymeric work becomes very small over the same range, leaving solvent viscosity to account for nearly all of the mechanical-power extraction at neutrality. Therefore, the increase in $V_{\mathrm{cr}}$ beyond $Wi=1$ is accompanied by a transition toward a solvent-dominated work balance and stronger ESC-localized shear deformation. This result does not establish a direct causal relation between an individual polymer pathway and the velocity deformation. It establishes that the large-$Wi$ polymeric pathway structure and the solvent-dominated deformation field emerge together in the fully coupled neutral eigenmode.

\begin{table}
  \begin{center}
  \begin{tabular}{ccrrrr}
    $Wi$ & Region
    & $W_{\mathrm{sol}}/\mathcal{W}_e$
    & $W_{\mathrm{sol}}/K_u$
    & $W_{\mathrm{sol},N}/K_u$
    & $W_{\mathrm{sol},S}/K_u$ \\[3pt]

    0.01 & ESC
    & $-4.584\times10^{-1}$ & $-1.234\times10^{3}$
    & $-4.458\times10^{1}$ & $-1.189\times10^{3}$ \\
         & Bulk
    & $-4.178\times10^{-2}$ & $-1.125\times10^{2}$
    & $-5.370\times10^{1}$ & $-5.876\times10^{1}$ \\
         & Total
    & $-5.002\times10^{-1}$ & $-1.346\times10^{3}$
    & $-9.827\times10^{1}$ & $-1.248\times10^{3}$ \\[3pt]

    0.1 & ESC
    & $-5.956\times10^{-1}$ & $-1.355\times10^{3}$
    & $-4.566\times10^{1}$ & $-1.310\times10^{3}$ \\
        & Bulk
    & $-5.171\times10^{-2}$ & $-1.176\times10^{2}$
    & $-5.483\times10^{1}$ & $-6.282\times10^{1}$ \\
        & Total
    & $-6.473\times10^{-1}$ & $-1.473\times10^{3}$
    & $-1.005\times10^{2}$ & $-1.372\times10^{3}$ \\[3pt]

    1 & ESC
    & $-9.140\times10^{-1}$ & $-1.550\times10^{3}$
    & $-4.142\times10^{1}$ & $-1.509\times10^{3}$ \\
      & Bulk
    & $-7.372\times10^{-2}$ & $-1.250\times10^{2}$
    & $-6.060\times10^{1}$ & $-6.442\times10^{1}$ \\
      & Total
    & $-9.877\times10^{-1}$ & $-1.675\times10^{3}$
    & $-1.020\times10^{2}$ & $-1.573\times10^{3}$ \\[3pt]

    3 & ESC
    & $-9.197\times10^{-1}$ & $-1.721\times10^{3}$
    & $-4.768\times10^{1}$ & $-1.673\times10^{3}$ \\
      & Bulk
    & $-7.935\times10^{-2}$ & $-1.484\times10^{2}$
    & $-6.645\times10^{1}$ & $-8.200\times10^{1}$ \\
      & Total
    & $-9.991\times10^{-1}$ & $-1.869\times10^{3}$
    & $-1.141\times10^{2}$ & $-1.755\times10^{3}$ \\[3pt]

    10 & ESC
    & $-9.128\times10^{-1}$ & $-2.072\times10^{3}$
    & $-6.288\times10^{1}$ & $-2.009\times10^{3}$ \\
       & Bulk
    & $-8.803\times10^{-2}$ & $-1.998\times10^{2}$
    & $-7.032\times10^{1}$ & $-1.295\times10^{2}$ \\
       & Total
    & $-1.001\times10^{0}$ & $-2.272\times10^{3}$
    & $-1.332\times10^{2}$ & $-2.139\times10^{3}$ \\[3pt]

    30 & ESC
    & $-9.055\times10^{-1}$ & $-2.472\times10^{3}$
    & $-8.281\times10^{1}$ & $-2.390\times10^{3}$ \\
       & Bulk
    & $-9.572\times10^{-2}$ & $-2.614\times10^{2}$
    & $-6.597\times10^{1}$ & $-1.954\times10^{2}$ \\
       & Total
    & $-1.001\times10^{0}$ & $-2.734\times10^{3}$
    & $-1.488\times10^{2}$ & $-2.585\times10^{3}$
  \end{tabular}
  \caption{Solvent work integrated over the ESC, bulk, and full domain
  for the $k=10$ mode. The first work column is normalized by the magnitude of the full-domain electric work $\mathcal{W}_e$; the remaining columns are normalized by
  the full-domain velocity measure $K_u$.}
  \label{tab:k10_solvent_work}
  \end{center}
\end{table}

\subsubsection{Eigenmode evolution at fixed voltage}
\label{sec:k10_spectrum}

\begin{figure}
\centerline{\includegraphics[scale=0.42]{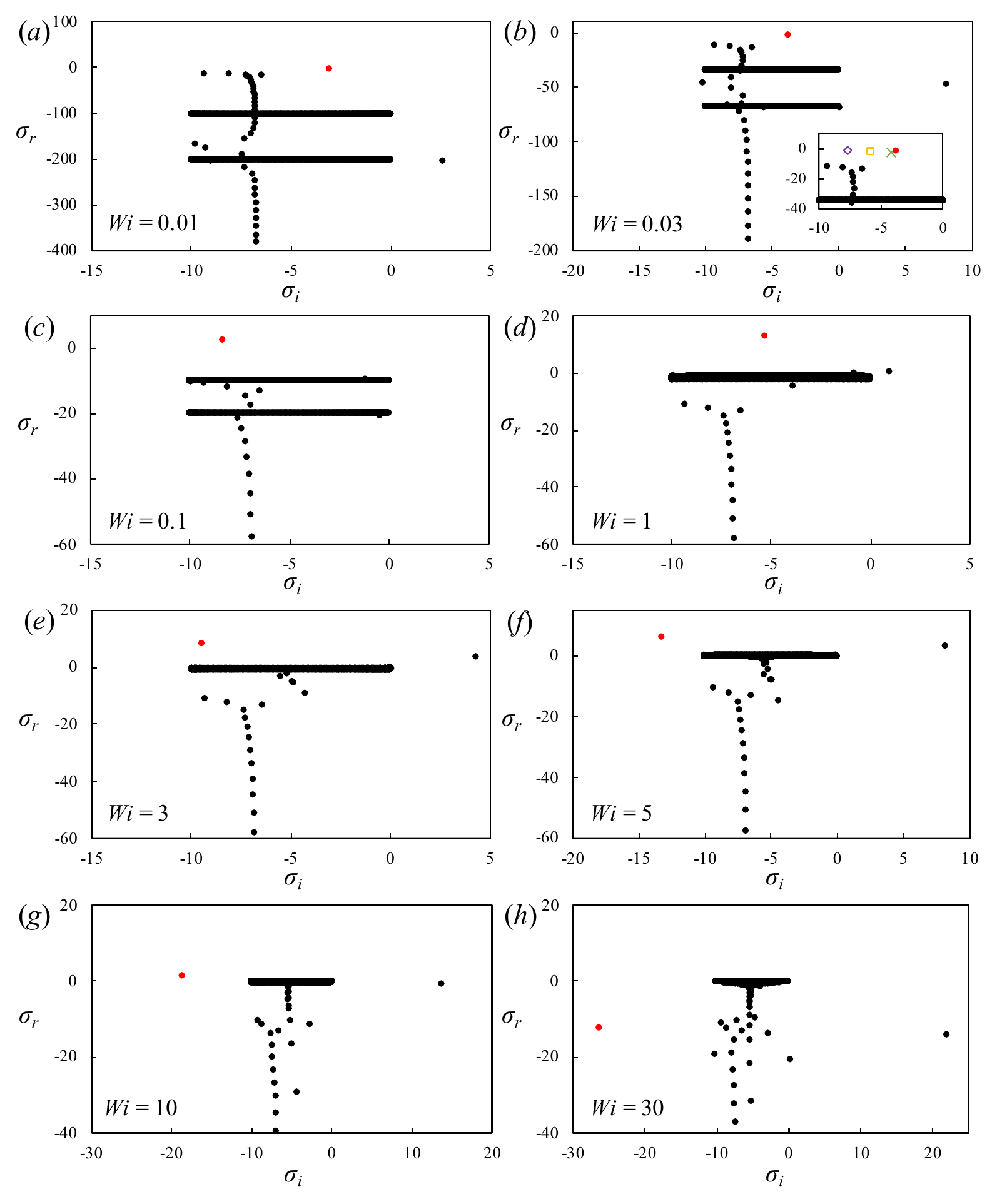}}
\caption{Eigenspectra of the electroconvective instability at $Pe=10$, $\beta=0.5$, $\varepsilon=10^{-4}$, $V=30$ ($Ra_E=105$), and $k=10$: ($a$) $Wi=0.01$, ($b$) $Wi=0.03$, ($c$) $Wi=0.1$, ($d$) $Wi=1$, ($e$) $Wi=3$, ($f$) $Wi=5$, ($g$) $Wi=10$, and ($h$) $Wi=30$. The red symbols identify the discrete eigenmode tracked as $Wi$ varies; at small $Wi$, this is the most unstable eigenmode associated with the A-mode branch. The inset in ($b$) tracks its detachment from the A-mode branch at $Wi=0.04$ (green cross), $0.06$ (orange square), and $0.08$ (purple diamond). The detached eigenmode moves toward larger growth rates up to $Wi=1$. For $Wi>1$, it migrates toward more negative $\sigma_i$ and smaller $\sigma_r$ and becomes stable by $Wi=30$.}
\label{fig:k10_spectrum}
\end{figure}

To further investigate how the instability evolves with $Wi$, we examine the eigenspectrum and identify the branches associated with the unstable modes.

Figure~\ref{fig:k10_spectrum} shows the evolution of the $k=10$ eigenspectrum with $Wi$ at the fixed voltage $V=30$, as used in \S\S~3.1--3.3. At small $Wi$ ($Wi=0.01$ and $0.03$), the spectrum exhibits a characteristic Y-shaped structure similar to previous studies of electroconvective instability obtained from the classical bulk analysis \citep{Li2021}, as well as analogous branch structures reported in inertial pressure-driven flows \citep{Schmid2001}. The right-hand branch is the wall-localized A-mode branch, associated with perturbations concentrated within the ESC layer and consistent with the mechanism identified in \S~3.2. At small $Wi$, the red symbol identifies the most unstable eigenmode associated with this branch. The left-hand branch is the center-mode branch (P-mode branch), while the lower branch is the stable-mode branch (S-mode branch) \citep{Schmid2001,Li2021}.

The inset in figure~\ref{fig:k10_spectrum}($b$) shows how the most unstable eigenmode identified by the red symbol detaches from the A-mode branch as $Wi$ increases from $0.03$ to $0.08$. Its positions at $Wi=0.04$, $0.06$, and $0.08$ demonstrate that the detached eigenvalue develops through continuous migration from the original branch. By $Wi=0.1$, the tracked eigenmode is clearly separated from the Y-shaped spectrum (figure~\ref{fig:k10_spectrum}$c$). Its growth rate continues to increase up to $Wi=1$ (figure~\ref{fig:k10_spectrum}$d$), consistent with the instability enhancement observed at fixed voltage in \S~3.1.

For $Wi>1$, the same discrete mode reverses its upward migration. It moves toward more negative $\sigma_i$ and smaller $\sigma_r$ through $Wi=3$, $5$, and $10$ (figures~\ref{fig:k10_spectrum}$e$--$g$). At $Wi=30$, the mode lies below the neutral axis and is therefore stable (figure~\ref{fig:k10_spectrum}$h$). The non-monotonic growth-rate response at fixed $V$ thus corresponds to the detachment, growth-rate evolution, and eventual stabilization of this detached discrete eigenmode.

In addition, two branches aligned at nearly constant negative growth rates are observed in the eigenspectrum. These correspond to the continuous-spectrum eigenvalues, as discussed in previous studies \citep{Khalid2021a}, satisfying

\begin{equation}
\sigma
=
-\frac{1}{Wi}
-ik\bar{U}
\qquad \text{and} \qquad
\sigma
=
-\frac{1}{\beta Wi}
-ik\bar{U}.
\end{equation}

These modes shift toward less negative growth rates as $Wi$ increases, but remain stable ($\sigma_r<0$) throughout the parameter range considered.

The eigenspectrum analysis explains the trends observed in figure~\ref{fig:fig2}. The non-monotonic dependence on $Wi$ arises from the evolution and eventual stabilization of the discrete wall-localized eigenmode that detaches from the A-mode branch. These results further support the conclusion that the instability is fundamentally controlled by ESC-localized electrohydrodynamics, with viscoelastic effects modifying the stability through changes in the polymeric response and eigenmode structure.

\subsection{Polymeric-work pathways and eigenmode evolution at low wavenumbers}
\label{sec:low_k}
As shown in figure~\ref{fig:fig2}, an additional dip appears in the low-wavenumber portion of the neutral curves over $3\leq Wi\leq30$, and its location shifts toward smaller $k$ as $Wi$ increases. To track the pathway evolution through this regime, we examine neutral modes at $k=5$ for $Wi=0.1$, $1$, and $3$, $k=3.5$ for $Wi=10$, $k=2.5$ for $Wi=30$, and $k=2$ for $Wi=50$ and $70$. This sequence covers the response preceding the dip, the modes selected near the dip, and the high-$Wi$ response after the dip disappears.

Because the constitutive pathways are evaluated from the fully coupled eigenmode, their work depends on which eigenspectral branch is selected. Section~3.6.1 therefore identifies the branch associated with the low-wavenumber dip before the pathway results are interpreted. Sections~3.6.2 and 3.6.3 examine the normal-stress-difference and shear-stress pathways, including the applicability of the asymptotic ordering and its recovery at higher $Wi$. Section~3.6.4 combines these contributions and compares the polymeric, solvent-viscous, and electric work rates in the complete mechanical-work balance.

\begin{figure}
\centerline{\includegraphics[scale=0.42]{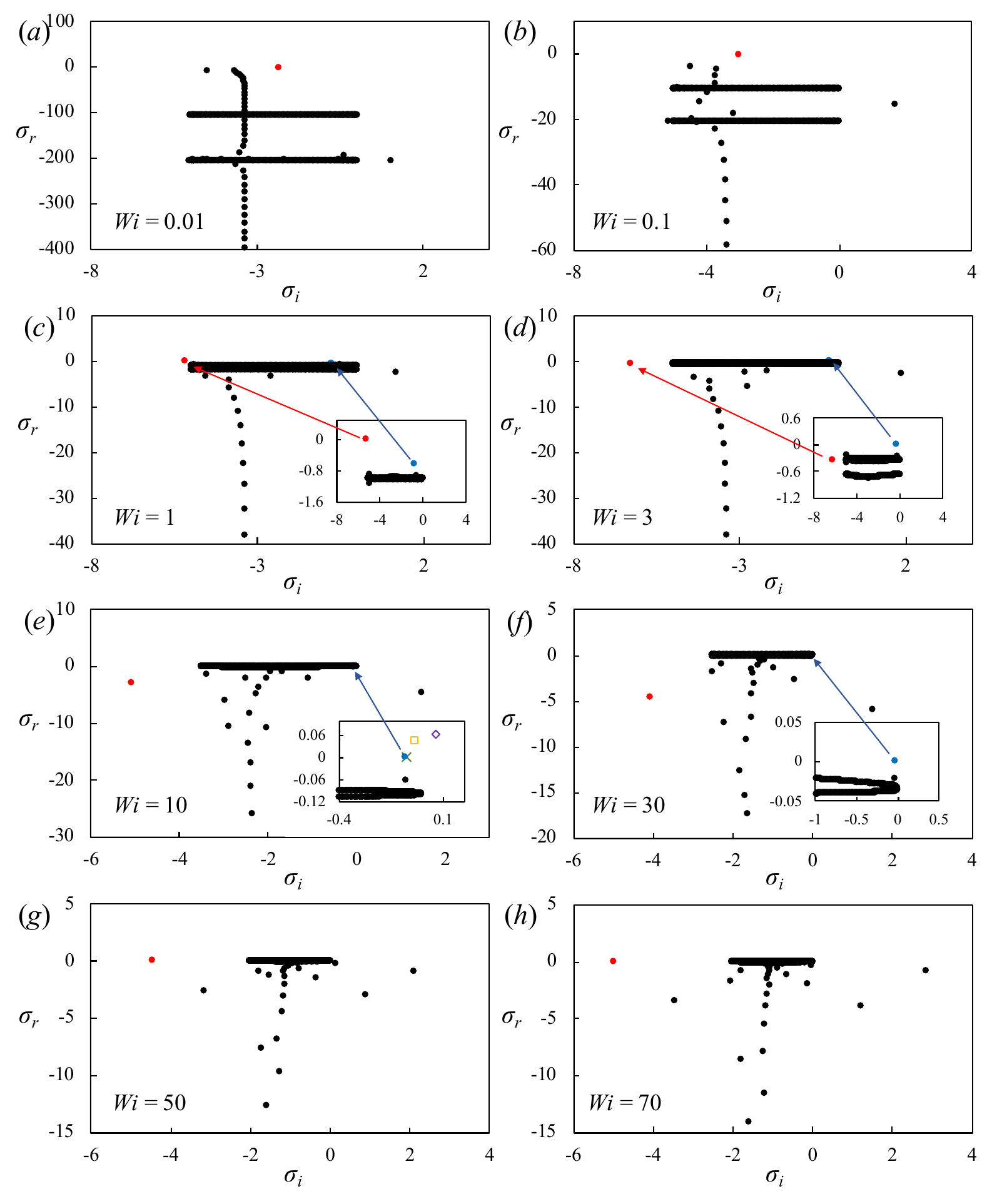}}
\caption{Evolution of the low-wavenumber eigenspectrum at the neutral stability boundary for $Pe=10$, $\beta=0.5$, and $\varepsilon=10^{-4}$. Panels ($a$)--($h$) correspond, respectively, to the $(Wi,V_{\mathrm{cr}})$ pairs $(0.01,30.17)$, $(0.1,29.91)$, $(1,26.76)$, $(3,27.52)$, $(10,27.58)$, $(30,31.76)$, $(50,51.26)$, and $(70,54.87)$. Each main spectrum is evaluated at the corresponding representative wavenumber and critical voltage on the neutral curve. The red symbols identify the eigenmode that separates from the A-mode branch and migrates toward more negative $\sigma_i$, whereas the blue symbols identify the slowly propagating mode near $\sigma_i=0$. The red and blue arrows connect these symbols to the corresponding eigenvalues in the magnified insets. The blue-symbol mode governs the instability onset over $3\leq Wi\leq30$. The inset in panel ($e$) shows the evolution of this mode at $Wi=10$ and $k=3.5$ from the neutral state $V_{\mathrm{cr}}=27.58$ to $V=28$, $29$, and $30$, denoted by the green cross, orange square, and purple diamond, respectively. At $Wi=50$ and $70$, the red-symbol mode originating from the A-mode branch is selected again.}
\label{fig:low_k_spectrum}
\end{figure}

\subsubsection{Eigenmode selection at $V=V_{\mathrm{cr}}$}
\label{sec:low_k_spectrum}
Figure~\ref{fig:low_k_spectrum} shows the eigenspectra evaluated at the neutral states selected along the low-wavenumber portion of the neutral curves. At $Wi=0.01$ and $0.1$, the leading neutral mode is the red-symbol eigenmode originating from the right-hand A-mode branch of the Y-shaped spectrum, consistent with the $k=10$ eigenspectra discussed in \S~\ref{sec:k10_spectrum} (figures~\ref{fig:low_k_spectrum}$a,b$).

At $Wi=1$, the red-symbol eigenmode has migrated toward more negative $\sigma_i$, following the evolution identified at $k=10$ (figure~\ref{fig:low_k_spectrum}$c$). A second eigenmode, marked by the blue symbol, is visible near $\sigma_i=0$, but it does not yet determine the instability threshold. At $Wi=3$, this slowly propagating blue-symbol mode becomes the leading neutral mode (figure~\ref{fig:low_k_spectrum}$d$). The same mode remains selected at $Wi=10$ and $30$ (figures~\ref{fig:low_k_spectrum}$e,f$). The low-wavenumber dip over $3\leq Wi\leq30$ is therefore associated with a change in the eigenmode governing instability onset.

The inset in figure~\ref{fig:low_k_spectrum}($e$) follows the blue-symbol mode at $Wi=10$ and $k=3.5$ as the voltage is increased above $V_{\mathrm{cr}}=27.58$. The eigenvalue crosses into $\sigma_r>0$ at $V=28$, $29$, and $30$. This voltage-dependent migration confirms that the mode selected at the neutral point becomes unstable above the critical voltage.

At $Wi=50$ and $70$, the slowly propagating mode near $\sigma_i=0$ no longer determines the instability threshold. The leading mode is again the red-symbol eigenvalue that originated from the A-mode branch and migrated toward more negative $\sigma_i$ as $Wi$ increased (figures~\ref{fig:low_k_spectrum}$g,h$). This return to the faster-propagating A-mode-derived eigenmode coincides with the disappearance of the low-wavenumber dip.

The phase speed $c=-\sigma_i/k$ distinguishes the two selected eigenmodes. The mode governing the dip over $3\leq Wi\leq30$ has $\sigma_i$ close to zero and consequently propagates slowly. The mode recovered at larger $Wi$ ($Wi=50$ and $70$) has a more negative $\sigma_i$ and a larger phase speed. This mode selection provides the eigenspectral basis for interpreting the normal- and shear-stress pathways in the following subsections.

\subsubsection{Normal-stress-difference pathways of the low-wavenumber mode}
\label{sec:low_k_normal}

\begin{figure}
\centerline{\includegraphics[scale=0.45]{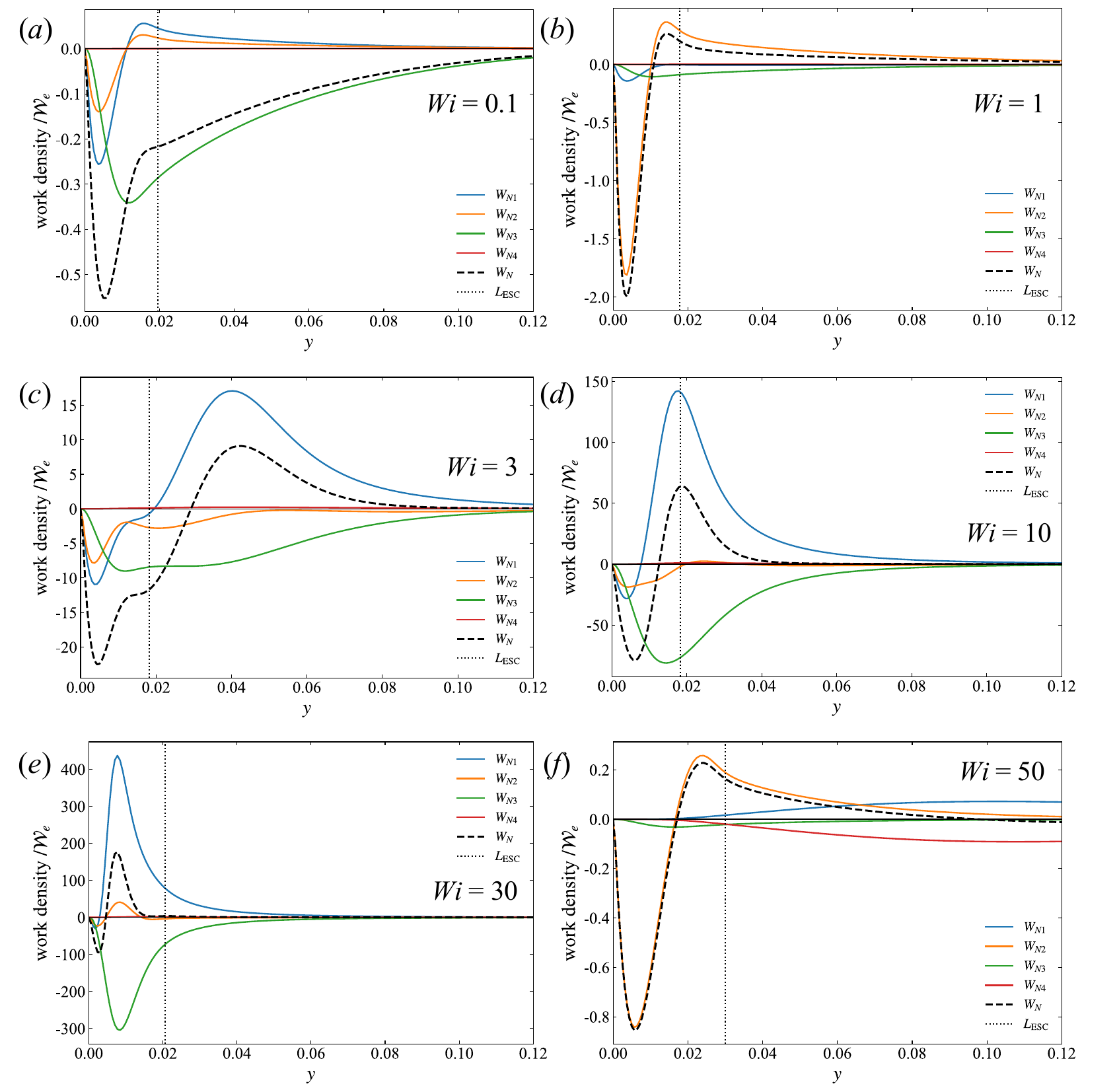}}
\caption{Normalized local work-density profiles of the normal-stress-difference pathways for representative neutral low-wavenumber modes at $Pe=10$, $\beta=0.5$, and $\varepsilon=10^{-4}$. Panels ($a$)--($f$) correspond to $(Wi,k)=(0.1,5)$, $(1,5)$, $(3,5)$, $(10,3.5)$, $(30,2.5)$, and $(50,2)$, respectively. All work densities are normalized by the magnitude of the full-domain electric work, $\mathcal{W}_e$, for the corresponding eigenmode.}
\label{fig:low_k_normal_work}
\end{figure}

Figure~\ref{fig:low_k_normal_work} shows the wall-normal profiles of the normalized normal-stress-difference pathway work densities, and their integrals over the ESC, bulk, and full domain are reported in table~\ref{tab:low_k_normal_work}. The corresponding phase efficiencies are reported in table~\ref{tab:low_k_phase} in Appendix~C.

At $Wi=0.1$, $N_3$ dominates in the ESC, bulk, and full domain, reproducing the Newtonian-like small-$Wi$ structure found at $k=10$ (figure~\ref{fig:low_k_normal_work}$a$). At $Wi=1$, $N_3$ remains the dominant negative contribution, and the other pathways become appreciable as the response departs from the small-$Wi$ limit (figure~\ref{fig:low_k_normal_work}$b$).

A different work balance appears over $3\leq Wi\leq30$, where the slowly propagating mode identified in \S~\ref{sec:low_k_spectrum} determines the instability threshold. The normal work is governed primarily by opposing $N_1$ and $N_3$ contributions (figures~\ref{fig:low_k_normal_work}$c$--$e$). At $Wi=3$, the negative $N_2$ and $N_3$ contributions exceed the positive $N_1$ contribution, giving $W_N<0$. At $Wi=10$, $W_{N1}/\mathcal W_e=2.856$ and $W_{N3}/\mathcal W_e=-2.432$.
Their cancellation, together with the smaller $N_2$ and $N_4$ contributions, gives $W_N/\mathcal W_e=0.210$. At $Wi=30$, $W_{N1}/\mathcal W_e=4.983$ and $W_{N3}/\mathcal W_e=-4.390$. Both contributions exceed the total electric-work scale in magnitude and act in opposite directions. Their strong cancellation reduces the net normal work to $W_N/\mathcal W_e=0.731$. The positive net normal work at $Wi=10$ and $30$ is therefore the residual of much larger opposing pathway contributions.

The prominence of $N_1$ and $N_3$ at $Wi=10$ and $30$ is consistent with the large-$Wi$ asymptotic ordering in equation~\eqref{eq:large_Wi_normal_ordering}. The signed work, however, differs substantially from that of the high-$Wi$, $k=10$ mode. For the $k=10$ mode, the full-domain phase efficiency of $N_3$ decreases to $-0.030$, $-0.006$, and $-0.002$ at $Wi=3$, $10$, and $30$, respectively. For the slowly propagating low-wavenumber mode, the corresponding values remain much larger in magnitude at $-0.792$, $-0.591$, and $-0.439$ (table~\ref{tab:low_k_phase}). Consequently, $N_3$ continues to perform substantial negative work and directly opposes the positive $N_1$ work. This phase difference produces the $N_1$ and $N_3$ balance observed over $3\leq Wi\leq30$.

At $Wi=50$, the faster mode identified in \S~\ref{sec:low_k_spectrum} is selected again. The $N_3$ phase efficiency decreases to $-0.005$, and its contribution to the full-domain normal work becomes negligible. The balance is then determined mainly by $W_{N1}/\mathcal W_e=0.0109$ and $W_{N4}/\mathcal W_e=-0.0142$, giving $W_N/\mathcal W_e=-0.0048$ (table~\ref{tab:low_k_normal_work}). The normal-stress pathway structure has therefore returned to the $N_1$ and $N_4$ balance observed for the high-$Wi$, $k=10$ mode. The results at $Wi=70$ and $100$ in table~\ref{tab:low_k_normal_work} show the same behavior.

The large opposing normal-work contributions on the slowly propagating low-wavenumber branch contrast with the high-$Wi$, $k=10$ result, for which the total polymeric work is very small relative to the electric work. The contribution of the shear pathways to this difference and the resulting balance among $W_p^{\mathrm{int}}$, $W_{\mathrm{sol}}^{\mathrm{int}}$, and $W_e$ are examined in \S~3.6.4.

\begin{table}
  \begin{center}
  \begin{tabular}{cccrrrrr}
    $Wi$ & $k$ & Region & $W_N/\mathcal{W}_e$ & $W_{N_1}/\mathcal{W}_e$ & $W_{N_2}/\mathcal{W}_e$ & $W_{N_3}/\mathcal{W}_e$ & $W_{N_4}/\mathcal{W}_e$ \\[3pt]

    0.01 & 5 & ESC
    & $-4.856\times10^{-3}$ & $ 8.308\times10^{-7}$
    & $ 5.699\times10^{-6}$ & $-4.863\times10^{-3}$
    & $ 1.696\times10^{-8}$ \\
    & & Bulk
    & $-1.377\times10^{-2}$ & $ 1.473\times10^{-5}$
    & $ 9.286\times10^{-6}$ & $-1.379\times10^{-2}$
    & $ 1.337\times10^{-7}$ \\
    & & Total
    & $-1.862\times10^{-2}$ & $ 1.556\times10^{-5}$
    & $ 1.499\times10^{-5}$ & $-1.865\times10^{-2}$
    & $ 1.507\times10^{-7}$ \\[3pt]

    0.1 & 5 & ESC
    & $-6.793\times10^{-3}$ & $-1.300\times10^{-3}$
    & $-7.076\times10^{-4}$ & $-4.787\times10^{-3}$
    & $ 1.595\times10^{-6}$ \\
    & & Bulk
    & $-1.174\times10^{-2}$ & $ 1.318\times10^{-3}$
    & $ 7.750\times10^{-4}$ & $-1.384\times10^{-2}$
    & $ 1.182\times10^{-5}$ \\
    & & Total
    & $-1.853\times10^{-2}$ & $ 1.862\times10^{-5}$
    & $ 6.743\times10^{-5}$ & $-1.863\times10^{-2}$
    & $ 1.341\times10^{-5}$ \\[3pt]

    1 & 5 & ESC
    & $-1.077\times10^{-2}$ & $-1.011\times10^{-3}$
    & $-8.412\times10^{-3}$ & $-1.360\times10^{-3}$
    & $ 7.757\times10^{-6}$ \\
    & & Bulk
    & $ 5.833\times10^{-3}$ & $ 3.664\times10^{-3}$
    & $ 1.082\times10^{-2}$ & $-6.824\times10^{-3}$
    & $-1.822\times10^{-3}$ \\
    & & Total
    & $-4.941\times10^{-3}$ & $ 2.653\times10^{-3}$
    & $ 2.404\times10^{-3}$ & $-8.184\times10^{-3}$
    & $-1.814\times10^{-3}$ \\[3pt]

    3 & 5 & ESC
    & $-2.775\times10^{-1}$ & $-8.982\times10^{-2}$
    & $-7.028\times10^{-2}$ & $-1.184\times10^{-1}$
    & $ 1.032\times10^{-3}$ \\
    & & Bulk
    & $ 1.824\times10^{-1}$ & $ 6.542\times10^{-1}$
    & $-7.892\times10^{-2}$ & $-4.140\times10^{-1}$
    & $ 2.116\times10^{-2}$ \\
    & & Total
    & $-9.505\times10^{-2}$ & $ 5.643\times10^{-1}$
    & $-1.492\times10^{-1}$ & $-5.324\times10^{-1}$
    & $ 2.220\times10^{-2}$ \\[3pt]

    10 & 3.5 & ESC
    & $-4.065\times10^{-1}$ & $ 7.892\times10^{-1}$
    & $-2.285\times10^{-1}$ & $-9.748\times10^{-1}$
    & $ 7.666\times10^{-3}$ \\
    & & Bulk
    & $ 6.165\times10^{-1}$ & $ 2.067$
    & $-5.771\times10^{-2}$ & $-1.457$
    & $ 6.405\times10^{-2}$ \\
    & & Total
    & $ 2.100\times10^{-1}$ & $ 2.856$
    & $-2.862\times10^{-1}$ & $-2.432$
    & $ 7.171\times10^{-2}$ \\[3pt]

    30 & 2.5 & ESC
    & $ 6.407\times10^{-1}$ & $ 3.923$
    & $ 1.325\times10^{-1}$ & $-3.439$
    & $ 2.402\times10^{-2}$ \\
    & & Bulk
    & $ 8.985\times10^{-2}$ & $ 1.060$
    & $-8.676\times10^{-2}$ & $-9.511\times10^{-1}$
    & $ 6.748\times10^{-2}$ \\
    & & Total
    & $ 7.306\times10^{-1}$ & $ 4.983$
    & $ 4.576\times10^{-2}$ & $-4.390$
    & $ 9.150\times10^{-2}$ \\[3pt]

    50 & 2 & ESC
    & $-6.428\times10^{-3}$ & $ 1.155\times10^{-4}$
    & $-5.724\times10^{-3}$ & $-6.625\times10^{-4}$
    & $-1.569\times10^{-4}$ \\
    & & Bulk
    & $ 1.596\times10^{-3}$ & $ 1.081\times10^{-2}$
    & $ 5.709\times10^{-3}$ & $-8.756\times10^{-4}$
    & $-1.405\times10^{-2}$ \\
    & & Total
    & $-4.832\times10^{-3}$ & $ 1.093\times10^{-2}$
    & $-1.535\times10^{-5}$ & $-1.538\times10^{-3}$
    & $-1.421\times10^{-2}$ \\[3pt]

    70 & 2 & ESC
    & $-5.491\times10^{-3}$ & $ 1.892\times10^{-4}$
    & $-4.914\times10^{-3}$ & $-5.047\times10^{-4}$
    & $-2.613\times10^{-4}$ \\
    & & Bulk
    & $-8.386\times10^{-5}$ & $ 1.011\times10^{-2}$
    & $ 4.855\times10^{-3}$ & $-5.333\times10^{-4}$
    & $-1.452\times10^{-2}$ \\
    & & Total
    & $-5.575\times10^{-3}$ & $ 1.030\times10^{-2}$
    & $-5.819\times10^{-5}$ & $-1.038\times10^{-3}$
    & $-1.478\times10^{-2}$ \\[3pt]

    100 & 2 & ESC
    & $-4.578\times10^{-3}$ & $ 2.791\times10^{-4}$
    & $-4.042\times10^{-3}$ & $-3.864\times10^{-4}$
    & $-4.275\times10^{-4}$ \\
    & & Bulk
    & $-2.249\times10^{-3}$ & $ 9.198\times10^{-3}$
    & $ 3.959\times10^{-3}$ & $-3.238\times10^{-4}$
    & $-1.508\times10^{-2}$ \\
    & & Total
    & $-6.827\times10^{-3}$ & $ 9.478\times10^{-3}$
    & $-8.332\times10^{-5}$ & $-7.102\times10^{-4}$
    & $-1.551\times10^{-2}$
\end{tabular}
  \caption{Normalized normal-stress pathway work integrated over the ESC, bulk, and full domain for the representative neutral low-wavenumber modes. The eigenfunctions are normalized by the magnitude of the full-domain electric work $\mathcal{W}_e$, and $W_N=\sum_{j=1}^{4}W_{Nj}$.}
  \label{tab:low_k_normal_work}
  \end{center}
\end{table}

\subsubsection{Shear-stress pathways of the low-wavenumber mode}
\label{sec:low_k_shear}

\begin{figure}
\centerline{\includegraphics[scale=0.45]{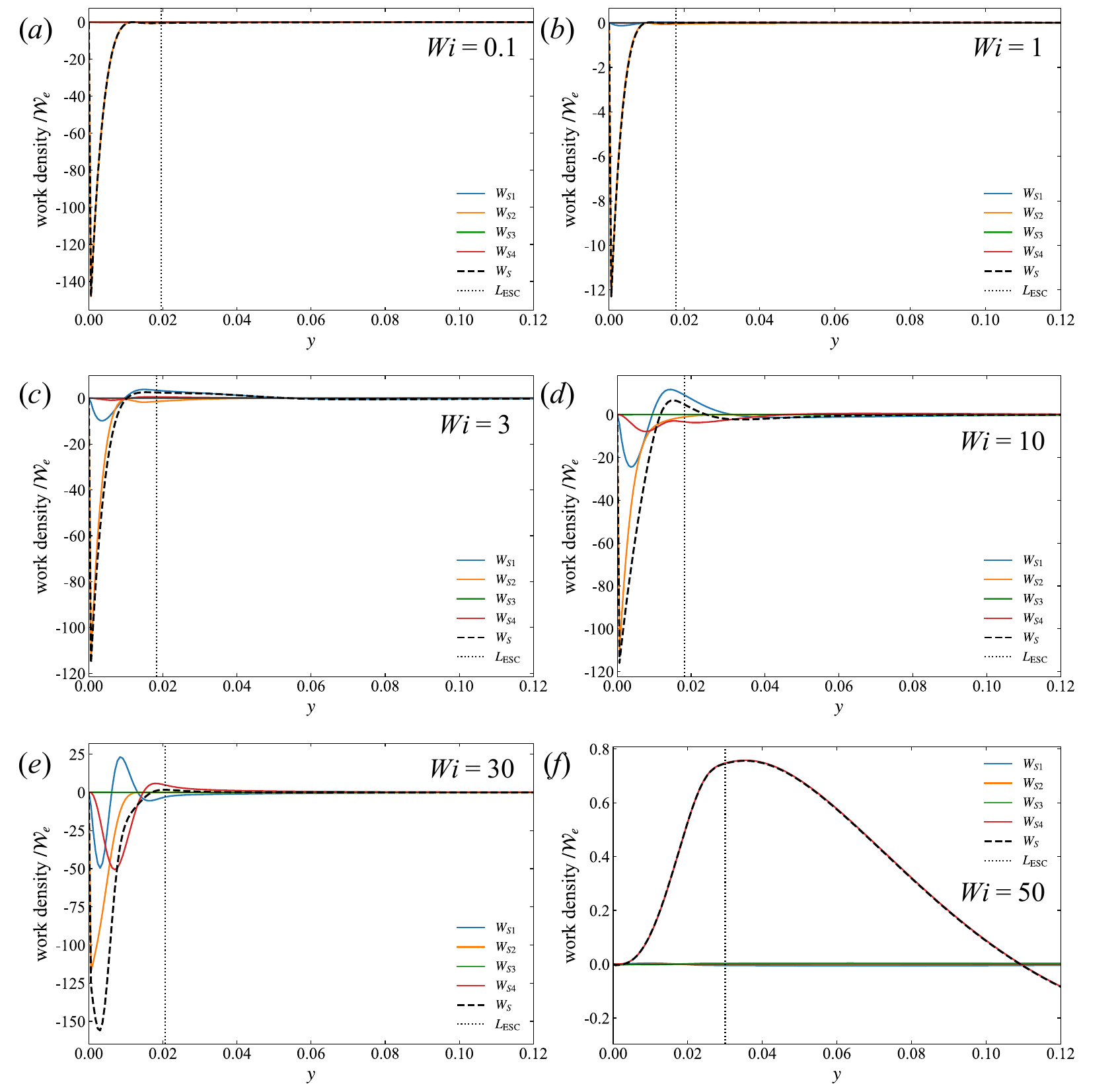}}
\caption{Normalized local work-density profiles of the shear-stress pathways for representative neutral low-wavenumber modes at $Pe=10$, $\beta=0.5$, and $\varepsilon=10^{-4}$. Panels ($a$)--($f$) correspond to $(Wi,k)=(0.1,5)$, $(1,5)$, $(3,5)$, $(10,3.5)$, $(30,2.5)$, and $(50,2)$, respectively. All work densities are normalized by the magnitude of the full-domain electric work, $\mathcal{W}_e$, for the corresponding eigenmode.}
\label{fig:low_k_shear_work}
\end{figure}

Figure~\ref{fig:low_k_shear_work} shows the wall-normal profiles of the normalized shear-stress pathway work densities, and their integrals over the ESC, bulk, and full domain are reported in table~\ref{tab:low_k_shear_work}. The corresponding phase efficiencies are reported in table~\ref{tab:low_k_phase} in Appendix~C.

At $Wi=0.1$ and $1$, the negative shear work is dominated by $S_2$, consistent with the Newtonian-like response found at $k=10$ (figures~\ref{fig:low_k_shear_work}$a,b$). Over $3\leq Wi\leq30$, $S_2$ remains a principal contribution, and $S_4$ also becomes important at $Wi=10$ and $30$. This result differs from the uniform large-$Wi$ ordering in equation~\eqref{eq:large_Wi_shear_ordering}, which predicts $\hat{\tau}_S^{(4)}=O(Wi)$ and $\hat{\tau}_S^{(2)}=O(Wi^{-1})$.

For a neutral or nearly neutral travelling mode with phase speed $c=-\sigma_i/k$, the local constitutive response is governed by
\begin{equation}
\Sigma(y)\simeq ik[\bar{U}(y)-c].
\label{eq:neutral_convective_response}
\end{equation}
The slowly propagating low-wavenumber mode admits a position $y_c$ satisfying
\begin{equation}
\bar{U}(y_c)=c
\label{eq:critical_layer_condition}
\end{equation}
within or adjacent to the ESC. Consequently, $|\Sigma|$ becomes small in the region where the electric forcing and polymeric work are concentrated. The condition $Wi|\Sigma|\gg1$ required by the large-$Wi$ expansion is therefore not satisfied uniformly, even when the nominal value of $Wi$ exceeds unity. The estimate $\hat{\tau}_S^{(2)}=O(Wi^{-1})$ does not apply uniformly in this mechanically active region, allowing $S_2$ to remain significant for $3\leq Wi\leq30$.

The persistence of $S_2$ in stress magnitude is accompanied by strong negative phase coupling. Its full-domain phase efficiencies are $-0.802$, $-0.855$, and $-0.936$ at $Wi=3$, $10$, and $30$, respectively (table~\ref{tab:low_k_phase}). At $Wi=3$, $W_{S2}/\mathcal W_e=-0.338$ accounts for nearly all of the total shear work. The positive $S_4$ contribution remains small at this condition (figure~\ref{fig:low_k_shear_work}$c$).

At $Wi=10$, both $S_2$ and $S_4$ perform negative work. Together with the smaller negative $S_1$ contribution, these pathways give $W_{S}/\mathcal W_e=-0.604$. At $Wi=30$, the negative contributions increase to $W_{S}/\mathcal W_e=-1.04$. The shear stress therefore extracts mechanical power at a rate comparable to the total electric-work input at $Wi=30$. 

At $Wi=50$, the $S_2$ contribution becomes negligible and $S_4$ provides the leading shear-stress contribution (figure~\ref{fig:low_k_shear_work}$f$). The negative shear-work structure of the slowly propagating mode has therefore disappeared. The relationship between this change, the recovered normal-stress balance, and the complete mechanical-work budget is examined in \S~3.6.4.

\begin{table}
  \begin{center}
  \begin{tabular}{cccrrrrr}
    $Wi$ & $k$ & Region & $W_S/\mathcal{W}_e$ & $W_{S1}/\mathcal{W}_e$ & $W_{S2}/\mathcal{W}_e$
    & $W_{S3}/\mathcal{W}_e$ & $W_{S4}/\mathcal{W}_e$ \\[3pt]

    0.01 & 5 & ESC
    & $-4.660\times10^{-1}$ & $-2.550\times10^{-5}$
    & $-4.660\times10^{-1}$ & $-5.282\times10^{-8}$
    & $-2.853\times10^{-5}$ \\
    & & Bulk
    & $-1.521\times10^{-2}$ & $ 6.043\times10^{-6}$
    & $-1.520\times10^{-2}$ & $ 3.089\times10^{-7}$
    & $-1.746\times10^{-5}$ \\
    & & Total
    & $-4.812\times10^{-1}$ & $-1.945\times10^{-5}$
    & $-4.812\times10^{-1}$ & $ 2.561\times10^{-7}$
    & $-4.600\times10^{-5}$ \\[3pt]

    0.1 & 5 & ESC
    & $-4.475\times10^{-1}$ & $-1.573\times10^{-3}$
    & $-4.459\times10^{-1}$ & $-7.351\times10^{-7}$
    & $-2.673\times10^{-5}$ \\
    & & Bulk
    & $-1.391\times10^{-2}$ & $ 1.068\times10^{-3}$
    & $-1.505\times10^{-2}$ & $ 2.582\times10^{-5}$
    & $ 4.856\times10^{-5}$ \\
    & & Total
    & $-4.614\times10^{-1}$ & $-5.046\times10^{-4}$
    & $-4.609\times10^{-1}$ & $ 2.509\times10^{-5}$
    & $ 2.183\times10^{-5}$ \\[3pt]

    1 & 5 & ESC
    & $-3.323\times10^{-2}$ & $-3.959\times10^{-4}$
    & $-3.294\times10^{-2}$ & $-1.960\times10^{-6}$
    & $ 1.091\times10^{-4}$ \\
    & & Bulk
    & $-6.665\times10^{-4}$ & $ 1.196\times10^{-3}$
    & $-1.460\times10^{-3}$ & $ 5.331\times10^{-5}$
    & $-4.561\times10^{-4}$ \\
    & & Total
    & $-3.390\times10^{-2}$ & $ 8.005\times10^{-4}$
    & $-3.440\times10^{-2}$ & $ 5.135\times10^{-5}$
    & $-3.470\times10^{-4}$ \\[3pt]

    3 & 5 & ESC
    & $-3.563\times10^{-1}$ & $-3.152\times10^{-2}$
    & $-3.223\times10^{-1}$ & $ 1.416\times10^{-4}$
    & $-2.659\times10^{-3}$ \\
    & & Bulk
    & $ 1.986\times10^{-2}$ & $ 2.854\times10^{-2}$
    & $-1.519\times10^{-2}$ & $ 9.361\times10^{-5}$
    & $ 6.413\times10^{-3}$ \\
    & & Total
    & $-3.365\times10^{-1}$ & $-2.974\times10^{-3}$
    & $-3.375\times10^{-1}$ & $ 2.353\times10^{-4}$
    & $ 3.754\times10^{-3}$ \\[3pt]

    10 & 3.5 & ESC
    & $-5.368\times10^{-1}$ & $-6.051\times10^{-2}$
    & $-3.965\times10^{-1}$ & $ 5.498\times10^{-4}$
    & $-8.037\times10^{-2}$ \\
    & & Bulk
    & $-6.728\times10^{-2}$ & $-2.816\times10^{-2}$
    & $-4.330\times10^{-3}$ & $-1.227\times10^{-3}$
    & $-3.356\times10^{-2}$ \\
    & & Total
    & $-6.041\times10^{-1}$ & $-8.867\times10^{-2}$
    & $-4.008\times10^{-1}$ & $-6.768\times10^{-4}$
    & $-1.139\times10^{-1}$ \\[3pt]

    30 & 2.5 & ESC
    & $-1.024$ & $-1.173\times10^{-1}$
    & $-5.594\times10^{-1}$ & $ 1.365\times10^{-4}$
    & $-3.477\times10^{-1}$ \\
    & & Bulk
    & $-2.054\times10^{-2}$ & $-7.307\times10^{-2}$
    & $-2.020\times10^{-4}$ & $-2.087\times10^{-3}$
    & $ 5.480\times10^{-2}$ \\
    & & Total
    & $-1.045$ & $-1.904\times10^{-1}$
    & $-5.596\times10^{-1}$ & $-1.951\times10^{-3}$
    & $-2.928\times10^{-1}$ \\[3pt]

    50 & 2 & ESC
    & $ 1.013\times10^{-2}$ & $-5.156\times10^{-6}$
    & $-1.979\times10^{-5}$ & $ 1.282\times10^{-5}$
    & $ 1.014\times10^{-2}$ \\
    & & Bulk
    & $-9.521\times10^{-4}$ & $-6.380\times10^{-4}$
    & $-1.553\times10^{-6}$ & $ 3.885\times10^{-4}$
    & $-7.013\times10^{-4}$ \\
    & & Total
    & $ 9.174\times10^{-3}$ & $-6.431\times10^{-4}$
    & $-2.134\times10^{-5}$ & $ 4.013\times10^{-4}$
    & $ 9.437\times10^{-3}$ \\[3pt]

    70 & 2 & ESC
    & $ 1.536\times10^{-2}$ & $-1.460\times10^{-5}$
    & $-7.446\times10^{-6}$ & $ 1.687\times10^{-5}$
    & $ 1.536\times10^{-2}$ \\
    & & Bulk
    & $-5.418\times10^{-3}$ & $-4.769\times10^{-4}$
    & $-3.577\times10^{-7}$ & $ 3.206\times10^{-4}$
    & $-5.262\times10^{-3}$ \\
    & & Total
    & $ 9.941\times10^{-3}$ & $-4.915\times10^{-4}$
    & $-7.803\times10^{-6}$ & $ 3.375\times10^{-4}$
    & $ 1.010\times10^{-2}$ \\[3pt]

    100 & 2 & ESC
    & $ 2.302\times10^{-2}$ & $-2.070\times10^{-5}$
    & $-2.786\times10^{-6}$ & $ 2.078\times10^{-5}$
    & $ 2.303\times10^{-2}$ \\
    & & Bulk
    & $-1.200\times10^{-2}$ & $-3.355\times10^{-4}$
    & $ 3.554\times10^{-8}$ & $ 2.574\times10^{-4}$
    & $-1.193\times10^{-2}$ \\
    & & Total
    & $ 1.102\times10^{-2}$ & $-3.562\times10^{-4}$
    & $-2.750\times10^{-6}$ & $ 2.782\times10^{-4}$
    & $ 1.110\times10^{-2}$
\end{tabular}
  \caption{Normalized shear-stress pathway work integrated over the ESC, bulk, and full domain for the representative neutral low-wavenumber modes. The eigenfunctions are normalized by the magnitude of the full-domain electric work $\mathcal{W}_e$, and $W_S=\sum_{j=1}^{4}W_{Sj}$.}
  \label{tab:low_k_shear_work}
  \end{center}
\end{table}

\subsubsection{Combined mechanical-work balance}
\label{sec:low_k_overall}

We next combine the normal-stress-difference and shear-stress contributions and examine their roles in the full-domain mechanical-work balance. Tables~\ref{tab:low_k_normal_work} and \ref{tab:low_k_shear_work} report $W_N/\mathcal W_e$ and $W_S/\mathcal W_e$, respectively. Their sum gives the total polymeric work, $W_p^{\mathrm{int}}/\mathcal W_e$, which is compared with the normalized solvent-viscous work below.

At small $Wi$, the polymer and solvent provide comparable mechanical-power extraction. At $Wi=1$, the polymeric extraction decreases sharply, and the solvent accounts for most of the dissipation, with $W_{\mathrm{sol}}^{\mathrm{int}}/\mathcal W_e=-0.961$. This loss of polymeric extraction coincides with the minimum of the neutral curve near $Wi\sim O(1)$.

The slowly propagating mode selected over $3\leq Wi\leq30$ produces a different work balance. At $Wi=3$, both normal and shear stresses extract mechanical power, giving $W_p^{\mathrm{int}}/\mathcal W_e=-0.432$. At $Wi=10$ and $30$, the normal-stress-difference work becomes positive, but it does not exceed the negative shear-stress work. Hence, the positive residual normal-stress-difference work offsets only part of the mechanical-power extraction produced by the shear-stress pathways. The polymer stress remains a substantial net sink of mechanical power throughout the low-wavenumber dip. The remaining electric-work input is dissipated by the solvent, for which $W_{\mathrm{sol}}^{\mathrm{int}}/\mathcal W_e=-0.569$, $-0.606$, and $-0.686$ at $Wi=3$, $10$, and $30$, respectively.

This behavior differs from that of the large-$Wi$, $k=10$ mode. For the $k=10$ mode, the positive shear work and negative normal work nearly cancel, producing a net polymeric contribution that is very small compared with the electric work. For the slowly propagating low-wavenumber mode, the negative shear work remains larger than the positive normal work. The selected eigenmode therefore retains substantial polymeric extraction.

To determine whether this contrast in polymeric work is accompanied by a corresponding difference in the velocity-deformation structure, we next examine the solvent-viscous dissipation. Table~\ref{tab:lowk_solvent_work} reports the solvent work normalized by $\mathcal W_e$ and $K_u$ for the selected low-wavenumber eigenmodes. Across all cases considered, shear deformation accounts for approximately $93.6$--$99.4\%$ of the full-domain solvent dissipation, and $89.0$--$98.3\%$ of this shear dissipation occurs within the ESC. However, its magnitude relative to $K_u$ depends strongly on the selected eigenmode.

At $Wi=1$, the full-domain value of $-W_{\mathrm{sol}}^{\mathrm{int}}/K_u$ is $872$. It decreases to $338$ at $Wi=3$ and remains $159$ and $169$ at $Wi=10$ and $30$, respectively. The slowly propagating mode selected over $3\leq Wi\leq30$ therefore does not exhibit the progressive increase in shear deformation found for the $k=10$ mode. Instead, it retains a substantial negative polymeric contribution and relatively weak solvent deformation with respect to $K_u$.

At $Wi=50$, selection returns to the faster A-mode-derived eigenmode identified in \S~\ref{sec:low_k_spectrum}, and the work partition changes abruptly. The net polymeric work becomes negligible, with $W_p^{\mathrm{int}}/\mathcal W_e=0.0043$, while $W_{\mathrm{sol}}^{\mathrm{int}}/\mathcal W_e=-1.004$ shows that solvent viscosity balances nearly all of the electric-work input. Simultaneously, $-W_{\mathrm{sol}}^{\mathrm{int}}/K_u$ increases from $169$ at $Wi=30$ to $972$ at $Wi=50$, corresponding to a factor of $5.77$. Because shear deformation accounts for more than $99\%$ of the solvent dissipation at $Wi=50$, this increase reflects the substantially stronger shear-deformation structure of the selected eigenmode.

The increase in $V_{\mathrm{cr}}$ from $31.76$ at $Wi=30$ to $51.26$ at $Wi=50$ therefore accompanies the disappearance of net polymeric extraction and the transition to a strongly shear-deforming, solvent-dominated eigenmode. The same solvent-dominated balance persists at $Wi=70$ and $100$. Having established these pathway transitions for the reference parameters, \S~3.7 examines the effects of $\varepsilon$, $Pe$, and $\beta$ on the non-monotonic stability response.

\begin{table}
  \begin{center}
  \begin{tabular}{ccrrrr}
    $Wi$ & Region
    & $W_{\mathrm{sol}}/\mathcal{W}_e$
    & $W_{\mathrm{sol}}/K_u$
    & $W_{\mathrm{sol},N}/K_u$
    & $W_{\mathrm{sol},S}/K_u$ \\[3pt]

    0.01 & ESC
    & $-4.711\times10^{-1}$ & $-6.281\times10^{2}$
    & $-6.484$ & $-6.216\times10^{2}$ \\
         & Bulk
    & $-2.902\times10^{-2}$ & $-3.869\times10^{1}$
    & $-1.839\times10^{1}$ & $-2.030\times10^{1}$ \\
         & Total
    & $-5.001\times10^{-1}$ & $-6.668\times10^{2}$
    & $-2.487\times10^{1}$ & $-6.419\times10^{2}$ \\[3pt]

    0.1 & ESC
    & $-4.901\times10^{-1}$ & $-6.426\times10^{2}$
    & $-6.540$ & $-6.360\times10^{2}$ \\
        & Bulk
    & $-3.003\times10^{-2}$ & $-3.937\times10^{1}$
    & $-1.851\times10^{1}$ & $-2.086\times10^{1}$ \\
        & Total
    & $-5.201\times10^{-1}$ & $-6.819\times10^{2}$
    & $-2.505\times10^{1}$ & $-6.569\times10^{2}$ \\[3pt]

    1 & ESC
    & $-9.087\times10^{-1}$ & $-8.244\times10^{2}$
    & $-6.662$ & $-8.178\times10^{2}$ \\
      & Bulk
    & $-5.244\times10^{-2}$ & $-4.757\times10^{1}$
    & $-2.141\times10^{1}$ & $-2.616\times10^{1}$ \\
      & Total
    & $-9.611\times10^{-1}$ & $-8.720\times10^{2}$
    & $-2.807\times10^{1}$ & $-8.439\times10^{2}$ \\[3pt]

    3 & ESC
    & $-5.118\times10^{-1}$ & $-3.046\times10^{2}$
    & $-2.469$ & $-3.022\times10^{2}$ \\
      & Bulk
    & $-5.669\times10^{-2}$ & $-3.374\times10^{1}$
    & $-1.805\times10^{1}$ & $-1.569\times10^{1}$ \\
      & Total
    & $-5.685\times10^{-1}$ & $-3.384\times10^{2}$
    & $-2.052\times10^{1}$ & $-3.178\times10^{2}$ \\[3pt]

    10 & ESC
    & $-5.476\times10^{-1}$ & $-1.435\times10^{2}$
    & $-7.099\times10^{-1}$ & $-1.428\times10^{2}$ \\
       & Bulk
    & $-5.833\times10^{-2}$ & $-1.528\times10^{1}$
    & $-9.443$ & $-5.841$ \\
       & Total
    & $-6.059\times10^{-1}$ & $-1.588\times10^{2}$
    & $-1.015\times10^{1}$ & $-1.486\times10^{2}$ \\[3pt]

    30 & ESC
    & $-6.515\times10^{-1}$ & $-1.602\times10^{2}$
    & $-5.256\times10^{-1}$ & $-1.597\times10^{2}$ \\
       & Bulk
    & $-3.433\times10^{-2}$ & $-8.441$
    & $-5.738$ & $-2.703$ \\
       & Total
    & $-6.859\times10^{-1}$ & $-1.686\times10^{2}$
    & $-6.263$ & $-1.624\times10^{2}$ \\[3pt]

    50 & ESC
    & $-9.010\times10^{-1}$ & $-8.724\times10^{2}$
    & $-3.028$ & $-8.693\times10^{2}$ \\
       & Bulk
    & $-1.034\times10^{-1}$ & $-1.001\times10^{2}$
    & $-4.265$ & $-9.581\times10^{1}$ \\
       & Total
    & $-1.004$ & $-9.724\times10^{2}$
    & $-7.293$ & $-9.651\times10^{2}$ \\[3pt]

    70 & ESC
    & $-8.953\times10^{-1}$ & $-9.240\times10^{2}$
    & $-3.423$ & $-9.206\times10^{2}$ \\
       & Bulk
    & $-1.091\times10^{-1}$ & $-1.126\times10^{2}$
    & $-3.905$ & $-1.087\times10^{2}$ \\
       & Total
    & $-1.004$ & $-1.037\times10^{3}$
    & $-7.328$ & $-1.029\times10^{3}$ \\[3pt]

    100 & ESC
    & $-8.910\times10^{-1}$ & $-9.700\times10^{2}$
    & $-3.834$ & $-9.661\times10^{2}$ \\
        & Bulk
    & $-1.132\times10^{-1}$ & $-1.232\times10^{2}$
    & $-3.519$ & $-1.197\times10^{2}$ \\
        & Total
    & $-1.004$ & $-1.093\times10^{3}$
    & $-7.354$ & $-1.086\times10^{3}$
  \end{tabular}
  \caption{Solvent work integrated over the ESC, bulk, and full domain
  along the selected low-wavenumber branch. The first work column is normalized by the magnitude of the full-domain electric work $\mathcal{W}_e$; the remaining columns are normalized by the full-domain velocity measure $K_u$.}
  \label{tab:lowk_solvent_work}
  \end{center}
\end{table}

\subsection{Robustness of the non-monotonic response across parameters}

\begin{figure}
  \centerline{\includegraphics[scale=0.5]{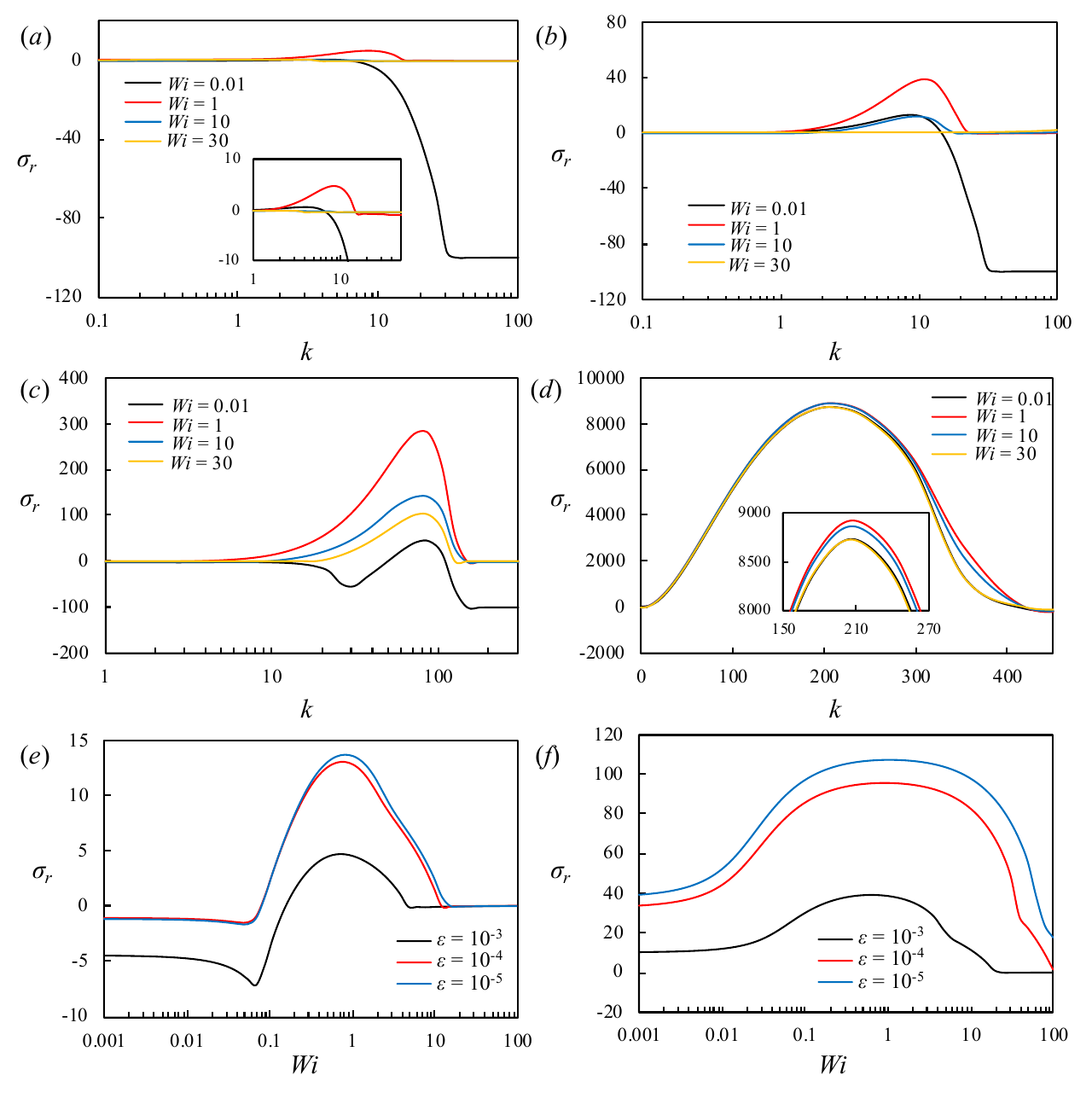}}
  \caption{Effect of the Debye length on the electroconvective instability
  at $Pe=10$ and $\beta=0.5$. The left and right columns correspond to
  $V=30$ and $V=40$, respectively. ($a,b$) Growth rate as a function of
  wavenumber at $\varepsilon=10^{-3}$ for representative values of $Wi$.
  ($c,d$) Corresponding results at $\varepsilon=10^{-5}$. The inset in ($a,d$) provides a magnified view
  near the maximum growth rate.
  ($e,f$) Growth rate as a function of $Wi$ at $k=10$ for
  $\varepsilon=10^{-3}$, $10^{-4}$, and $10^{-5}$. Decreasing
  $\varepsilon$ increases the growth rate and shifts the unstable modes
  toward larger wavenumbers. At both applied potentials, the growth rate
  exhibits a maximum at $Wi\sim O(1)$ for all Debye lengths considered,
  demonstrating that the non-monotonic viscoelastic response is robust
  to variations in $\varepsilon$.}
  \label{fig:fig_rev12}
\end{figure}

The analysis presented above shows that the non-monotonic dependence of the growth rate on the Weissenberg number is associated with systematic changes in the magnitude, phase relation, and spatial distribution of polymer-stress and solvent work across the different $Wi$ regimes. We now examine whether this non-monotonic response persists under variations in the system parameters.

The Debye length controls the thickness of the space charge layer and thus the spatial extent over which electric forcing is localized. As $\varepsilon$ decreases (i.e., increasing ion concentration), the ESC becomes thinner and the electric field becomes more strongly concentrated near the wall. This leads to a stronger localization of the perturbation charge density and electric body force, resulting in larger instability growth rates. These trends are observed at both $V=30$ and $V=40$
(figure~\ref{fig:fig_rev12}$a$--$d$).

Despite these changes in spatial structure, the growth rate consistently exhibits a maximum at $Wi\sim O(1)$ for all values of $\varepsilon$ considered at both applied potentials (figure~\ref{fig:fig_rev12}$e,f$). Therefore, the Debye length modifies the spatial localization of the
instability but does not alter its non-monotonic dependence on $Wi$.

\begin{figure}
  \centerline{\includegraphics[scale=0.5]{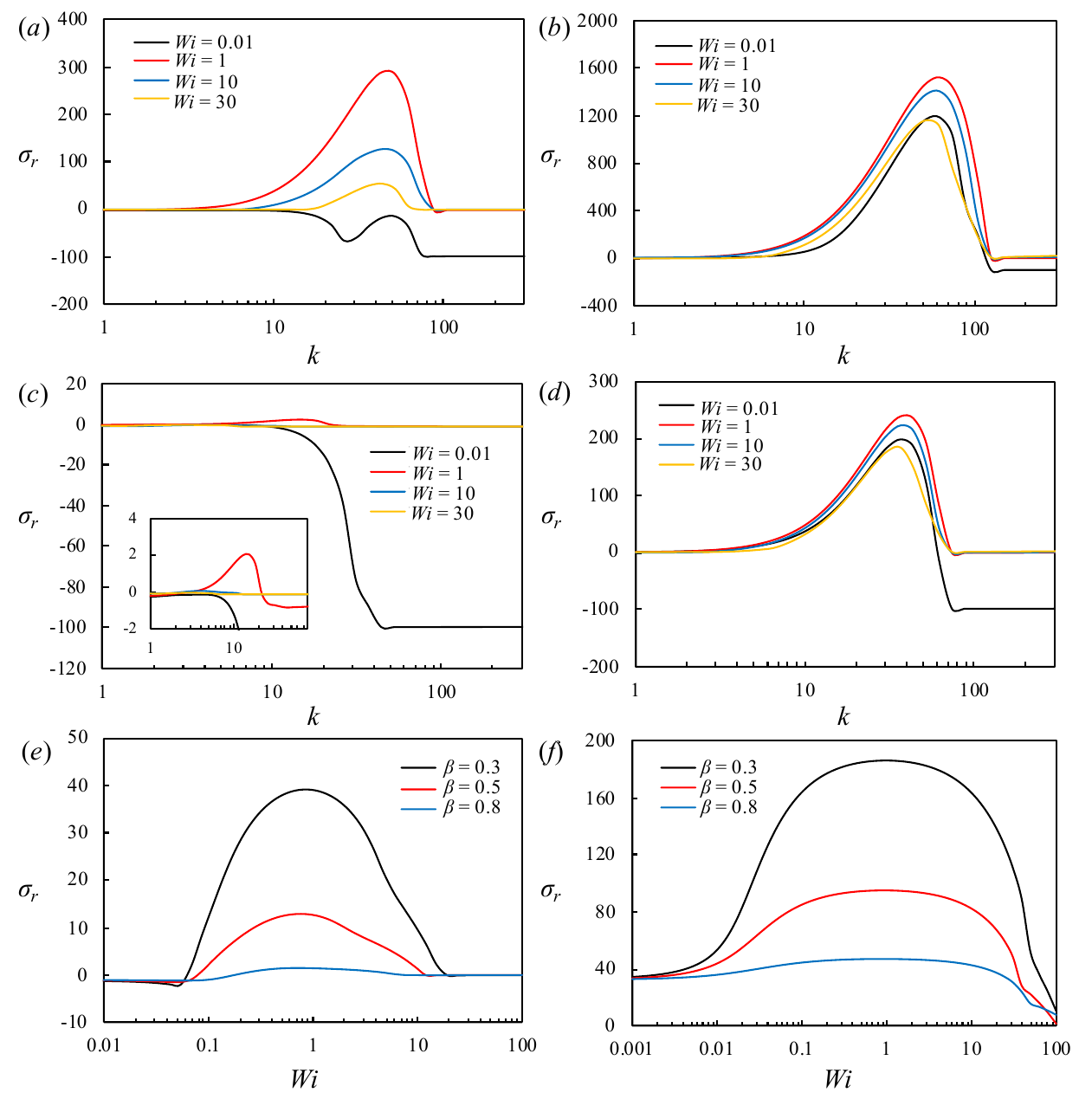}}
  \caption{Effect of the viscosity ratio on the electroconvective
  instability at $Pe=10$ and $\varepsilon=10^{-4}$. The left and right
  columns correspond to $V=30$ ($Ra_E=105$) and $V=40$ ($Ra_E=187$),
  respectively. ($a,b$) Growth rate as a function of wavenumber at
  $\beta=0.3$ for representative values of $Wi$. ($c,d$) Corresponding
  results at $\beta=0.8$. The inset in ($c$) provides a magnified view
  near the maximum growth rate. ($e,f$) Growth rate as a function of
  $Wi$ at $k=10$ for $\beta=0.3$, $0.5$, and $0.8$. In all cases, the
  growth rate reaches a maximum at $Wi\sim O(1)$, demonstrating that the
  non-monotonic viscoelastic response persists across the viscosity
  ratios and applied potentials considered.}
  \label{fig:fig_rev13}
\end{figure}

The viscosity ratio $\beta$ controls the relative contribution of
polymeric stress to the total stress. As $\beta$ decreases, the polymeric
contribution increases, leading to stronger viscoelastic effects and
larger magnitudes of the polymeric stress. Accordingly, decreasing
$\beta$ substantially increases the growth rate and broadens the range of
unstable wavenumbers at both $V=30$ and $V=40$
(figure~\ref{fig:fig_rev13}$a$--$d$).

Despite these quantitative differences, the growth rate reaches its
maximum at $Wi\sim O(1)$ for all values of $\beta$ at both applied
potentials (figure~\ref{fig:fig_rev13}$e,f$). The growth rate increases
toward this intermediate-$Wi$ maximum and subsequently decreases as $Wi$
increases further. Thus, $\beta$ controls the magnitude of the
viscoelastic modification but does not eliminate the non-monotonic
dependence of the instability on $Wi$.

\begin{figure}
  \centerline{\includegraphics[scale=0.5]{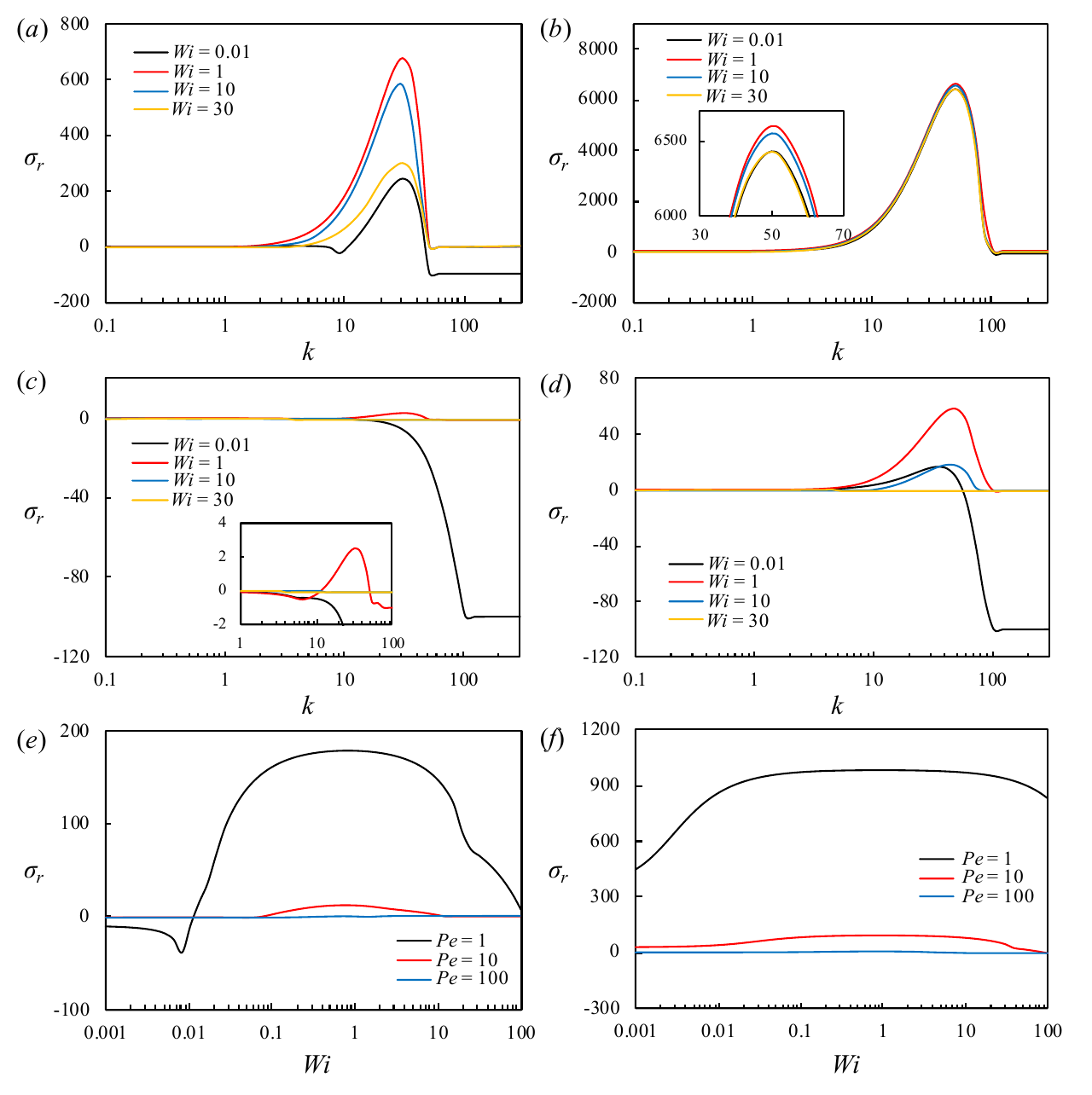}}
  \caption{Effect of the Peclet number on the electroconvective
  instability at $\beta=0.5$ and $\varepsilon=10^{-4}$. The left and
  right columns correspond to $V=30$ ($Ra_E=105$) and $V=40$
  ($Ra_E=187$), respectively. ($a,b$) Growth rate as a function of
  wavenumber at $Pe=1$ for representative values of $Wi$.
  ($c,d$) Corresponding results at $Pe=100$. The insets in ($b$) and
  ($c$) provide magnified views near the growth-rate maxima.
  ($e,f$) Growth rate as a function of $Wi$ at $k=10$ for $Pe=1$, $10$,
  and $100$. Increasing $Pe$ substantially reduces the instability
  growth rate, but the intermediate-$Wi$ enhancement persists over the
  range of Peclet numbers and applied potentials considered.}
  \label{fig:fig_rev14}
\end{figure}

The Peclet number $Pe$ characterizes the relative importance of advective
to diffusive transport. As $Pe$ increases, convective transport becomes
more dominant, which tends to suppress the instability \citep{Li2021}.
At both $V=30$ and $V=40$, increasing $Pe$ substantially reduces the
growth rate and narrows the range of unstable wavenumbers
(figure~\ref{fig:fig_rev14}$a$--$d$). As a result, stronger convective
effects require a higher applied potential to produce a comparable
instability growth rate, reflecting the increased stabilizing influence
of convection.

Despite these quantitative changes, the growth rate retains its
non-monotonic dependence on $Wi$
(figure~\ref{fig:fig_rev14}$e,f$). The intermediate-$Wi$ enhancement is
pronounced at smaller and moderate $Pe$ and becomes substantially weaker
as $Pe$ increases. Thus, $Pe$ strongly modifies the magnitude and
critical conditions of the instability but does not eliminate its
non-monotonic response to viscoelasticity.

Across all parameters considered---Debye length, viscosity ratio, and
Peclet number---variations in these parameters modify the spatial
localization, characteristic wavenumber, magnitude, and critical
conditions of the instability. Despite these changes, the growth rate
consistently remains non-monotonic in $Wi$ and reaches a maximum at
$Wi\sim O(1)$, confirming that this viscoelastic response extends beyond
the reference parameter set.

\section{Conclusions}

In this study, we investigated the viscoelastic modulation of
electroconvective instability using linear stability analysis,
asymptotic pathway ordering, eigenspectrum analysis, and a
mechanical-work-rate formulation. The polymeric work was separated into
normal-stress-difference and shear-stress contributions and further
decomposed into their constitutive pathways. This framework distinguishes
the stress magnitude, its phase relation with the corresponding
deformation, and the spatial region in which mechanical power is
transferred.

The linear stability results show that the growth rate is maximized and
the critical voltage is minimized near $Wi\sim O(1)$. The instability
remains rooted in ESC-localized electrohydrodynamics. At $k=10$, the
unstable eigenvalue originates from the wall-localized A-mode branch of
the Y-shaped eigenspectrum. As $Wi$ increases, this mode separates from
the branch, reaches its maximum growth rate near $Wi\sim O(1)$, and
subsequently migrates toward smaller growth rates before becoming stable
at large $Wi$. 

The mechanical-work analysis identifies a transition in the polymeric
response across the different $Wi$ regimes. For $Wi\ll1$, the polymer
stress recovers Newtonian viscous dissipation at leading order. The
negative polymeric work is dominated by the mechanically effective
$N3$ and $S2$ pathways, which extract mechanical power from the
perturbation. As $Wi$ approaches unity, their phase efficiencies decrease
substantially, causing the strongly negative small-$Wi$ polymeric work to
collapse. This reduction coincides with the minimum in the critical
voltage near $Wi\sim O(1)$.

At large $Wi$, the dominant pathway structure changes. The normal work is governed primarily by the opposing $N1$ and $N4$ contributions, and the shear work is increasingly controlled by $S4$. For the $k=10$ mode, the negative normal work and positive shear work nearly cancel, leaving a net polymeric contribution that is small compared with the electric work. The subsequent increase in the critical voltage therefore cannot be attributed to a recovery of net polymeric dissipation. Instead, the reorganization of the polymeric-work balance occurs together with the development of stronger ESC-localized shear deformation relative to the overall velocity magnitude. These two changes arise within the same fully coupled eigenmode, in which $Wi$ simultaneously modifies the velocity-gradient and polymer-stress fields that determine the solvent and polymeric work, respectively. Solvent viscosity consequently accounts for nearly all of the mechanical-power extraction at the large-$Wi$ neutral states. A dominant polymeric pathway therefore identifies the largest work contribution within this coupled response, not an independent forcing mechanism or the sole cause of the increase in the critical voltage.

The low-wavenumber modes reveal an additional dependence on eigenmode
selection. Over $3\leq Wi\leq30$, the dip in the neutral curve is
associated with a slowly propagating mode near $\sigma_i=0$. For this
mode, negative shear work exceeds the positive normal work, and the
polymer stress remains a substantial net sink of mechanical power. This
balance differs from that of the large-$Wi$, $k=10$ mode and is accompanied
by relatively weak solvent deformation relative to the overall velocity magnitude. At
$Wi\geq50$, the slowly propagating mode no longer controls the
instability threshold, the faster mode originating from the A-mode branch
is selected again, and the low-wavenumber dip disappears. This branch
change coincides with negligible net polymeric work and a sharp increase
in shear deformation relative to the overall velocity magnitude, leaving solvent
viscosity to balance nearly all of the electric-work input. These results
demonstrate that pathway work depends strongly on the selected
eigenspectral branch and cannot be inferred from stress magnitude alone.

The non-monotonic dependence on $Wi$ persists under variations in Debye
length, viscosity ratio, and Peclet number and at both applied potentials
examined. These parameters modify the spatial localization,
characteristic wavenumber, magnitude, and critical conditions of the
instability. Despite these changes, the growth rate consistently reaches
a maximum near $Wi\sim O(1)$. No critical viscosity ratio is observed
within the investigated range $\beta=0.3$--$0.8$, although decreasing
$\beta$ amplifies the variation in growth rate. Whether
this behaviour persists in the polymer-dominated limit
$\beta\rightarrow0$, where solvent-viscous regularization vanishes,
requires a separate stability analysis.

Finally, the large-$Wi$ increase in $V_{\mathrm{cr}}$ should be interpreted as a higher threshold for the linear instability of the quiescent concentration-polarization state, not as a general suppression of electroconvection. The post-onset dynamics may depart from this linear trend once finite-amplitude electroconvective vortices have developed. Nonlinear simulations have shown that polymer elasticity can destabilize an established electroconvective circulation and alter its time-averaged ion transport \citep{Li2019}. Extending the present full-domain formulation to nonlinear direct numerical simulations would therefore be valuable for determining whether the large-$Wi$ linear restabilization persists at finite amplitude and how post-onset flow
transitions affect overlimiting transport.

Overall, the present results establish that viscoelastic
electroconvection is governed by the coupled evolution of an
ESC-localized instability, polymeric-work pathways, solvent deformation,
and eigenspectral branch selection.

\section*{Appendix A. Numerical method and validation of the linear stability analysis}
This appendix summarizes the numerical implementation and validates the linear stability formulation through comparisons with published results and a grid convergence test.

\subsection*{A.1 Numerical method}
The numerical implementation follows the finite-difference procedure described by ~\citet{Kim2026}. 

The steady one-dimensional Nernst–Planck and Poisson equations are discretized on a non-uniform wall-normal grid generated using a tangent–hyperbolic mapping. The grid points are concentrated near $y=0$ and $2$ to resolve the electric double layers and the ESC. Second-order central finite differences are applied on this grid. The resulting nonlinear base-state equations are solved using Newton iteration with a numerically assembled Jacobian until the maximum relative residual is less than $10^{-6}$.

The linearized perturbation equations are discretized on the same non-uniform grid using second-order central finite differences. After imposing the perturbation boundary conditions, the discrete equations are assembled into the generalized eigenvalue problem
\begin{equation}
\boldsymbol{L}\hat{\boldsymbol{q}}
=
\sigma\boldsymbol{M}\hat{\boldsymbol{q}},
\end{equation}
where $\hat{\boldsymbol{q}}$ contains the discretized velocity, ion-concentration, electric-potential, and conformation perturbations. The complete eigenspectrum and corresponding eigenvectors are computed using \texttt{scipy.linalg.eig} in Python~\citep{Virtanen2020}.

To establish the accuracy and robustness of the linear stability formulation, comparisons with previous studies and a grid-convergence test are performed. The electrokinetic formulation in the absence of imposed flow was previously validated by~\citet{Kim2026} through comparisons of neutral stability curves and perturbation eigenfunctions with the reference solutions of~\citet{Li2022}. The validation here provides three additional tests: (i) validation against the Newtonian limit of electroconvective instability, (ii) validation against inertialess viscoelastic Poiseuille flow without electric-field effects, and (iii) grid convergence of the complete viscoelastic electroconvective eigenspectrum.

First, electroconvective instability for Newtonian fluids was validated in our recent work~\citep{Kim2026} through comparisons of neutral stability curves and perturbation eigenfunctions with the reference solutions reported by~\citet{Li2022}, showing excellent agreement. This confirms the correctness of the electrokinetic formulation in the absence of viscoelastic effects.

Next, the viscoelastic formulation is further validated through comparisons with (i) the Newtonian limit of electroconvective instability ($Wi \rightarrow 0$ and $\beta \rightarrow 1$), (ii) inertialess viscoelastic Poiseuille flow without electric-field effects, and (iii) grid convergence tests.

\begin{figure}
  \centerline{\includegraphics[scale=0.188]{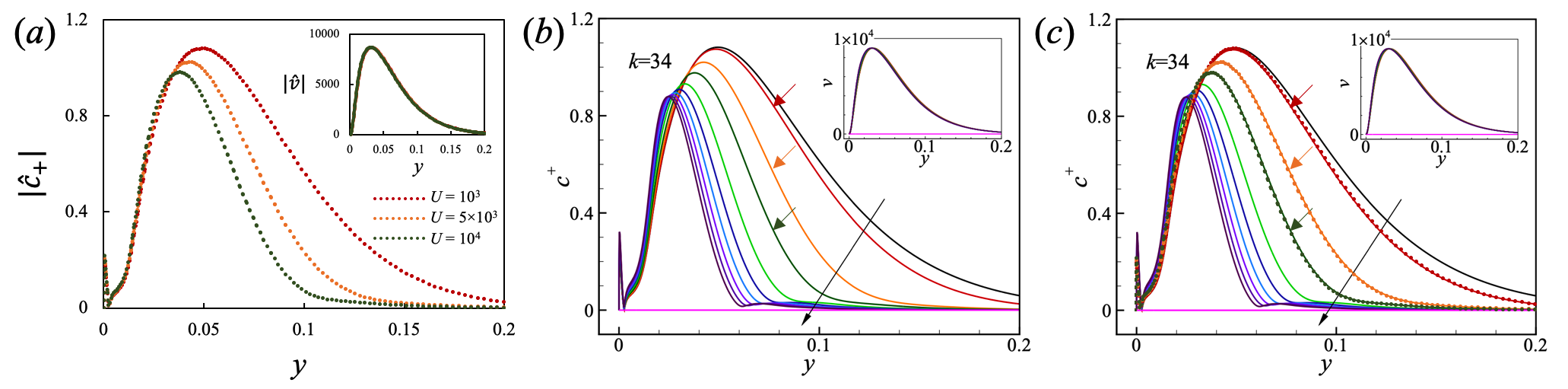}} 
  \caption{Validation of the Newtonian limit through comparison of perturbation eigenfunctions with the results of~\citet{Li2021}. ($a$) Eigenfunctions of the perturbed cation concentration obtained from the present formulation at $V=25$, $\varepsilon=10^{-4}$, $k=34$ and $\beta=0.999$ for $U=10^3, 5\times 10^3$, and $10^4$. The inset shows the corresponding perturbed normal velocity eigenfunctions. ($b$) Corresponding eigenfunctions reported by ~\citet{Li2021}. The arrows denote the cases $U=10^3$ (dark red), $5 \times 10^3$ (orange), and $10^4$ (dark green) for clarity. ($c$) Direct superposition of the present and published results on common axes. Close agreement is observed for all cases, confirming that the present formulation correctly reproduces the Newtonian electroconvective instability in the limit $Wi \rightarrow 0$ and $\beta \rightarrow 1$.
}
\label{fig:figli}
\end{figure}

\subsection*{A.2 Newtonian-limit validation of electroconvective instability}
To validate the Newtonian limit of the present formulation, perturbation eigenfunctions are compared with the results reported by~\citet{Li2021}. Following the nondimensionalization adopted by~\citet{Li2021}, the governing equations are rescaled using $H^2/D$ as the characteristic time scale $t_c$ and $\delta V_T^2/(\mu D)$ as the characteristic velocity scale $U_0$. The comparison is performed at $V=25$, $\varepsilon =10^{-4}$, $k=34$, $Pe=0.35$, and $D = 2/3$ for three imposed-flow conditions: $U=10^3$, $5\times10^3$, and $10^4$. Figure~\ref{fig:figli} compares the perturbation eigenfunctions obtained from the present formulation at $Wi_0 = \lambda U_0/H = 0.0001$ and $\beta = 0.999$ (figure~\ref{fig:figli}$a$) with those reported by ~\citet{Li2021} (figure~\ref{fig:figli}$b$). Their direct superposition in figure~\ref{fig:figli}($c$) shows close agreement for all three imposed-flow conditions, confirming that the present formulation correctly reproduces the Newtonian electroconvective instability in the limit $Wi \rightarrow 0$ and $\beta \rightarrow 1$.

\subsection*{A.3 Viscoelastic eigenspectrum validation}

To validate the viscoelastic formulation independently of electrokinetic effects, the electric-field terms are removed from the governing equations, and the resulting eigenspectrum is compared with the inertialess viscoelastic Poiseuille-flow results reported by~\citet{Khalid2021a}.

\begin{figure}
  \centerline{\includegraphics[scale=0.189]{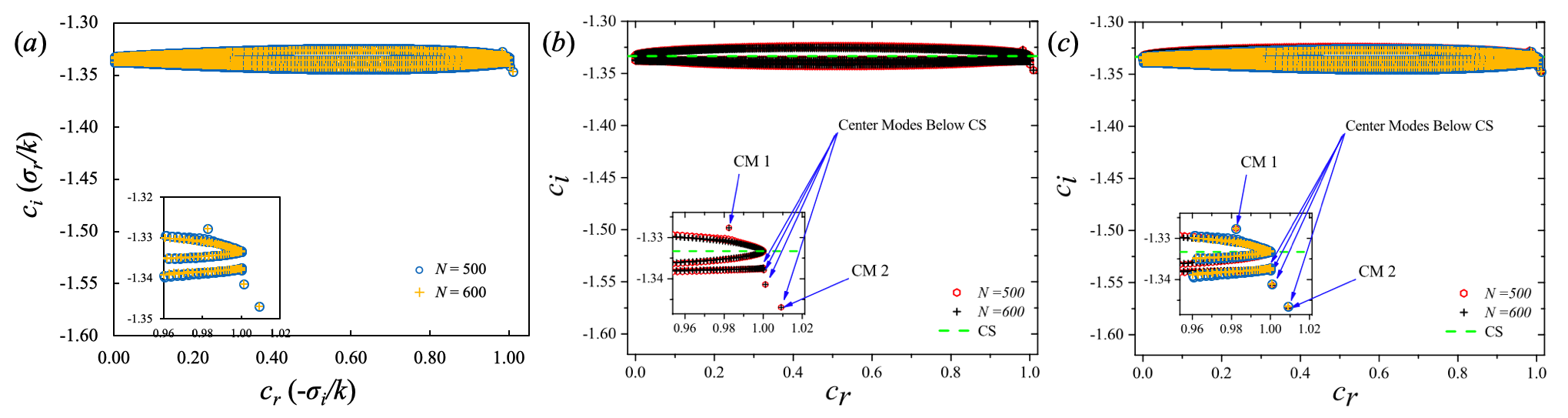}} 
  \caption{Comparison of the viscoelastic eigenspectrum at $Wi=1$, $\beta = 0.997$ and $k=0.75$: ($a$) present finite-difference formulation with wall clustering, ($b$) Chebyshev spectral results reproduced from ~\citet{Khalid2021a}, and ($c$) Direct superposition of the two eigenspectra on common axes. The overall branch structure, including the center-mode branches and characteristic bulging region, shows close agreement between the two formulations.
}
\label{fig:figkhalid}
\end{figure}

The overall structure of the eigenspectrum at $Wi=1$, $\beta=0.997$, and $k=0.75$, including its characteristic branches and their distribution in the complex plane, shows close agreement with the reference solution. Figure~\ref{fig:figkhalid}($a$) presents the eigenspectrum obtained from the present formulation, figure~\ref{fig:figkhalid}($b$) shows the corresponding spectrum reported by~\citet{Khalid2021a}, and figure~\ref{fig:figkhalid}($c$) directly superimposes the two eigenspectra on common axes.

A direct comparison requires accounting for the difference in temporal formulation. In the present study, perturbations are expressed as $\hat{q}(y)e^{ikx+\sigma t}$, whereas the reference study used the form $\hat{q}(y)e^{ik(x-ct)}$, where $c$ is the complex eigenvalue whose real part $c_r$ represents the wave speed and whose imaginary part $c_i$ determines the temporal growth rate through $kc_i=\sigma_r$. Under this transformation, the imaginary part of the eigenvalue in the reference formulation corresponds to the real part of the eigenvalue in the present formulation.

In both results, two nearly horizontal branches at constant $c_i$ are observed, corresponding to the continuous-spectrum eigenvalues
\begin{equation}
c_i=\frac{\sigma_r}{k}=-\frac{1}{kWi}
\quad \text{and} \quad
-\frac{1}{\beta kWi},
\end{equation}
associated with inertialess viscoelastic plane Poiseuille flow~\citep{Khalid2021a}. In addition, several discrete center modes are observed near the continuous spectrum, consistent with the previously reported eigenspectrum structure.

Furthermore, a characteristic bulging structure of the spectrum is observed in both computations despite the use of different numerical methods (Chebyshev spectral discretization in the reference study and finite-difference discretization with wall clustering in the present work). This feature has previously been attributed to numerical artifacts associated with discretization of the continuous spectrum, and its appearance in both computations is consistent with that interpretation.

\begin{figure}
  \centerline{\includegraphics[scale=0.34]{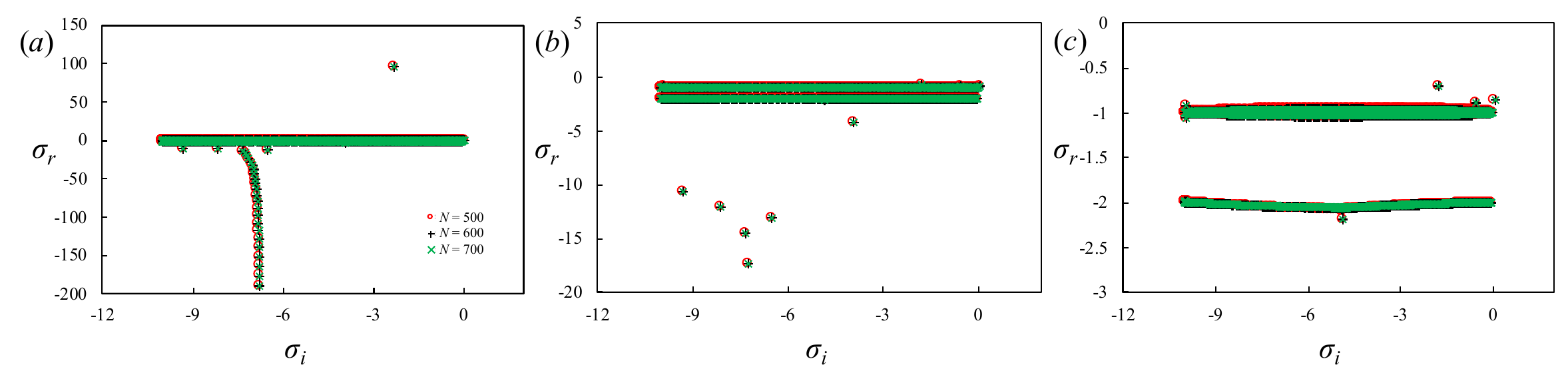}} 
  \caption{Grid convergence test of the viscoelastic eigenspectrum at $Pe=10$, $\beta=0.5$, $\varepsilon =10^{-4}$, $Ra_E=187$ ($V=40$), and $k=10$ using $N = 500, 600$, and $700$: ($a$) overall eigenspectrum showing the most unstable mode and Y-shaped branch structure, ($b$) magnified view near the Y-branch bifurcation region, and ($c$) enlarged view of the continuous-spectrum eigenvalue branches. The eigenvalue distributions obtained using different grid resolutions show excellent agreement, confirming numerical convergence of the present formulation.
}
\label{fig:figmesh}
\end{figure}

\subsection*{A.4 Grid convergence test}

To verify the numerical convergence of the computed eigenspectrum, a grid refinement study is performed at $Pe=10$, $\beta=0.5$, $\varepsilon =10^{-4}$, $Ra_E=187$ ($V=40$), and $k=10$ using multiple grid resolutions ($N=500$, $600$, and $700$).

The resulting eigenspectra exhibit nearly identical eigenvalue distributions across all grid resolutions, showing that the solution is insensitive to further grid refinement. Figure~\ref{fig:figmesh}($a$) compares the overall eigenspectrum, including the most unstable mode and the characteristic Y-shaped branch structure. Figure~\ref{fig:figmesh}($b$) provides a magnified view near the Y-branch bifurcation region, while figure~\ref{fig:figmesh}($c$) shows a further magnification of the continuous-spectrum eigenvalue branches.

In all regions of the spectrum, the eigenvalues obtained using different grid resolutions collapse onto the same distributions, confirming the numerical convergence of the present formulation.

\section*{Appendix B. Derivation of the mechanical-work and constitutive-pathway relations}
\label{app:work_derivation}

\subsection*{B.1 Complex-amplitude convention}

For a perturbation quantity written as
\begin{equation}
q'(x,y,t)=\hat q(y)\exp(ikx+\sigma t),
\end{equation}
the streamwise derivative is represented by $ik$. Complex conjugation acts on this Fourier factor as well as on the amplitude:
\begin{equation}
(ik\hat q)^*=-ik\hat q^*.
\label{eq:app_complex_conjugation}
\end{equation}
The time- and streamwise-averaged product of two real perturbation fields contains a common factor $1/2$. Because this factor multiplies every term in the mechanical-power balance, it is omitted throughout.

The perturbation continuity equation and its complex conjugate are
\begin{subequations}
\begin{align}
ik\hat u+D\hat v&=0,\\
-ik\hat u^*+D\hat v^*&=0.
\end{align}
\label{eq:app_incompressibility}
\end{subequations}
Consequently,
\begin{equation}
D\hat v=-ik\hat u,
\qquad
D\hat v^*=ik\hat u^*.
\label{eq:app_incompressibility_identities}
\end{equation}

\subsection*{B.2 Componentwise mechanical-power balance}

For any symmetric stress perturbation $\hat{\boldsymbol\tau}$, the two components of its divergence are
\begin{equation}
\bnabla_k\bcdot\hat{\boldsymbol\tau}
=
\begin{pmatrix}
ik\hat\tau_{xx}+D\hat\tau_{xy}\\
ik\hat\tau_{xy}+D\hat\tau_{yy}
\end{pmatrix}.
\label{eq:app_stress_divergence}
\end{equation}
Multiplication by $\widehat{\boldsymbol u}^{*}=(\hat u^*,\hat v^*)$ and integration over $y_1\leq y\leq y_2$ gives
\begin{align}
\Real\int_{y_1}^{y_2}
\widehat{\boldsymbol u}^{*}\bcdot
(\bnabla_k\bcdot\hat{\boldsymbol\tau})\,\mathrm{d}y
={}&
\Real\int_{y_1}^{y_2}
\left[
ik\hat u^*\hat\tau_{xx}
+\hat u^*D\hat\tau_{xy}
+ik\hat v^*\hat\tau_{xy}
+\hat v^*D\hat\tau_{yy}
\right]\mathrm{d}y \nonumber\\
={}&
\Real\left[
\hat u^*\hat\tau_{xy}
+\hat v^*\hat\tau_{yy}
\right]_{y_1}^{y_2}
-\Real\int_{y_1}^{y_2}
(\bnabla_k\widehat{\boldsymbol u})^*:
\hat{\boldsymbol\tau}\,\mathrm{d}y.
\label{eq:app_stress_integration}
\end{align}
The first term is the work rate exerted by traction at the boundaries of the selected interval, while the second is internal stress--deformation work.

For pressure,
\begin{align}
W_{\mathrm{pr}}
&=-\Real\int_{y_1}^{y_2}
\left(ik\hat u^*\hat p+\hat v^*D\hat p\right)\mathrm{d}y \nonumber\\
&=-\Real[\hat v^*\hat p]_{y_1}^{y_2}
+\Real\int_{y_1}^{y_2}
\left(D\hat v^*-ik\hat u^*\right)\hat p\,\mathrm{d}y \nonumber\\
&=-\Real[\hat v^*\hat p]_{y_1}^{y_2},
\label{eq:app_pressure_work}
\end{align}
where the last equality follows from equation~\eqref{eq:app_incompressibility}.

For the solvent stress $\hat{\boldsymbol\tau}_s=2\beta\hat{\mathsfbi E}$,
\begin{equation}
W_{\mathrm{sol}}^{\mathrm{int}}
=-2\beta\int_{y_1}^{y_2}
\hat{\mathsfbi E}^*:\hat{\mathsfbi E}\,\mathrm{d}y.
\label{eq:app_solvent_internal_tensor}
\end{equation}
The non-zero components of the strain-rate tensor are
\begin{equation}
\hat E_{xx}=ik\hat u,\qquad
\hat E_{yy}=D\hat v,\qquad
\hat E_{xy}=\hat E_{yx}=\frac{1}{2}(D\hat u+ik\hat v),
\end{equation}
and hence
\begin{equation}
W_{\mathrm{sol}}^{\mathrm{int}}
=-\beta\int_{y_1}^{y_2}
\left[
2k^2|\hat u|^2
+2|D\hat v|^2
+|D\hat u+ik\hat v|^2
\right]\mathrm{d}y
\leq0.
\label{eq:app_solvent_internal_components}
\end{equation}
The corresponding surface work is
\begin{equation}
W_{\mathrm{sol}}^{\mathrm{surf}}
=\beta\Real\left[
\hat u^*(D\hat u+ik\hat v)
+2\hat v^*D\hat v
\right]_{y_1}^{y_2}.
\label{eq:app_solvent_surface}
\end{equation}

Applying equation~\eqref{eq:app_stress_integration} to
$\hat{\boldsymbol\tau}_p=(1-\beta)\hat{\mathsfbi A}/Wi$ gives
\begin{align}
W_p^{\mathrm{surf}}
&=\frac{1-\beta}{Wi}
\Real\left[
\hat u^*\hat A_{xy}
+\hat v^*\hat A_{yy}
\right]_{y_1}^{y_2},
\label{eq:app_polymer_surface}\\
W_p^{\mathrm{int}}
&=\Real\int_{y_1}^{y_2}
\left[
ik\hat u^*\hat\tau_{p,xx}
-(D\hat u)^*\hat\tau_{p,xy}
+ik\hat v^*\hat\tau_{p,xy}
-(D\hat v)^*\hat\tau_{p,yy}
\right]\mathrm{d}y.
\label{eq:app_polymer_internal_components}
\end{align}

The linearized electric force is
\begin{equation}
\hat{\boldsymbol f}_e
=\frac{Ra_E}{Pe}
\left[
(D^2\bar\phi)\bnabla_k\hat\phi
+(D^2-k^2)\hat\phi\,D\bar\phi\,\boldsymbol e_y
\right],
\end{equation}
so that
\begin{align}
W_e
=\frac{Ra_E}{Pe}\Real\int_{y_1}^{y_2}
\big[
&ik\hat u^*(D^2\bar\phi)\hat\phi
+\hat v^*(D^2\bar\phi)D\hat\phi \nonumber\\
&+\hat v^*(D^2-k^2)\hat\phi\,D\bar\phi
\big]\mathrm{d}y.
\label{eq:app_electric_work}
\end{align}
The electric contribution is retained as body-force work and is not integrated into separate electric surface and internal terms.

Adding equations~\eqref{eq:app_pressure_work}, \eqref{eq:app_solvent_internal_tensor}, \eqref{eq:app_polymer_surface}, \eqref{eq:app_polymer_internal_components}, and \eqref{eq:app_electric_work} recovers equation~\eqref{eq:mechanical_work_balance}. These identities hold for any interval. At the no-slip walls, the surface terms vanish. At an internal ESC--bulk interface, equal and opposite tractions give equal and opposite regional surface-work contributions.

\subsection*{B.3 Normal--shear decomposition of the polymer work}

Using $D\hat v^*=ik\hat u^*$ from equation~\eqref{eq:app_incompressibility_identities} in equation~\eqref{eq:app_polymer_internal_components}, the two normal-stress terms become
\begin{equation}
ik\hat u^*\hat\tau_{p,xx}
-(D\hat v)^*\hat\tau_{p,yy}
=ik\hat u^*
(\hat\tau_{p,xx}-\hat\tau_{p,yy}).
\end{equation}
Furthermore,
\begin{equation}
-(D\hat u)^*\hat\tau_{p,xy}
+ik\hat v^*\hat\tau_{p,xy}
=-(D\hat u-ik\hat v)^*\hat\tau_{p,xy}
=-\hat\gamma^*\hat\tau_{p,xy}.
\end{equation}
Therefore,
\begin{equation}
W_p^{\mathrm{int}}
=\Real\int_{y_1}^{y_2}
\left[
ik\hat u^*(\hat\tau_{p,xx}-\hat\tau_{p,yy})
-\hat\gamma^*\hat\tau_{p,xy}
\right]\mathrm{d}y
=W_N+W_S.
\label{eq:app_normal_shear_decomposition}
\end{equation}

\subsection*{B.4 Derivation of the constitutive pathways}

With $\Lambda=\sigma+ik\bar{U}+Wi^{-1}$, equations~\eqref{eq:perturbAxx}--\eqref{eq:perturbAyy} may be written as
\begin{subequations}
\begin{align}
\Lambda\hat A_{xx}
&=2g\hat A_{xy}
+2\bar A_{xy}D\hat u
-2\bar A_{xx}D\hat v
-\hat vD\bar A_{xx},
\\
\Lambda\hat A_{xy}
&=g\hat A_{yy}
+\bar A_{yy}D\hat u
-\hat vD\bar A_{xy}
+\bar A_{xx}ik\hat v,
\\
\Lambda\hat A_{yy}
&=2\bar A_{xy}ik\hat v
+2\bar A_{yy}D\hat v.
\end{align}
\label{eq:app_conformation_components}
\end{subequations}
Subtracting the third equation from the first yields
\begin{align}
\Lambda(\hat A_{xx}-\hat A_{yy})
={}&2g\hat A_{xy}
+2\bar A_{xy}(D\hat u-ik\hat v)\nonumber\\
&-2(\bar A_{xx}+\bar A_{yy})D\hat v
-\hat vD\bar A_{xx},
\end{align}
which gives $N_1$--$N_4$ in equation~\eqref{eq:normal_pathways}. The second equation directly gives $S_1$--$S_4$ in equation~\eqref{eq:shear_pathways}. Division by $\Lambda$, followed by multiplication by $(1-\beta)/Wi$, produces the pathway stresses in equation~\eqref{eq:pathway_stresses}.

\subsection*{B.5 Small-\texorpdfstring{$Wi$}{Wi} expansion}

For $Wi|\Sigma|\ll1$,
\begin{equation}
\Lambda^{-1}
=Wi-Wi^2\Sigma+Wi^3\Sigma^2+O(Wi^4).
\label{eq:app_small_response}
\end{equation}
The base conformation and its gradients are
\begin{gather}
\bar A_{xx}=1+2Wi^2g^2,\qquad
\bar A_{xy}=Wig,\qquad
\bar A_{yy}=1,\\
D\bar A_{xx}=4Wi^2gg',\qquad
D\bar A_{xy}=Wig',
\qquad g'=Dg=-2.
\label{eq:app_base_conformation}
\end{gather}
The wall-normal conformation perturbation follows directly from equation~\eqref{eq:app_conformation_components}:
\begin{align}
\hat A_{yy}
&=\frac{2Wig\,ik\hat v+2D\hat v}{\Lambda}\nonumber\\
&=2WiD\hat v
+Wi^2\left(2gik\hat v-2\Sigma D\hat v\right)
+O(Wi^3).
\label{eq:app_small_Ayy}
\end{align}
Substitution into the shear-conformation equation gives
\begin{align}
\hat A_{xy}
={}&Wi\hat\gamma
+Wi^2\left[
-\Sigma\hat\gamma
+2gD\hat v
-g'\hat v
\right]
+O(Wi^3)\nonumber\\
={}&Wi\hat\gamma
+Wi^2\left[
-\Sigma\hat\gamma
+2gD\hat v
+2\hat v
\right]
+O(Wi^3).
\label{eq:app_small_Axy}
\end{align}
For completeness, the streamwise component is
\begin{equation}
\hat A_{xx}
=-2WiD\hat v
+Wi^2\left[
2\Sigma D\hat v
+2g\hat\gamma
+2gD\hat u
\right]
+O(Wi^3).
\label{eq:app_small_Axx}
\end{equation}

The normal pathway sources are consequently
\begin{subequations}
\begin{align}
N_1&=2gWi\hat\gamma+O(Wi^2),\\
N_2&=2gWi(D\hat u-ik\hat v),\\
N_3&=-4D\hat v-4Wi^2g^2D\hat v,\\
N_4&=-4Wi^2gg'\hat v.
\end{align}
\label{eq:app_small_normal_sources}
\end{subequations}
Multiplication by $(1-\beta)/(Wi\Lambda)$ gives
\begin{subequations}
\begin{align}
\hat\tau_N^{(1)}
&=2(1-\beta)gWi\hat\gamma+O(Wi^2),\\
\hat\tau_N^{(2)}
&=2(1-\beta)gWi(D\hat u-ik\hat v)+O(Wi^2),\\
\hat\tau_N^{(3)}
&=(1-\beta)\left[-4D\hat v+4Wi\Sigma D\hat v\right]+O(Wi^2),\\
\hat\tau_N^{(4)}
&=-4(1-\beta)Wi^2gg'\hat v+O(Wi^3).
\end{align}
\label{eq:app_small_normal_stresses}
\end{subequations}
Thus, $N_3$ is leading, the $N_3$ response correction and $N_1/N_2$ are first subleading, and $N_4$ is second subleading.

The corresponding shear sources are
\begin{subequations}
\begin{align}
S_1&=2gWiD\hat v+O(Wi^2),\\
S_2&=D\hat u,\\
S_3&=-Wig'\hat v,\\
S_4&=ik\hat v+2Wi^2g^2ik\hat v.
\end{align}
\label{eq:app_small_shear_sources}
\end{subequations}
Hence,
\begin{subequations}
\begin{align}
\hat\tau_S^{(1)}
&=2(1-\beta)gWiD\hat v+O(Wi^2),\\
\hat\tau_S^{(2)}
&=(1-\beta)\left[D\hat u-Wi\Sigma D\hat u\right]+O(Wi^2),\\
\hat\tau_S^{(3)}
&=-(1-\beta)Wig'\hat v+O(Wi^2),\\
\hat\tau_S^{(4)}
&=(1-\beta)\left[ik\hat v-Wi\Sigma ik\hat v\right]+O(Wi^2).
\end{align}
\label{eq:app_small_shear_stresses}
\end{subequations}
Adding these expressions gives
\begin{subequations}
\begin{align}
\hat\tau_{p,xx}-\hat\tau_{p,yy}
&=4(1-\beta)ik\hat u
+4(1-\beta)Wi
\left[-\Sigma ik\hat u+gD\hat u\right]
+O(Wi^2),\\
\hat\tau_{p,xy}
&=(1-\beta)\hat\gamma
+(1-\beta)Wi
\left[-\Sigma\hat\gamma+2gD\hat v-g'\hat v\right]
+O(Wi^2).
\end{align}
\label{eq:app_small_total_stresses}
\end{subequations}
Substitution into equation~\eqref{eq:app_normal_shear_decomposition} yields the leading Newtonian-like work densities in equation~\eqref{eq:small_Wi_work}.

\subsection*{B.6 Large-\texorpdfstring{$Wi$}{Wi} expansion}

For $Wi|\Sigma|\gg1$,
\begin{equation}
\Lambda^{-1}
=\frac{1}{\Sigma}
-\frac{1}{Wi\Sigma^2}
+\frac{1}{Wi^2\Sigma^3}
+O(Wi^{-3}).
\label{eq:app_large_response}
\end{equation}
Equation~\eqref{eq:app_conformation_components} first gives
\begin{equation}
\hat A_{yy}
=\frac{2gik}{\Sigma}Wi\hat v
+\left(
\frac{2D\hat v}{\Sigma}
-\frac{2gik\hat v}{\Sigma^2}
\right)
+O(Wi^{-1}),
\label{eq:app_large_Ayy}
\end{equation}
and subsequent substitution into the shear equation gives
\begin{equation}
\hat A_{xy}
=\frac{2g^2ik}{\Sigma}Wi^2\hat v
-\frac{g'}{\Sigma}Wi\hat v
+O(1)
=\frac{2g^2ik}{\Sigma}Wi^2\hat v
+\frac{2}{\Sigma}Wi\hat v
+O(1).
\label{eq:app_large_Axy}
\end{equation}

The leading normal sources are
\begin{subequations}
\begin{align}
N_1
&=\frac{4g^3ik}{\Sigma}Wi^2\hat v+O(Wi),\\
N_2
&=2gWi(D\hat u-ik\hat v),\\
N_3
&=-4g^2Wi^2D\hat v+O(1),\\
N_4
&=-4gg'Wi^2\hat v.
\end{align}
\label{eq:app_large_normal_sources}
\end{subequations}
Therefore,
\begin{subequations}
\begin{align}
\hat\tau_N^{(1)}
&=(1-\beta)Wi\frac{4g^3ik}{\Sigma^2}\hat v+O(1),\\
\hat\tau_N^{(2)}
&=2(1-\beta)\frac{g}{\Sigma}(D\hat u-ik\hat v)+O(Wi^{-1}),\\
\hat\tau_N^{(3)}
&=-4(1-\beta)Wi\frac{g^2}{\Sigma}D\hat v+O(1),\\
\hat\tau_N^{(4)}
&=-4(1-\beta)Wi\frac{gg'}{\Sigma}\hat v+O(1).
\end{align}
\label{eq:app_large_normal_stresses}
\end{subequations}
The leading total normal-stress difference is thus
\begin{equation}
\hat\tau_{p,xx}-\hat\tau_{p,yy}
=(1-\beta)Wi
\left[
\frac{4g^3ik}{\Sigma^2}\hat v
-\frac{4g^2}{\Sigma}D\hat v
-\frac{4gg'}{\Sigma}\hat v
\right]
+O(1).
\label{eq:app_large_total_normal}
\end{equation}

For the shear sources,
\begin{subequations}
\begin{align}
S_1
&=\frac{2g^2ik}{\Sigma}Wi\hat v+O(1),\\
S_2&=D\hat u,\\
S_3&=-Wig'\hat v,\\
S_4&=2g^2ikWi^2\hat v+ik\hat v.
\end{align}
\label{eq:app_large_shear_sources}
\end{subequations}
Their stress contributions satisfy
\begin{subequations}
\begin{align}
\hat\tau_S^{(1)}
&=2(1-\beta)\frac{g^2ik}{\Sigma^2}\hat v+O(Wi^{-1}),\\
\hat\tau_S^{(2)}
&=(1-\beta)\frac{D\hat u}{Wi\Sigma}+O(Wi^{-2}),\\
\hat\tau_S^{(3)}
&=-(1-\beta)\frac{g'}{\Sigma}\hat v+O(Wi^{-1}),\\
\hat\tau_S^{(4)}
&=2(1-\beta)Wi\frac{g^2ik}{\Sigma}\hat v+O(1).
\end{align}
\label{eq:app_large_shear_stresses}
\end{subequations}
Thus,
\begin{equation}
\hat\tau_{p,xy}
=2(1-\beta)Wi\frac{g^2ik}{\Sigma}\hat v+O(1),
\label{eq:app_large_total_shear}
\end{equation}
with $S_1$ and $S_3$ supplying the first subleading response and $S_2$ entering at $O(Wi^{-1})$.

\section*{Appendix C. Tables of pathway phase efficiencies}
\label{app:work_efficiency}

\begin{table}
  \begin{center}
  \resizebox{\textwidth}{!}{%
  \begin{tabular}{ccrrrrrrrrrr}
    $Wi$ & Region
    & $C_N$ & $C_{N1}$ & $C_{N2}$ & $C_{N3}$ & $C_{N4}$
    & $C_S$ & $C_{S1}$ & $C_{S2}$ & $C_{S3}$ & $C_{S4}$ \\[3pt]

    0.01 & ESC
    & $-0.986$ & $-0.014$ & $-0.006$ & $-1.000$ & $ 1.000$
    & $-1.000$ & $-0.041$ & $-1.000$ & $-0.009$ & $ 0.230$ \\
         & Bulk
    & $-0.998$ & $ 0.052$ & $ 0.023$ & $-1.000$ & $ 0.026$
    & $-0.999$ & $ 0.030$ & $-0.939$ & $ 0.016$ & $-0.056$ \\
         & Total
    & $-0.992$ & $ 0.007$ & $ 0.006$ & $-1.000$ & $ 0.106$
    & $-1.000$ & $-0.017$ & $-0.997$ & $ 0.014$ & $-0.029$ \\[3pt]

    0.1 & ESC
    & $-0.767$ & $-0.561$ & $-0.332$ & $-0.738$ & $ 0.741$
    & $-0.735$ & $-0.519$ & $-0.732$ & $ 0.170$ & $ 0.185$ \\
        & Bulk
    & $-0.552$ & $ 0.959$ & $ 0.496$ & $-0.794$ & $-0.109$
    & $-0.719$ & $ 0.897$ & $-0.717$ & $ 0.043$ & $-0.115$ \\
        & Total
    & $-0.667$ & $-0.049$ & $ 0.013$ & $-0.770$ & $-0.046$
    & $-0.734$ & $-0.013$ & $-0.731$ & $ 0.054$ & $-0.089$ \\[3pt]

    1 & ESC
    & $-0.403$ & $-0.236$ & $-0.466$ & $-0.104$ & $ 0.082$
    & $-0.100$ & $-0.040$ & $-0.102$ & $ 0.248$ & $ 0.173$ \\
      & Bulk
    & $ 0.311$ & $ 0.497$ & $ 0.607$ & $-0.135$ & $-0.440$
    & $-0.053$ & $ 0.344$ & $-0.078$ & $-0.176$ & $-0.057$ \\
      & Total
    & $-0.069$ & $ 0.287$ & $ 0.026$ & $-0.123$ & $-0.416$
    & $-0.095$ & $ 0.181$ & $-0.101$ & $-0.147$ & $-0.036$ \\[3pt]

    3 & ESC
    & $-0.204$ & $-0.110$ & $-0.430$ & $-0.028$ & $-0.014$
    & $ 0.008$ & $ 0.032$ & $-0.028$ & $ 0.303$ & $ 0.251$ \\
      & Bulk
    & $ 0.168$ & $ 0.517$ & $ 0.564$ & $-0.032$ & $-0.520$
    & $-0.012$ & $-0.068$ & $-0.003$ & $ 0.098$ & $-0.010$ \\
      & Total
    & $-0.014$ & $ 0.461$ & $ 0.011$ & $-0.030$ & $-0.491$
    & $-0.002$ & $-0.039$ & $-0.026$ & $ 0.117$ & $ 0.020$ \\[3pt]

    10 & ESC
    & $-0.058$ & $ 0.039$ & $-0.350$ & $-0.006$ & $-0.062$
    & $ 0.227$ & $-0.145$ & $-0.007$ & $ 0.395$ & $ 0.330$ \\
       & Bulk
    & $ 0.044$ & $ 0.572$ & $ 0.529$ & $-0.007$ & $-0.570$
    & $-0.027$ & $-0.450$ & $ 0.012$ & $ 0.418$ & $-0.026$ \\
       & Total
    & $-0.008$ & $ 0.520$ & $ 0.003$ & $-0.006$ & $-0.526$
    & $ 0.026$ & $-0.390$ & $-0.006$ & $ 0.415$ & $ 0.029$ \\[3pt]

    30 & ESC
    & $-0.015$ & $ 0.098$ & $-0.258$ & $-0.002$ & $-0.103$
    & $ 0.373$ & $-0.396$ & $-0.003$ & $ 0.487$ & $ 0.385$ \\
       & Bulk
    & $-0.002$ & $ 0.591$ & $ 0.489$ & $-0.002$ & $-0.581$
    & $-0.070$ & $-0.581$ & $ 0.016$ & $ 0.537$ & $-0.070$ \\
       & Total
    & $-0.010$ & $ 0.517$ & $-0.002$ & $-0.002$ & $-0.515$
    & $ 0.026$ & $-0.541$ & $-0.001$ & $ 0.527$ & $ 0.026$
  \end{tabular}%
  }
  \caption{Integrated phase efficiencies of the combined normal- and
  shear-stress contributions and their individual pathways over the ESC,
  bulk, and full domain for the $k=10$ mode. Positive and negative values
  denote net mechanical-power input and extraction, respectively.}
  \label{tab:k10_phase_efficiency}
  \end{center}
\end{table}

\begin{table}
  \begin{center}
  \resizebox{\textwidth}{!}{%
  \begin{tabular}{cccrrrrrrrrrr}
    $Wi$ & $k$ & Region
    & $C_N$ & $C_{N1}$ & $C_{N2}$ & $C_{N3}$ & $C_{N4}$
    & $C_S$ & $C_{S1}$ & $C_{S2}$ & $C_{S3}$ & $C_{S4}$ \\[3pt]

    0.01 & 5 & ESC
    & $-0.959$ & $ 0.002$ & $ 0.014$ & $-1.000$ & $ 1.000$
    & $-1.000$ & $-0.064$ & $-1.000$ & $-0.062$ & $-0.133$ \\
    & & Bulk
    & $-0.998$ & $ 0.059$ & $ 0.027$ & $-1.000$ & $ 0.229$
    & $-0.999$ & $ 0.024$ & $-0.957$ & $ 0.015$ & $-0.003$ \\
    & & Total
    & $-0.988$ & $ 0.024$ & $ 0.020$ & $-1.000$ & $ 0.251$
    & $-1.000$ & $-0.030$ & $-0.998$ & $ 0.012$ & $-0.009$ \\[3pt]

    0.1 & 5 & ESC
    & $-0.668$ & $-0.340$ & $-0.178$ & $-0.961$ & $ 0.960$
    & $-0.960$ & $-0.415$ & $-0.958$ & $-0.088$ & $-0.118$ \\
    & & Bulk
    & $-0.814$ & $ 0.555$ & $ 0.223$ & $-0.975$ & $ 0.202$
    & $-0.926$ & $ 0.429$ & $-0.930$ & $ 0.126$ & $ 0.009$ \\
    & & Total
    & $-0.753$ & $ 0.003$ & $ 0.009$ & $-0.972$ & $ 0.223$
    & $-0.959$ & $-0.080$ & $-0.958$ & $ 0.118$ & $ 0.004$ \\[3pt]

    1 & 5 & ESC
    & $-0.660$ & $-0.400$ & $-0.631$ & $-0.194$ & $ 0.171$
    & $-0.191$ & $-0.155$ & $-0.191$ & $-0.073$ & $ 0.185$ \\
    & & Bulk
    & $ 0.296$ & $ 0.610$ & $ 0.736$ & $-0.290$ & $-0.568$
    & $-0.064$ & $ 0.302$ & $-0.202$ & $ 0.053$ & $-0.036$ \\
    & & Total
    & $-0.137$ & $ 0.311$ & $ 0.086$ & $-0.268$ & $-0.558$
    & $-0.184$ & $ 0.123$ & $-0.192$ & $ 0.050$ & $-0.026$ \\[3pt]

    3 & 5 & ESC
    & $-0.928$ & $-0.568$ & $-0.640$ & $-0.892$ & $ 0.398$
    & $-0.763$ & $-0.330$ & $-0.800$ & $ 0.502$ & $-0.159$ \\
    & & Bulk
    & $ 0.293$ & $ 0.668$ & $-0.801$ & $-0.768$ & $ 0.495$
    & $ 0.100$ & $ 0.265$ & $-0.846$ & $ 0.030$ & $ 0.049$ \\
    & & Total
    & $-0.103$ & $ 0.496$ & $-0.716$ & $-0.792$ & $ 0.506$
    & $-0.506$ & $-0.015$ & $-0.802$ & $ 0.070$ & $ 0.025$ \\[3pt]

    10 & 3.5 & ESC
    & $-0.258$ & $ 0.493$ & $-0.623$ & $-0.964$ & $ 0.939$
    & $-0.639$ & $-0.160$ & $-0.860$ & $ 0.582$ & $-0.620$ \\
    & & Bulk
    & $ 0.462$ & $ 0.501$ & $-0.369$ & $-0.470$ & $ 0.288$
    & $-0.127$ & $-0.175$ & $-0.588$ & $-0.234$ & $-0.072$ \\
    & & Total
    & $ 0.072$ & $ 0.499$ & $-0.547$ & $-0.591$ & $ 0.311$
    & $-0.441$ & $-0.164$ & $-0.855$ & $-0.110$ & $-0.191$ \\[3pt]

    30 & 2.5 & ESC
    & $ 0.272$ & $ 0.689$ & $ 0.156$ & $-0.710$ & $ 0.626$
    & $-0.660$ & $-0.135$ & $-0.937$ & $ 0.087$ & $-0.754$ \\
    & & Bulk
    & $ 0.083$ & $ 0.185$ & $-0.986$ & $-0.185$ & $ 0.137$
    & $-0.030$ & $-0.944$ & $-0.178$ & $-0.568$ & $ 0.081$ \\
    & & Total
    & $ 0.213$ & $ 0.436$ & $ 0.049$ & $-0.439$ & $ 0.172$
    & $-0.468$ & $-0.201$ & $-0.936$ & $-0.372$ & $-0.258$ \\[3pt]

    50 & 2 & ESC
    & $-0.047$ & $ 0.068$ & $-0.462$ & $-0.005$ & $-0.082$
    & $ 0.398$ & $-0.043$ & $-0.005$ & $ 0.228$ & $ 0.468$ \\
    & & Bulk
    & $ 0.010$ & $ 0.567$ & $ 0.621$ & $-0.005$ & $-0.516$
    & $-0.005$ & $-0.716$ & $-0.003$ & $ 0.628$ & $-0.004$ \\
    & & Total
    & $-0.016$ & $ 0.526$ & $-0.001$ & $-0.005$ & $-0.488$
    & $ 0.044$ & $-0.636$ & $-0.005$ & $ 0.595$ & $ 0.046$ \\[3pt]

    70 & 2 & ESC
    & $-0.031$ & $ 0.090$ & $-0.417$ & $-0.003$ & $-0.098$
    & $ 0.451$ & $-0.143$ & $-0.003$ & $ 0.285$ & $ 0.484$ \\
    & & Bulk
    & $-0.0005$ & $ 0.563$ & $ 0.624$ & $-0.003$ & $-0.508$
    & $-0.025$ & $-0.721$ & $-0.001$ & $ 0.632$ & $-0.025$ \\
    & & Total
    & $-0.016$ & $ 0.514$ & $-0.003$ & $-0.003$ & $-0.473$
    & $ 0.040$ & $-0.644$ & $-0.003$ & $ 0.596$ & $ 0.041$ \\[3pt]

    100 & 2 & ESC
    & $-0.019$ & $ 0.111$ & $-0.371$ & $-0.002$ & $-0.115$
    & $ 0.482$ & $-0.241$ & $-0.002$ & $ 0.342$ & $ 0.496$ \\
    & & Bulk
    & $-0.012$ & $ 0.564$ & $ 0.626$ & $-0.002$ & $-0.504$
    & $-0.048$ & $-0.724$ & $ 0.000$ & $ 0.635$ & $-0.048$ \\
    & & Total
    & $-0.016$ & $ 0.503$ & $-0.005$ & $-0.002$ & $-0.461$
    & $ 0.037$ & $-0.649$ & $-0.002$ & $ 0.597$ & $ 0.038$
  \end{tabular}%
  }
  \caption{Integrated phase efficiencies of the combined normal- and
  shear-stress contributions and their individual pathways over the
  ESC, bulk, and full domain for the representative neutral
  low-wavenumber modes.}
  \label{tab:low_k_phase}
  \end{center}
\end{table}


\begin{thebibliography}{}
\expandafter\ifx\csname natexlab\endcsname\relax
\def\natexlab#1{#1}\fi
\expandafter\ifx\csname selectlanguage\endcsname\relax
\def\selectlanguage#1{\relax}\fi

\bibitem[Alkhadra $et$ $al$. (2022)] {Alkhadra2022}
 {\sc Alkhadra, M.A., Su,X., Suss, M.E., Tian, H., Guyes, E.N., Shocron, A.N., Conforti, K.M., de Souza, J.P., Kim, N., Tedesco, M. $et$ $al$.} 2022 {Electrochemical methods for water purification, ion separations, and energy conversion.} {\it Chem. Rev.} {\bf122} (16), 13547--13635.

\bibitem[Alves $et$ $al$. (2016)] {Alves2016}
 {\sc Alves, L.S.B., Hirata, S.C. \& Ouarzazi, M.N.} 2016 {Linear onset of convective instability for Rayleigh-Bénard-Couette flows of viscoelastic fluids.} {\it J. Non-Newton. Fluid Mech.} {\bf231}, 70--90.


\bibitem[Benzi $et$ $al$. (2010)] {Benzi2010}
 {\sc Benzi, R., Ching, E.S.C. \& De Angelis, E.} 2010 {Effect of polymer additives on heat transport in turbulent thermal convection.} {\it Phys. Rev. Lett.} {\bf104}, 024502.

\bibitem[Benzi $et$ $al$. (2012)] {Benzi2012}
 {\sc Benzi, R., Ching, E.S.C. \& Chu, V.W.S.} 2012 {Heat transport by laminar boundary layer flow with polymers.} {\it J. Fluid Mech.} {\bf696}, 330--344.

\bibitem[Benzi $et$ $al$. (2016)] {Benzi2016}
 {\sc Benzi, R., Ching, E.S.C. \& De Angelis, E.} 2016 {Turbulent Rayleigh-Bénard convection with polymers: Understanding how heat flux is modified.} {\it Phys. Rev. E} {\bf94}, 063110.

\bibitem[Cai $et$ $al$. (2023)]{Cai2023}
{\sc Cai, J., Zhao, D. \& Li, G.} 2023 {Direct numerical simulation of electroconvection under a uniform magnetic field.} {\it Phys. Rev. Fluids} {\bf 8}, 113701.

\bibitem[Demekhin $et$ $al$. (2013)] {Demekhin2013}
 {\sc Demekhin, E.A., Nikitin, N.V. \& Shelistov, V.S.} 2013 {Direct numerical simulation of electrokinetic instability and transition to chaotic motion.} {\it Phys. Fluids} {\bf25}, 122001.

 \bibitem[Dubief \& Terrapon (2020)] {Dubief2020}
 {\sc Dubief, Y. \& Terrapon, V.E.} 2020 {Heat transfer enhancement and reduction in low-Rayleigh number natural convection flow with polymer additives.} {\it Phys. Fluids} {\bf32}, 033103.

\bibitem[Dydek $et$ $al$. (2011)] {Dydek2011}
 {\sc Dydek, E.V., Zaltzman, B., Rubinstein, I., Deng, D.S., Mani, A. \& Bazant, M.Z.} 2011 {Overlimiting current in a microchannel.} {\it Phys. Rev. Lett.} {\bf107}, 118301.
 
\bibitem[Fleury $et$ $al$. (1993)] {Fleury1993}
 {\sc Fleury, V., Chazalviel, J.N. \& Rosso, M.} 1993 {Coupling of drift, diffusion, and electroconvection, in the vicinity of growing electrodeposits.} {\it Phys. Rev. E} {\bf48} (2), 1279.

\bibitem[Garg $et$ $al$. (2018)] {Garg2018}
 {\sc Garg, P., Chaudhary, I., Khalid, M, Shankar, V. \& Subramanian, G.} 2018 {Viscoelastic pipe flow is linearly unstable.} {\it Phys. Rev. Lett} {\bf121}, 024502.

\bibitem[Han $et$ $al$. (2016)] {Han2016}
 {\sc Han, S.I., Hwang, K.S., Kwak, R. \& Lee, J.H.} 2016 {Microfluidic paper-based biomolecule preconcentrator based on ion concentration polarization.} {\it Lab Chip} {\bf16} (12), 2219--2227.

\bibitem[Huth $et$ $al$. (1995)] {Huth1995}
 {\sc Huth, J.M., Swinney, H.L., McCormick, W.D., Kuhn, A. \& Argoul, F.} 1995 {Role of convection in thin-layer electrodeposition.} {\it Phys. Rev. E} {\bf51} (4), 3444.

\bibitem[Khalid $et$ $al$. (2021{\it a})] {Khalid2021a}
 {\sc  Khalid, M, Shankar, V. \& Subramanian, G.} 2021{\it a} {Continuous pathway between the elasto-inertial and elastic turbulent states in viscoelastic channel flow.} {\it Phys. Rev. Lett.} {\bf127}, 134502.

 \bibitem[Khalid $et$ $al$. (2021{\it b})] {Khalid2021b}
 {\sc  Khalid, M, Chaudhary, I., Garg, P., Shankar, V. \& Subramanian, G.} 2021{\it b} {The centre-mode instability of viscoelastic plane Poiseuille flow.} {\it J. Fluid Mech.} {\bf915}, A43.

  \bibitem[Khalid $et$ $al$. (2025)] {Khalid2025}
 {\sc  Khalid, M, Badoni, A., Dutta, D., Naidu, P., Subramanian, G. \& Shankar, V.} 2025 {Role of finite extensibility on the centre-mode instability in viscoelastic channel flow.} {\it J. Fluid Mech.} {\bf1009}, A28.

 \bibitem[Kim $et$ $al$. (2016)] {Kim2016}
 {\sc Kim, B., Kwak, R., Kwon, H.J., Pham, V.S., Kim, M., Al-Anzi, B. \& Han, J.} 2016 {Purification of high salinity brine by multi-stage ion concentration polarization desalination.} {\it Sci. Rep.} {\bf6}, 31850.

 \bibitem[Kim \& Kwak (2026)] {Kim2026}
 {\sc Kim, M. \& Kwak, R.} 2026 {Electroconvective instability near an ion-selective membrane: Scaling law from extended space charge dynamics.} {\it J. Membr. Sci.} {\bf738}, 124846.

\bibitem[Kim $et$ $al$. (2010)] {Kim2010}
 {\sc Kim, S.J., Ko, S.H., Kang, K.H. \& Han, J.} 2010 {Direct seawater desalination by ion concentration polarization.} {\it Nat. Nanotechnol.} {\bf5} (4), 297--301.

 \bibitem[Kwak $et$ $al$. (2011)] {Kwak2011}
 {\sc Kwak, R., Kim, S.J. \& Han, J.} 2011 {Continuous-flow biomolecule and cell concentrator by ion concentration polarization.} {\it Anal. Chem.} {\bf83} (19), 7348--7355.

\bibitem[Kwak $et$ $al$. (2013)] {Kwak2013}
 {\sc Kwak, R., Guan, G., Peng, W.K. \& Han, J.} 2013 {Microscale electrodialysis: Concentration profiling and vortex visualization.} {\it Desalination} {\bf308}, 138--146.

 \bibitem[Kwak $et$ $al$. (2016)] {Kwak2016}
 {\sc Kwak, R., Kang, J.Y. \& Kim, T.S.} 2016 {Spatiotemporally defining biomolecule preconcentration by merging ion concentration polarization.} {\it Anal. Chem.} {\bf88} (1), 988--996.

\bibitem[Li $et$ $al$. (2019)] {Li2019}
 {\sc Li, G., Archer, L.A. \& Koch, D.L.} 2019 {Electroconvection in a Viscoelastic Electrolyte.} {\it Phys. Rev. Lett.} {\bf122}, 124501.

 \bibitem[Li $et$ $al$. (2021)] {Li2021}
 {\sc Li, G., Townsend, A., Archer, L.A. \& Koch, D.L.} 2021 {Suppression of electroconvective and morphological instabilities by an imposed cross flow of the electrolyte.} {\it Phys. Rev. Fluids} {\bf6}, 033701.

\bibitem[Li $et$ $al$. (2022)] {Li2022}
 {\sc Li, G., Townsend, A., Archer, L.A. \& Koch, D.L.} 2022 {Electroconvection near an ion-selective surface with Butler–Volmer kinetics.} {\it J. Fluid Mech.} {\bf930}, A26.

\bibitem[Mani \& Wang (2020)] {Mani2020}
 {\sc Mani, A. \& Wang, K.M.} 2020 {Electroconvection near electrochemical interfaces: experiments, modeling, and computation.} {\it Annu. Rev. Fluid Mech.} {\bf52}, 509--529.

 \bibitem[Martinez-Mardones $et$ $al$. (2000)] {Martinez-Mardones2000}
 {\sc Martinez-Mardones, J., Tiemann, R. \& Walgraef, D.} 2000 {Rayleigh–Bénard convection in binary viscoelastic fluid.} {\it Physica A} {\bf283}, 233--236.

\bibitem[Nikonenko $et$ $al$. (2010)] {Nikonenko2010}
 {\sc Nikonenko V.V., Pismenskaya, N.D., Belova, E.I., Sistat, P., Huguet, P., Pourcelly, G. \& Larchet, C.} 2010 {Intensive current transfer in membrane systems: Modelling, mechanisms and application in electrodialysis.} {\it Adv. Colloid Interface Sci} {\bf160}, 101--123.

\bibitem[Nikonenko $et$ $al$. (2016)] {Nikonenko2016}
 {\sc Nikonenko V.V., Vasil’eva, V.I., Akberova, E.M., Uzdenova, A.M., Urtenov, M.K., Kovalenko, A.V., Pismenskaya, N.P., Mareev, S.A. \& Pourcelly, G.} 2016 {Competition between diffusion and electroconvection at an ion-selective surface in intensive current regimes.} {\it Adv. Colloid Interface Sci.} {\bf235}, 233--246.

\bibitem[Qin \& Bau (2011)]{Qin2011}
{\sc Qin, M. \& Bau, H.H.} 2011 {When MHD-based microfluidics is equivalent to pressure-driven flow.} {\it Microfluid. Nanofluid.} {\bf 10}, 287--300.
 
\bibitem[Rubinstein \& Zaltzman (2000)] {Rubinstein2000}
 {\sc Rubinstein, I. \& Zaltzman, B.} 2000 {Electro-osmotically induced convection at a permselective membrane.} {\it Phys. Rev. E} {\bf62}, 2238--2251.

\bibitem[Rubinstein \& Zaltzman (2001)] {Rubinstein2001}
 {\sc Rubinstein, I. \& Zaltzman, B.} 2001 {Electro-osmotic slip of the second kind and instability in concentration polarization at electrodialysis membranes.} {\it Math. Models Methods Appl. Sci.} {\bf11}, 263--300.

\bibitem[Rubinstein \& Zaltzman (2010{\it a})] {Rubinstein2010a}
 {\sc Rubinstein, I. \& Zaltzman, B.} 2010{\it a} {Extended space charge in concentration polarization.} {\it Adv. Colloid Interface Sci.} {\bf159}, 117--129.

\bibitem[Rubinstein \& Zaltzman (2010{\it b})] {Rubinstein2010b}
 {\sc Rubinstein, I. \& Zaltzman, B.} 2010{\it b} {Dynamics of extended space charge in concentration polarization.} {\it Phys. Rev. E} {\bf81}, 061502.

\bibitem[Schmid \& Henningson (2001)] {Schmid2001}
 {\sc Schmid, P.J. \& Henningson, D.S.} 2001 {\it Stability and Transition in Shear Flows.} {Applied Mathematical Sciences} {\bf142} Springer Science \& Business Media. Berlin.

\bibitem[Sharma $et$ $al$.(2023)] {Sharma2023}
 {\sc Sharma, A., Mukherjee, A., Warren, A., Jin, S., Li, G., Koch, D.L. \& Archer, L.A.} 2023 {Electroconvective flow in liquid electrolytes containing oligomer additives.} {\it Langmuir} {\bf39} (1), 92--100.

\bibitem[Sim $et$ $al$.(2024)] {Sim2024}
 {\sc Sim, G., Pishnamazi, M., Seo, D., Kong, S.R., Lee, J., Park, Y. \& Chae, S.R.} 2024 {Recent advances in electrodialysis technologies for recovering critical minerals from unconventional sources.} {\it Chem. Eng. J.} {\bf497} (3), 154640.
 
\bibitem[Strathmann (2010)] {Strathmann2010}
 {\sc Strathmann, H.} 2010 {Electrodialysis, a mature technology with a multitude of new applications.} {\it Desalination} {\bf264} (3), 268--288.

\bibitem[Tikekar $et$ $al$. (2018)] {Tikekar2018}
 {\sc Tikekar, M.D., Li, G., Archer, L.A. \& Koch, D.L.} 2018 {Electroconvection and morphological instabilities in potentiostatic electrodeposition across liquid electrolytes with polymer additives.} {\it J. Electrochem. Soc.} {\bf165}, A3697.

\bibitem[Urtenov $et$ $al$.(2013)] {Urtenov2013}
 {\sc Urtenov, M.K., Uzdenova, A.M., Kovalenko, A.V., Nikonenko, V.V., Pismenskaya, N.D., Vasil’eva, V.I., Sistat, P.  \& Pourcelly, G.} 2013 {Basic mathematical model of overlimiting transfer enhanced by electroconvection in flow-through electrodialysis membrane cells.} {\it J. Membr. Sci.} {\bf447}, 190--202.

\bibitem[Virtanen $et$ $al$. (2020)]{Virtanen2020}
{\sc Virtanen, P., Gommers, R., Oliphant, T.E., Haberland, M., Reddy, T.,
Cournapeau, D., Burovski, E., Peterson, P., Weckesser, W., Bright, J.
$et$ $al$.} 2020 {SciPy 1.0: Fundamental algorithms for scientific
computing in Python.} {\it Nat. Methods} {\bf17}, 261--272.

\bibitem[Wei $et$ $al$. (2012)] {Wei2012}
 {\sc Wei, P., Ni, R. \& Xia, K.-Q.} 2012 {Enhanced and reduced heat transport in turbulent thermal convection with polymer additives.} {\it Phys. Rev. E} {\bf86}, 016325.

\bibitem[Xie $et$ $al$. (2015)] {Xie2015}
 {\sc Xie, Y.-C., Huang, S.-D., Funfschilling, D., Li, X.-M., Ni, R. \& Xia, K.-Q.} 2015 {Effects of polymer additives in the bulk of turbulent thermal convection.} {\it J. Fluid Mech.} {\bf784}, R3.

 \bibitem[Xu $et$ $al$. (2025{\it a})] {Xu2025a}
 {\sc Xu, F., Liu, X.-S., Li, X.-M. \& Xia, K.-Q.} 2025{\it a} {Restoration of axisymmetric flow structure in turbulent thermal convection by polymer additives.} {\it Phys. Rev. Lett.} {\bf134}, 084001.

\bibitem[Xu $et$ $al$. (2025{\it b})] {Xu2025b}
 {\sc Xu, C., Zhang, C., Brandt, L., Song, J. \& Shishkina, O.} 2025{\it b} {Direct numerical simulations of turbulent Rayleigh--Bénard convection with polymer additives.} {\it J. Fluid Mech.} {\bf1014}, A22.

\bibitem[Yao $et$ $al$. (2025)] {Yao2025}
 {\sc Yao, Z., Lu, C., Zhang, M., Luo, K. \& Li, H.} 2025 {Non-modal analysis of Rayleigh–Bénard convection with and without bounded shear flows in viscoelastic fluids.} {\it J. Fluid Mech.} {\bf1011}, A37.
 
\bibitem[Zaltzman \& Rubinstein (2007)] {Zaltzman2007}
 {\sc Zaltzman, B. \& Rubinstein, I.} 2007 {Electro-osmotic slip and electroconvective instability.} {\it J. Fluid Mech.} {\bf579}, 173--226.

\bibitem[Zheng \& Archer (2021)] {Zheng2021}
 {\sc Zheng, J. \& Archer, L.A.} 2021 {Controlling electrochemical growth of metallic zinc electrodes: Toward affordable rechargeable energy storage systems.} {\it Sci. Adv.} {\bf7} (2), eabe0219.

 


\end{thebibliography}



\end{document}